\documentclass[11pt]{article}
\usepackage[top=1in, bottom=1.25in, left=1.25in, right=1.25in]{geometry}
\usepackage{amsmath, amsthm, amsfonts, amssymb, amscd, bm, enumerate}

\usepackage{amsmath, amsfonts, amssymb, mathrsfs}

\usepackage{bm} 
\usepackage{cool} 
\usepackage{mathtools}
\usepackage{cases}
\usepackage{bbm}

\counterwithin{equation}{section}
\makeatletter 
\providecommand*{\diff}%
	{\@ifnextchar^{\DIfF}{\DIfF^{}}}
\def\DIfF^#1{%
	\mathop{\mathrm{\mathstrut d}}%
	\nolimits^{#1}\gobblespace}
\def\gobblespace{%
	\futurelet\diffarg\opspace}
\def\opspace{%
	\let\DiffSpace\!%
	\ifx\diffarg(%
		\let\DiffSpace\relax
	\else
		\ifx\diffarg[%
			\let\DiffSpace\relax
		\else
			\ifx\diffarg\{%
				\let\DiffSpace\relax
			\fi\fi\fi\DiffSpace}

\usepackage{xcolor}
\definecolor{oxford_blue}{RGB}{14,31,71}
\usepackage{graphicx}
\usepackage{wrapfig}
\newcommand\figscaleapp{0.66}
\usepackage{booktabs}
\usepackage{multirow}
\usepackage{tablefootnote}
\usepackage{multicol}
\usepackage{cite}
\usepackage[numbers]{natbib}
\usepackage[labelfont=bf]{caption} 
\usepackage{subcaption} 
\usepackage{enumerate, enumitem}
\usepackage[colorlinks]{hyperref}
\usepackage{etoolbox} 
\usepackage{url}
\usepackage{textcomp} 

\theoremstyle{remark}
\newtheorem{remark}{Remark}

\makeatletter
\newcommand\carr{\@setfontsize\carr\@viipt\@viipt}
\makeatother

\begin{document}

\title{
Hedging option books using neural-SDE market models
}

\author{Samuel N. Cohen \and Christoph Reisinger \and Sheng Wang \and
Mathematical Institute, University of Oxford \\
\texttt{ \{samuel.cohen, christoph.reisinger, sheng.wang\} }\\ \texttt{@maths.ox.ac.uk}
}

\maketitle

\begin{abstract}
We study the capability of arbitrage-free neural-SDE market models to yield effective strategies for hedging options. In particular, we derive sensitivity-based and minimum-variance-based hedging strategies using these models and examine their performance when applied to various option portfolios using real-world data. Through backtesting analysis over typical and stressed market periods, we show that neural-SDE market models achieve lower hedging errors than Black--Scholes delta and delta-vega hedging consistently over time, and are less sensitive to the tenor choice of hedging instruments. In addition, hedging using market models leads to similar performance to hedging using Heston models, while the former tends to be more robust during stressed market periods.
\end{abstract}



\section{Introduction}

Hedging is critical for managing market risks of option books; in particular, the smaller hedging errors are, the smaller bid-ask spreads market makers are can offer, improving liquidity. In general, to hedge an option dynamically, one starts with identifying its risk factors (e.g.\ the price and volatility of the underlying asset), then evaluates the sensitivities of the option value with respect to these risk factors (normally known as ``Greeks''), and finally trades liquid instruments so that the sensitivities of the combined portfolio to \emph{some} risk factors vanish. In order to calculate sensitivities, we require a pricing model that evaluates an option's price from its risk factors. We note that this is true even for vanilla options, so while models are typically redundant for determining the current value of vanilla options, as these are observed on the market, they are needed for finding the hedge ratios. 
Neural-SDE factor-based models, introduced in our previous work \cite{cohen2021mktmdl, cohen2022estimating}, are extremely well suited for this task. While there is a vast literature on \emph{martingale} models for option pricing and their associated hedging strategies, this paper explores the capacity of the neural-SDE market models for producing effective hedging strategies for portfolios of vanilla European options in real markets. Though we also discuss hedge ratios for exotics, such as barrier and lookback options, we leave empirical assessment of these for future works.

A typical option price model describes the dynamics of the underlying price by a stochastic process. The Black--Scholes model \cite{BlackScholes1973} assumes constant volatility, such that continuously rebalanced delta hedging using the underlying asset should in theory completely remove risk from the option exposure, leading to a complete market. However, due to model misspecifications (for instance, the assumption of constant volatility) and the impossibility of continuous rebalancing in practice, there is residual risk for delta-hedged options. The resulting residual risk may be reduced by the addition of a gamma or vega hedge, which requires taking positions in a second option on the same underlying. The introduction of stochastic volatilities and/or jumps in models such as Heston \cite{Heston1993}, SABR \cite{Hagan2002, Hagan2014}, Merton \cite{merton1976option} and more generally exponential L\'{e}vy models \cite{cont2003financial}, renders the market incomplete, making it impossible to eliminate option risk by solely hedging with the underlying asset, even under continuously rebalancing. The hedging strategies associated with these models then differ\footnote{Nevertheless, different option pricing models may yield the same hedge ratios (and consequently hedging performance) if they are scale-invariant; see the in-depth analysis by Alexander and Nogueira \cite{alexander2007model}.}, depending on what risk factors are included and how their dynamics are modelled.

The neural-SDE market model, introduced in our previous work \cite{cohen2021mktmdl, cohen2022estimating}, consists of an SDE system for states that represent the underlying asset price and a small number of \emph{factors}. These factors are chosen, in a data-driven manner, for a statistically accurate dynamic representation of the option surface and the minimisation of dynamic and static arbitrage. Consequently, hedging an option using a neural-SDE market model means neutralising the sensitivities of the option's modelled price to the underlying (i.e.\ delta hedging) and the \emph{factors}. As we shall see later, there are at least two notable differences in the derivation of hedging strategies between the neural-SDE market models in
\cite{cohen2021mktmdl, cohen2022estimating} and traditional martingale models:
\begin{enumerate}[leftmargin=*, label=(\roman*)]
    \item The market model leads to a straightforward computation of price sensitivities to the factors; specifically, the \emph{normalised} call option price is linear in the factors (as shown in \eqref{eq:call_price_transformation} below), so the first partial derivative with respect to a given factor is simply the factor weight. In contrast, under a martingale model, computing sensitivities generally involves numerical integration or Monte--Carlo simulation to compute Greeks, due to the lack of analytical formulas for option prices.
    \item A model needs to be calibrated to market data before using it for hedging. The neural-SDE market models are calibrated with the \emph{joint} objectives of (a) calibrating (exactly, in principle) to the cross sections of option prices, which contain forward-looking information, and (b) maximising the likelihood of having observed the historical time-series of the underlying and \emph{liquid options}. These two objectives, however, cannot be fulfilled simultaneously, in a consistent manner\footnote{There are attempts to estimate some parameters using time series of the underlying and calibrating the remaining parameters to option prices (for example, Bates \cite{bates2000post} and Broadie et al. \cite{broadie2007model}). However, this approach remains inconsistent, as it still relies on daily recalibration of the model, which leads to time-varying calibrated parameters, whereas in theory the model parameters are assumed constant over time.}, when calibrating martingale models, and there is not a straightforward mechanism of factoring in historical option prices. Choosing one objective over the other can lead to significantly different parameter values, and therefore raises concerns of model misspecifications (see 
    \cite{alexander2012does}).
\end{enumerate}

We now discuss the difference between our approach and other published works on the use of neural networks for pricing and hedging options; see Ruf and Wang \cite{ruf2020neural} for an overview of a large body of the literature. While most works focus on first outputting option prices and then deriving hedging strategies as sensitivities, some directly output hedging strategies from neural networks, such as Buehler et al. \cite{buehler2019} and Ruf and Wang \cite{ruf2021hedging}. In constructing neural-SDE market models, the neural networks act as function approximators to the drift and diffusion coefficients of the SDE, in order to give a flexible class of models. As a result, neural-SDE market models largely reduce the ``black-box'' nature of neural network models (see the discussion by \cite{cohen2021blackbox}) and enhance model interpretability while still retaining their computational advantages.

In this paper, we will derive sensitivity-based and minimum-variance(MV)-based hedging strategies using neural-SDE market models, and examine their performance when applied to various portfolios of EURO STOXX 50 index options over typical and stressed market periods. We compare these market model hedging strategies with Black--Scholes (BS) and Heston delta and delta-vega hedging, along with considering the use of different hedging instruments (i.e.\ vanilla options) and rebalancing frequencies. When only hedging with the underlying, our empirical results show that neural-SDE MV-based delta hedging outperforms BS and Heston delta hedging on average, and is notably more effective during stressed market periods (e.g.\ COVID-19 outbreak). When hedging with an additional vanilla option, neural-SDE hedging strategies yield similar performance to hedging using Heston models, while their performance is less sensitive to the tenor of the hedging options than hedging using Black--Scholes models. Neural-SDE hedging is shown to outperform BS hedging consistently over time. Moreover, the MV-based variant of neural-SDE hedges (but not the sensitivity-based variant) outperforms Heston hedging for most portfolios, while the analysis of outright options indicates that Heston hedging gives smaller hedging errors specifically for short-dated options.

The rest of the paper is structured as follows. In Section \ref{sec:hedging_problem}, we formulate the option hedging problem and introduce sensitivity-based and minimum-variance(MV)-based hedging approaches. Thereafter, we derive hedge ratios with these two approaches using the neural-SDE market model framework in Section \ref{sec:nSDE_hedging}, and investigate the hedging performance relative to Black--Scholes and Heston delta and delta-vega hedging, in Section \ref{eq:numeric_hedging} for EURO STOXX 50 data. Finally, we offer conclusion from our empirical observations in Section \ref{sec:conclusion}.

\section{Option hedging}
\label{sec:hedging_problem}

Suppose an option pricing model considers a set of $d+1$ risk factors $\{S, \xi_1, \dots, \xi_d\}$. An agent sells an option (or a portfolio of options) with price $V$ modelled as a function of the risk factors. To ensure that the agent's position is neutral to all $d+1$ risk factors (at least instantaneously), it is typically necessary to hedge the sold option using $d+1$ risky instruments. We assume that hedges are rebalanced at discrete time intervals of length $\Delta t$, called the \emph{rebalancing horizon}. Therefore, we consider the hedged portfolio
\begin{equation}
    \Pi(S, \xi_1, \dots, \xi_d) = V(S, \xi_1, \dots, \xi_d) - X^S S - \sum_{i=1}^d  X^{C_i} C_{i}(S, \xi_1, \dots, \xi_d),
    \label{eq:hedged_portfolio}
\end{equation}
where $X^S$ is the amount (or \emph{hedge ratio}) held of the underlying asset with value $S$, and $X^{C_i}$ is the amount of the $i$-th liquidly tradable call\footnote{We use liquidly traded European style vanilla options as hedging instruments. Though we only consider call options here, by put-call parity, put options can also be used.} option with value $C_{i}$. We observe that $V$ and $\Pi$ are deterministic functions of the risk factors $(S, \xi_1, \dots, \xi_d)$, while by substituting the stochastic processes of the risk factors $(S_t, \xi_{1t}, \dots, \xi_{dt})$, we obtain processes $V_t$ and $\Pi_t$. 

In a one-period hedging problem, suppose at time $t$ the agent holds this hedged portfolio (with all $X^S, X^{C_i}$ constant), then closes it after $\Delta t$, and obtains a hedging error of
\begin{equation}
    \mathcal{E}_t(\Delta t) = \Pi_{t + \Delta t} - \Pi_t.
    \label{eq:hedging_error}
\end{equation}
For simplicity, we assume zero interest and dividend yield\footnote{The agent could invest the residual cash from the hedged portfolio \eqref{eq:hedged_portfolio} in a risk-free asset. If, for example, the risk-free asset has non-zero interest yield $r_t$, then the hedging error in \eqref{eq:hedging_error} would be revised to $\mathcal{E}_t (\Delta t) = \Pi_t e^{\int_t^{t+\Delta t} r_s \diff s} - \Pi_{t+\Delta t}$ instead.}. To reduce the variations of the hedging error in \eqref{eq:hedging_error}, the agent finds a hedging strategy defined by $X = (X^S, X^{C_1}, \dots, X^{C_d})$ such that the hedged portfolio $\Pi$ is \emph{instantaneously} invariant with respect to each risk factor $f \in \{S, \xi_1, \dots, \xi_d\}$. In the following sections, we introduce sensitivity-based and minimum-variance-based approaches to derive hedging strategies.

\subsection{Delta hedging}

Usually an agent will not fully hedge all risk factors imposed by a model, likely due to high transaction costs and trading limits of adding extra hedges. In particular, in a delta hedging strategy, we use the underlying asset as the only hedging instrument, and exclude the use of other options as hedging instruments. Assuming broadly markets described by continuous semi-martingales, by It\^{o}'s lemma,
\begin{equation*}
\diff \Pi = \left( \pderiv{V}{S} - X^S \right) \diff S + \sum_{i=1}^d [\cdots] \diff \xi_i + [\cdots] \diff t.
\end{equation*}
To eliminate $\Pi$'s instantaneous exposure to $S$, we set the coefficient of $\diff S$ to zero and obtain $X^S = \partial V / \partial S$, which is the standard delta of the option.

When the dynamics of the other risk factors $\xi = (\xi_1, \dots, \xi_d)$ imposed by the model interact with that of the underlying $S$, the standard delta might not be the most efficient hedge ratio. An alternative derivation of hedging strategies has been introduced to \emph{locally} minimise the conditional variance process of the hedging error \eqref{eq:hedging_error} by F{\"o}llmer and Schweizer \cite{follmer1990hedging} and Schweizer \cite{schweizer1991option}. The resulting hedge ratios are known as minimum variance (MV) hedge ratios. Bakshi et al. \cite{bakshi1997empirical} and Poulsen et al. \cite{poulsen2009risk} provide an intuitive and tractable derivation based on reducing the instantaneous covariance of a delta-hedged portfolio with the underlying to zero, i.e.\
\begin{equation*}
0 = \langle \diff \Pi, \diff S \rangle = \langle \diff V - X^S_\mathrm{mv} \diff S, \diff S \rangle = \langle \diff V, \diff S \rangle - X^S_\mathrm{mv} \langle \diff S, \diff S \rangle,
\end{equation*}
where we use $X^S_\mathrm{mv}$ to denote the MV delta hedge ratio, and $\langle \cdot, \cdot \rangle$ to denote the quadratic co-variation. Using It\^{o}'s lemma on $V$, we obtain
\begin{equation*}
    \langle \diff V, \diff S \rangle = \left\langle \pderiv{V}{S} \diff S + \sum_{i=1}^d \pderiv{V}{\xi_i} \diff \xi_i, \diff S \right\rangle = \pderiv{V}{S} \langle \diff S, \diff S \rangle + \sum_{i=1}^d \pderiv{V}{\xi_i} \langle \diff \xi_i, \diff S \rangle.
\end{equation*}
Therefore,
\begin{equation}
    X^S_\mathrm{mv} = \frac{\langle \diff V, \diff S \rangle}{ \langle \diff S, \diff S \rangle} = \pderiv{V}{S} + \sum_{i=1}^d \pderiv{V}{\xi_i} \frac{\langle \diff \xi_i, \diff S \rangle}{\langle \diff S, \diff S \rangle} = X^S + \sum_{i=1}^d \pderiv{V}{\xi_i} \frac{\langle \diff \xi_i, \diff S \rangle}{\langle \diff S, \diff S \rangle}.
    \label{eq:XS_mv1}
\end{equation}
Hence, the MV delta hedge ratio $X^S_\mathrm{mv}$ is the sensitivity-based delta hedge ratio $X^S$ with adjustments that take into account sensitivities to other risk factors, weighted by the correlation between the risk factor and the underlying price. In particular, if the underlying-factor correlations are all zero, the MV hedge ratio is identical to the sensitivity-based hedge ratio (see discussion by Alexander and Nogueira \cite{alexander2007model}).

\begin{remark}
The prevailing presence of the leverage effect \cite{black1976studies} in equity markets, i.e.\ negative correlation between stock return and volatility, has encouraged practitioners to adjust sensitivity-based deltas by either following the MV approach \eqref{eq:XS_mv1} or, in a model-free manner, adding other Greeks weighted by statistically regressed coefficients. For example, Hull and White \cite{hull2017optimal} and Ruf and Wang \cite{ruf2021hedging} have demonstrated the excellent hedging performance on S\&P 500 index options of the statistically adjusted Black--Scholes delta.
\end{remark}

\subsection{Delta-factor hedging}

We now generalise to hedging other risk factors. Suppose we want to eliminate the exposures to $S$ and the first $d'$ factors (i.e.\ $\xi_1, \dots, \xi_{d'}$), without loss of generality, using the underlying asset and $d'$ options $\{C_i\}_{i=1,\dots,d'}$ as the hedging instruments.

We use $f \in \{S, \xi_1, \dots, \xi_{d'} \}$ to denote a risk factor. By It\^{o}'s lemma,
\begin{equation*}
\diff \Pi = \sum_f \pderiv{\Pi}{f} \diff f + [\cdots] \diff t = \sum_f \left( \pderiv{V}{f} - \pderiv{S}{f} X^S - \sum_{i=1}^{d'} \pderiv{C_i}{f} X^{C_i} \right) \diff f + [\cdots] \diff t.
\end{equation*}
Similarly to the delta hedging approaches discussed previously, when hedging each risk factor $f$, there are two alternatives to deriving the hedge ratios, either using sensitivities,
\begin{subequations}
\begin{equation}
    \pderiv{\Pi}{f} = 0 
    \quad \Rightarrow \quad 
    \begin{bmatrix}
    \partial S / \partial f \\
    \partial C_1 / \partial f \\
    \vdots \\
    \partial C_{d'} / \partial f
    \end{bmatrix}^\top 
    \begin{bmatrix}
    X^S \\ X^{C_1} \\ \vdots \\ X^{C_{d'}} 
    \end{bmatrix}  = 
    \pderiv{V}{f},
    \label{eq:hedge_pd}
\end{equation}
or using a minimum variance criterion in terms of the risk factors, 
\begin{equation}
    \langle \diff \Pi, \diff f \rangle = 0
    \quad \Rightarrow \quad 
    \begin{bmatrix}
    \langle \diff S, \diff f \rangle \\
    \langle \diff C_1, \diff f \rangle \\
    \vdots \\
    \langle \diff C_{d'}, \diff f \rangle
    \end{bmatrix}^\top 
    \begin{bmatrix}
    X^S \\ X^{C_1} \\ \vdots \\ X^{C_{d'}} 
    \end{bmatrix}  = 
    \langle \diff V, \diff f \rangle.
    \label{eq:hedge_mv}
\end{equation}
\end{subequations}
Either condition leads to a \textit{linear} equation with regard to the hedging strategy $X$, because both the partial derivative operator and the quadratic co-variation operator are linear. Consequently, the delta-factor hedging problem can be formulated as solving a $(d'+1)$-dimensional linear system of equations.

\begin{remark}
Alternatively, we can use a minimum variance criterion on the hedging instruments rather than the risk factors, i.e.\ $\langle \diff \Pi, \diff C_i \rangle = 0$, which also leads to a linear condition with regard to the hedging strategy. We compare the difference between this approach and the two criteria above under a specific neural-SDE market model in Section \ref{sec:choice_hedge_tenor}.
\end{remark}

\section{Hedging with neural-SDE market models}
\label{sec:nSDE_hedging}

We first briefly recall the modelling components of the neural-SDE market model introduced in our previous work \cite{cohen2021mktmdl, cohen2022estimating}, and then derive corresponding hedging strategies. Specifically, suppose we work with $d$ factors $\xi = (\xi_1, \dots, \xi_d)$. 
Let $\tilde{\xi} = (\ln S, \xi)$, then we model its dynamics by the following SDE\footnote{We have assumed that the drift and diffusion coefficients are independent of $S$, in the sense that the underlying asset's log-return and volatility surfaces (represented by $\xi$) are unaffected by its price level. Nevertheless, it is possible, and simple, to add $S$ as an additional argument and train the model with neural nets (as is done in \cite{cohen2021mktmdl}).}:
\begin{equation}
    \diff \tilde{\xi}_t = \mu(\xi_t) \diff t + \sigma(\xi_t) \diff W_t,
    \label{eq:xi_sde}
\end{equation}
where $W =[W_{1} ~\cdots ~W_{d+1}]^\top \in \mathbb{R}^{d+1}$ are standard independent Brownian motions under the real-world measure $\mathbb{P}$. We denote by $L^p_\text{loc}(\mathbb{R}^d)$ the space of all $\mathbb{R}^d$-valued, progressively measurable, and locally $p$-integrable (in $t$, $\mathbb{P}$-a.s.) processes, and assume that $\mu \in L^1_\text{loc}(\mathbb{R}^{d+1})$ and $\sigma \in L^2_\text{loc}(\mathbb{R}^{d+1})$. In the SDE model calibration step, we represent the drift and diffusion functions by neural networks, the reason why the \emph{neural-SDE} is so named.

Now we associate $\tilde{\xi}$ with option prices. We denote by $C_t(T,K)$ the market price at time $t$ of an European call option with expiry $T$ and strike $K$. In addition, we reparametrise the call price function $C_t(\cdot,\cdot)$ by introducing $\tau=T-t$ to be the time-to-expiry and $m = \ln (K/S_t)$ the moneyness. We assume these prices, in the underlying numeraire, can be written as time independent linear functions of the factors, that is,
\begin{subequations}
\begin{align}
    C_t(T,K) & = S_t \tilde{c}_t (\tau, m), \label{eq:call_price_transformation21} \\ 
    \tilde{c}_t(\tau,m) & = G_0(\tau, m) + \sum_{i=1}^d G_i(\tau, m) \xi_{it},
\label{eq:call_price_transformation22}
\end{align}
\label{eq:call_price_transformation}
\end{subequations}
where $G_i \in C^{1,2}(\mathcal{R}_\text{liq})$ for $i=0,\dots,d$ are called \emph{price basis functions}. We fix a constant compact set $\mathcal{R}_\text{liq} \subset \{(\tau, m) \in \mathbb{R}^2 \}$, which represents the range of time-to-expiries and moneynesses where options are liquidly traded. We call $\tilde{c}$ the normalised call option price.

The neural-SDE market model rules out static, model-free arbitrage on the underlying factors as follows. Considering a finite set of option specifications $\mathcal{L}_\text{liq} = \{ (\tau_j, m_j) \}_{j=1,\dots,N} \subset \mathcal{R}_\text{liq}$, called liquid lattice, we have their normalised prices $\mathbf{c}_t = [\tilde{c}_t(\tau_1, m_1) ~\cdots ~\tilde{c}_t(\tau_N, m_N)]^\top$. 
Noting that absence of static arbitrage can be expressed as a system of linear inequalities, using the construction in our previous work \cite{Cohen2020}, we write static arbitrage constraints in the form $\mathbf{A} \mathbf{c}_t \geq \widehat{\mathbf{b}}$, where $\mathbf{A} = (A_{ij}) \in \mathbb{R}^{R \times N}$ and $\widehat{\mathbf{b}} = (\hat{b}_j) \in \mathbb{R}^{R}$ are a known constant matrix and vector, and $R$ is the number of static arbitrage constraints. Our derivation is mainly based on Carr et al. \cite{CGMY2003}, Carr and Madan \cite{Carr2005}, Davis and Hobson \cite{davis2007}, and Cousot \cite{cousot2007}. Given the factor representation \eqref{eq:call_price_transformation22}, we have $\mathbf{c}_t = \mathbf{G}_0 + \mathbf{G}^\top \xi_t$, where we define $\mathbf{G}$ as the $d\times N$ matrix with $i$-th row $\mathbf{G}_i = (G_{ij})_{j=1}^N \in \mathbb{R}^N$ for $i=0,\dots, d$, and $G_{ij} = G_i(\tau_j, m_j)$. Consequently, the market model allows no static arbitrage among options on the liquid lattice $\mathcal{L}_\text{liq}$ if $\xi_t$ satisfies, for all $t$,
\begin{equation}
\mathbf{A} \mathbf{G}^\top \xi_t \geq \mathbf{b} := \widehat{\mathbf{b}} - \mathbf{A G}_0^\top.
\label{eq:factor_static_arbitrage}
\end{equation}
We learn functions $\mu$, $\sigma$ in \eqref{eq:xi_sde} to ensure $\xi_t$ satisfies this inequality at all times (see \cite{cohen2021mktmdl} for details).

\subsection{Delta hedging}

Next, we derive the hedge ratios with the sensitivity-based and the MV-based approaches for a call option with maturity $T^*$ and strike price $K^*$. For ease of notation, we omit time dependence in the following derivations, and use the same symbol for 
the option's value function and the realisation of its value, if it is clear from the context which is being used.

Let $V = C(T^*,K^*)$ be the option to hedge. Using \eqref{eq:call_price_transformation21}, in a sensitivity-based delta hedging strategy, the hedge ratio of the underlying asset is
\begin{equation}
    X^S = \pderiv{V}{S} = \pderiv{}{S} \left(S \tilde{c}\right) = \left( \tilde{c} - \pderiv{\tilde{c}}{m} \right) (\tau^*, m^*), \text{ where } \tau^*=T^*-t, m^* = \ln (K^*/S).
    \label{eq:XS}
\end{equation}
It is worth noting that the sensitivity-based delta $X^S$ is independent of the dynamics of $S$ and $\xi$ given in the model \eqref{eq:xi_sde}. In addition, if the prices of the $T^*$-maturity options, for a continuum of strike prices, are available, the evaluation of $\partial \tilde{c} / \partial m$ is model-free, hence $X^S$ could be computed independently of any models. This is indeed consistent with the conclusion by Alexander and Nogueria \cite{alexander2007model} that hedge ratios for European claims under scale-invariant models are model-free, provided  a perfect fit to market prices.

In contrast to sensitivity-based deltas, deriving MV deltas relies on the dynamics in \eqref{eq:xi_sde}. In particular, we assume, without loss of generality, that $\sigma = (\sigma_{i,j}) \in \mathbb{R}^{(d+1) \times (d+1)}$ is a lower triangular matrix, so that we have,
\begin{equation*}
    \langle \diff S, \diff S \rangle = \sigma_{1,1}^2 S^2 \diff t, \quad  \langle \diff \xi_i, \diff S \rangle = \sigma_{1,1} \sigma_{i+1,1} S \diff t.
\end{equation*}
In addition,
\begin{equation}
    \pderiv{V}{\xi_i} = \pderiv{}{\xi_i} \left(S \tilde{c}\right) = S \pderiv{}{\xi_i} \left( G_0 + \sum_{j=1}^d G_j \xi_j \right) = S G_i.
    \label{eq:Vxi}
\end{equation}
Hence, the MV delta hedge ratio in \eqref{eq:XS_mv1} under the market model can be computed as
\begin{equation}
    X^S_\mathrm{mv} = X^S + \sum_{i=1}^d \frac{\sigma_{i+1,1}(\xi)}{\sigma_{1,1}(\xi)} G_i (\tau^*, m^*).
\end{equation}

\subsection{Delta-factor hedging}

Now we evaluate the two alternative hedging equations \eqref{eq:hedge_pd} and \eqref{eq:hedge_mv} under the market models given by \eqref{eq:xi_sde} and \eqref{eq:call_price_transformation}. For an asset (underlying or option) $U$, we introduce the following notation:
\begin{equation}
\begin{aligned}
    \Delta(U) & := \pderiv{U}{S}, \quad \Delta_\mathrm{mv}(U) := \Delta (U) + \sum_{i=1}^d \Delta^i(U) \frac{\sigma_{i+1,1}}{\sigma_{1,1} S}, \\
    \Delta^j(U) & := \pderiv{U}{\xi_j}, \quad \Delta^j_\mathrm{mv}(U) := \Delta(U) \sigma_1 \sigma_{j+1}^\top S + \sum_{i=1}^d \Delta^i(U) \sigma_{i+1} \sigma_{j+1}^\top,
\end{aligned}
\label{eq:omega}
\end{equation}
where $\sigma_i$ is the $i$-th row of the diffusion matrix $\sigma \in \mathbb{R}^{(d+1)\times (d+1)}$. In particular, we call $\Delta^j$ ($\Delta^j_\mathrm{mv}$) the sensitivity-based (MV-based) \emph{$\xi_j$-exposure}. To hedge $V$ using $d'$ options with the strategy $X=(X^S, X^{C_1}, \dots, X^{C_{d'}})$ following the sensitivity-based approach, we set the partial sensitivity of the hedged portfolio $\Pi$ with regard to a risk factor $f$ to zero and establish the linear equation
\begin{equation}
    \pderiv{\Pi}{f} = 0 
    \Rightarrow 
    \begin{cases}
    X^S + \sum_{i=1}^{d'} \Delta(C_{i}) X^{C_i}= \Delta(V), & \text{ if } f = S; \\
    \Delta^j(S) X^S + \sum_{i=1}^{d'} \Delta^j(C_i) X^{C_i} = \Delta^j(V), & \text{ if } f = \xi_j, \text{ for } j = 1,\dots,d'. 
    \end{cases}
    \label{eq:hedge_pd_mktmdl}
\end{equation}
Alternatively, with MV-based hedging, where there is zero instantaneous co-variation between the hedged portfolio $\Pi$ and the risk factor $f$, we have
\begin{equation}
    \langle \diff \Pi, \diff f \rangle = 0
    \Rightarrow
    \begin{cases}
    X^S + \sum_{i=1}^{d'} \Delta_\textrm{mv}(C_{i}) X^{C_i}= \Delta_\textrm{mv}(V), & \text{ if } f = S; \\
    \Delta_\textrm{mv}^j(S) X^S + \sum_{i=1}^{d'} \Delta^j_\textrm{mv} (C_i) X^{C_i} = \Delta^j_\textrm{mv} (V), & \text{ if } f = \xi_j, \text{ for } j = 1,\dots,d'.
    \end{cases}
    \label{eq:hedge_mv_mktmdl}
\end{equation}

Both \eqref{eq:hedge_pd_mktmdl} and \eqref{eq:hedge_mv_mktmdl} indicate that, for each risk factor $f$, the aggregated $f$-exposure of all hedging instruments equals the $f$-exposure of the option to be hedged. In addition, the only difference between them is that the corresponding $f$-exposure is computed based on zero-sensitivity and minimum-variance, respectively. 

\begin{remark}
The sensitivity-based hedging equation \eqref{eq:hedge_pd_mktmdl} does not exploit statistical information established by the SDE model \eqref{eq:xi_sde}, and completely relies on the factor representation of option prices given by \eqref{eq:call_price_transformation}. This means it is a simpler approach to hedging, which does not use ``machine learning'' techniques to a significant extent.
\end{remark}

\begin{remark} \label{rmk:independent_factor}
When one risk factor $f$ is independent of all other risk factors, its MV-based hedging condition \eqref{eq:hedge_mv_mktmdl} becomes equivalent to its sensitivity-based hedging condition \eqref{eq:hedge_pd_mktmdl}. To see this, suppose $f = S$ is independent of $\xi$, then $\sigma_{i+1,1} = 0$, which yields $\Delta_\textrm{mv}(U) = \Delta (U)$ according to \eqref{eq:omega}; if $f = \xi_j$ is independent of $S$ and the other factors, $\Delta^j_\textrm{mv} (U) = \Delta^j (U) \sigma_{j+1, j+2}^2 \propto \Delta^j (U)$.
\end{remark}

\subsection{A two-factor market model}
\label{sec:specific_mdl}

Here we consider a two-factor market model (i.e. $d=2$) as in \cite{cohen2022estimating}\footnote{We assume independent Brownian motions that drive the randomness in $S$ and $\xi$ in the main text of \cite{cohen2022estimating} for simplicity, but build a model for $S$ and $\xi$ jointly with a full $\mathbb{R}^{(d+1)\times(d+1)}$ covariance matrix, in Appendix B.2 of \cite{cohen2022estimating}. Since we would like to see how MV-based delta hedging behaves differently from its sensitivity-based counterpart using market models, we need a market model that captures the correlation between $S$ and $\xi$, according to Remark \ref{rmk:independent_factor}. Hence, here we allow a fully specified diffusion matrix (by neural networks) when building a market model.} for modelling historical price data of EURO STOXX 50 index options and DAX options and therefore constructing a realistic option market simulator for computing VaR. 

We focus on studying the following three problems:

\begin{enumerate}[leftmargin=*, label=(\roman*)]
	\item Does MV-based delta hedging using neural-SDE market models perform better than sensitivity-based delta hedging and MV-based delta hedging using Heston models? This poses two questions: (a) is MV-based delta hedging superior to sensitivity-based delta hedging, given that the MV approach takes into account possible correlation between $S$ and other modelled risk factors? and (b) do neural-SDE market models predict
more accurate joint dynamics (of $S$ and other modelled risk factors) than Heston models, hence producing better MV-based deltas?
    \item In both the numerical experiments with Heston-SLV synthetic data \cite{cohen2021mktmdl} and with real world data \cite{cohen2022estimating}, the first factor $\xi_1$ behaves similarly to the volatility process of the underlying. In fact, as shall be seen in Section \ref{sec:hedging_bs}, vegas of vanilla options have qualitatively similar geometry over strikes and expiries to their $\xi_1$-exposures. Hence, we want to examine whether hedging an option's $\xi_1$-exposure achieves comparable performance with hedging its vega, while the former has computational advantages. The calculation of an option's vega relies on the model specified for the underlying and its volatility process, and, in general, is computationally costly due to lack of analytical formulas and non-trivial model calibration. In contrast, the sensitivity-based $\xi_1$-exposure, i.e.\ $SG_1$, is easily decoded from historical time series of option price data using a fast PCA-based algorithm (see Algorithm 1 in \cite{cohen2021mktmdl}).
    \item Will hedging MV-based $\xi_1$-exposures perform better than hedging sensitivity-based $\xi_1$-exposures? Since MV-based hedging exploits the correlation between factors, captured by the SDE model \eqref{eq:xi_sde}, we are therefore assessing whether estimating a joint dynamic model for the factors is beneficial for hedging.
\end{enumerate}

Assuming the hedging instruments available are the underlying asset $S$ and one liquid call option $C_1$, we derive their hedge ratios using a two-factor market model with the objective of hedging exposures to $S$ and $\xi_1$ (i.e. $d'=1$). With sensitivity-based hedging, we solve \eqref{eq:hedge_pd_mktmdl} and obtain
\begin{equation}
    X^S = \left( \tilde{c} - \pderiv{\tilde{c}}{m} \right) (\tau^*, m^*) - X^{C_1} \left( \tilde{c} - \pderiv{\tilde{c}}{m} \right) (\tau_1, m_1), \quad X^{C_1} = \frac{G_1(\tau^*, m^*)}{G_1(\tau_1, m_1)}.
    \label{eq:hedge_ratio_pd}
\end{equation}
And with MV-based hedging, we solve \eqref{eq:hedge_mv_mktmdl} and obtain
\begin{equation}
    X^S_\mathrm{mv} = \left( \tilde{c} - \pderiv{\tilde{c}}{m} \right) (\tau^*, m^*) - X^{C_1}_\textrm{mv} \left( \tilde{c} - \pderiv{\tilde{c}}{m} \right) (\tau_1, m_1), \quad X^{C_1}_\textrm{mv} = \frac{\mathcal{G}(V)}{\mathcal{G}(C_1)},
    \label{eq:hedge_ratio_mv}
\end{equation}
where, for a call option $U$ with time-to-expiry $\tau_U$ and moneyness $m_U$,
\begin{equation}
\mathcal{G}(U) := \sum_{i=1}^d G_i(\tau_U, m_U) \sigma_{i+1} \left[ \frac{\sigma^\top_2}{\sigma_1 \sigma_2^\top} -\frac{\sigma^\top_1}{\sigma_1 \sigma_1^\top} \right].
\label{eq:mv_G}
\end{equation}

\subsection{Hedge ratios for more option types}

The hedge ratios \eqref{eq:hedge_ratio_pd} and \eqref{eq:hedge_ratio_mv} derived in the previous section apply to European vanilla call options. Nevertheless, the sensitivity-based and MV-based hedging equations \eqref{eq:hedge_pd_mktmdl} and \eqref{eq:hedge_mv_mktmdl} can, in principle, be universally applied to a much wider range of option types, and therefore yield corresponding hedge ratios.

To derive the hedge ratios for a generic option $V$, we need to evaluate $\Delta(V)$ and $\Delta^j(V)$, for which the definitions are given in \eqref{eq:omega}. This will be straightforward if $V = h(S, C_1, \dots, C_n)$ where $\{C_1, \dots, C_n\}$ is a series of $n$ European vanilla call options:
\begin{equation*}
\pderiv{V}{S} = \pderiv{h}{S} + \sum_{i=1}^n \pderiv{h}{C_i} \Delta (C_i), \quad \pderiv{V}{\xi_j} = \sum_{i=1}^n \pderiv{h}{C_i} \Delta^j (C_i).
\end{equation*}
We discuss a few examples as follows.

\paragraph*{European put option.}

For a European put option with maturity $T$ and strike price $K$, there is a model-free no-arbitrage relationship with the European call option with exactly the same maturity and strike price, namely put-call parity given by $P(K) = C(K) - S + K$, where we use $P(K)$ and $C(K)$ to denote prices of the put and the call (without explicitly showing their dependence on $T$ for ease of notation). Therefore, we have
\begin{equation*}
\pderiv{P(K)}{S} = \Delta(C(K)) - 1, \quad
\pderiv{P(K)}{\xi_j} = \Delta^j (C(K)).
\end{equation*}

\paragraph*{Binary options.}

Let $BC(K)$ denote a binary call option with strike price $K$. It is known that binary calls can be synthesised using an infinite number of vertical spreads of European calls:
\begin{equation*}
BC(K) = \lim_{n \rightarrow + \infty} n \left( C(K) - C(K + \frac{1}{n}) \right) = \pderiv{C(K)}{K}.
\end{equation*}
Assuming that $C \in \mathcal{C}^2$ (i.e.\ all the second-order partial derivatives exist and are continuous),
\begin{align*}
\pderiv{BC(K)}{S} & = \pderiv{\Delta (C(K))}{K} = \frac{1}{K} \left( \pderiv{\tilde{c}}{m} - \pderiv[2]{\tilde{c}}{m} \right), \\
\pderiv{BC(K)}{\xi_j} & = \pderiv{\Delta^j (C(K))}{K} = - \frac{S}{K} \left( G_j - \pderiv{G_j}{m} \right).
\end{align*}
Thereafter, the corresponding partial derivatives for a binary put option $BP(K)$ can be derived easily using the model-free no-arbitrage put-call parity $BP(K) = 1 - BC(K)$.

\paragraph*{Exotic options.}

Carr, Ellis and Gupta \cite{carr1999static} have developed static hedges for various exotic options such as (single and multiple) barrier options and lookback options using European vanilla puts and calls, with assumptions on the shape of the local volatility smiles\footnote{However, these assumptions will not generally hold for neural-SDE market models. It remains to be investigated empirically how the violation of these assumptions impacts hedging performance using neural-SDE market models.}.

For example, a down-and-out call option\footnote{A down-and-out call option is a vanilla call option if its barrier has not been hit by its expiry date. However it becomes worthless if its barrier is hit at any time before it expires.}, denoted as $DOC(K,B)$, with strike $K$ and barrier $B < K$ can be exactly replicated by a portfolio of calls and puts
\begin{equation*}
DOC(K,B) = C(K) - KB^{-1} P(B^2 K^{-1}).
\end{equation*}
The replicating portfolio matches the terminal payoff (i.e.\ long a call $C(K)$) and the payoff along the barrier (i.e.\ short additional $KB^{-1}$ puts struck at $B^2 K^{-1}$). Using put-call parity, we can further re-write $DOC(K,B)$ as a portfolio of vanilla calls. We refer readers to \cite{carr1999static} for how to construct vanilla call portfolios that statically replicate down-and-out, double knockouts, roll-down, ratchet, and lookback options.

In general, if an option can be statically hedged\footnote{Nevertheless, the static hedging relation is usually derived with certain model assumptions, so that the replicating portfolios may only \emph{approximate} the option's payoff. The impact of the approximation errors on hedging performance needs to be examined with empirical tests.} by European vanilla options, then we can derive its partial derivatives with regard to $S$ and $\xi$ and therefore its hedge ratios. There is vast literature on static hedging of exotic options, to name a few recent developments, \cite{boyarchenko2020static}, \cite{kim2021static} and \cite{kirkby2019static}.

\subsection{Hedging with Black--Scholes and Heston models}
\label{sec:hedging_bs}

In the empirical studies that follow, we use Black--Scholes and Heston models as two benchmark models for comparing hedging performance. Now we derive the delta(-vega) hedge ratios using the two benchmark models.

\paragraph*{Black--Scholes model.}

The delta and vega of a call option $U = C_t(T,K)$ derived using the Black--Scholes model are given by
\begin{equation}
    \Delta_\mathrm{bs}(U) = \Phi(d_1), ~\mathcal{V}_\mathrm{bs} (U) = S \sqrt{T-t} \phi(d_1), \text{ where } d_1 = \frac{\ln(S/K)}{\sigma \sqrt{T-t}} + \frac{1}{2} \sigma \sqrt{T-t}.
    \label{eq:bs_greeks}
\end{equation}
Therefore, neutralising the delta and vega exposures of $V$ yields the hedge ratios
\begin{equation}
    X^S_\mathrm{bs} = \Delta_\mathrm{bs}(V) - X^{C_1}_\mathrm{bs} \cdot \Delta_\mathrm{bs}(C_1), \quad X^{C_1}_\mathrm{bs} = \frac{\mathcal{V}_\mathrm{bs} (V)}{\mathcal{V}_\mathrm{bs} (C_1)}.
    \label{eq:hedge_ratio_bs}
\end{equation}

\begin{remark}
There remains the choice of which volatility should be used as $\sigma$ in the formulas in \eqref{eq:bs_greeks}, as the implied volatility of the option to hedge ($V$) is usually different from the IV of the option used as the hedging instrument ($C_1$). For example, in computing the hedge ratios in \eqref{eq:hedge_ratio_bs}, should one use $V$'s IV in the deltas/vegas of both $V$ and $C_1$? In our empirical experiments that follow, we choose to use an option's own IV for computing its delta/vega. As it turns out, this has limited impact due to the stability of the Black--Scholes model \cite{rebonato2005volatility}.
\end{remark}

\paragraph*{Heston model.}

Heston models are constructed from the SDE
\begin{equation}
\begin{aligned}
    \diff S_t & = r_t S_t \diff t + \sqrt{\nu_t} S_t \diff W_t^S, \\
    \diff \nu_t & = k (\theta - \nu_t) \diff t + \sigma \sqrt{\nu_t} \diff W_t^\nu, ~
    \diff \langle W_t^S, W_t^\nu \rangle = \rho \diff t, 
\end{aligned}
\label{eq:heston_sde}
\end{equation}%
We denote the delta and vega of a call option $U$ using a Heston model as $\Delta_\textrm{h} (U)$ and $\mathcal{V}_\textrm{h} (U)$, which can be derived as the partial derivatives of $U$, evaluated using the Heston model, with respective to $S_0$ and $\nu_0$, respectively. We refer readers to Heston \cite{Heston1993} and Albrecher et al. \cite{albrecher2007little} for a formula for the call option price, from which one could compute call option delta and vega (numerically). In particular, Alexander and Nogueria \cite{alexander2007model} derive the MV-based delta using Heston models
\begin{equation}
\Delta_\textrm{h(mv)} = \Delta_\textrm{h} + \mathcal{V}_\textrm{h} \frac{\rho \sigma}{S_0}.
\label{eq:heston_delta_mv}
\end{equation}
In fact, one could derive \eqref{eq:heston_delta_mv} from \eqref{eq:XS_mv1}.

Calibration can affect the hedging results considerably, yet Alexander and Kaeck \cite{alexander2012does} find that daily recalibration of the Heston model to option prices ``clearly improves its hedging performance... for all hedging strategies considered (delta, delta-gamma, and delta-vega) and for both rebalancing frequencies (one day and one week)''. Hence, we compute hedge ratios using Heston models with daily recalibrated parameters. We give the objective function and time series of calibrated parameters in Appendix \ref{apd:heston_calib}.

In Figure \ref{fig:vega_xi1}, we show the Black--Scholes vegas, the Heston vegas and the market model $\xi_1$-exposures for a few traded call options that have selected time-to-expiries (roughly 1M, 3M, 6M, 1Y and 2Y) as of 2nd January 2019. BS and Heston vegas and $\xi_1$-exposures have qualitatively similar behaviour over strikes and expiries. For a fixed strike price, they increase over expiries; for a fixed expiry, they first increase over strikes, reach a peak near the ATM strike level, and then decline. 

\begin{figure}[!ht]
    \centering
    \includegraphics[scale=\figscaleapp]{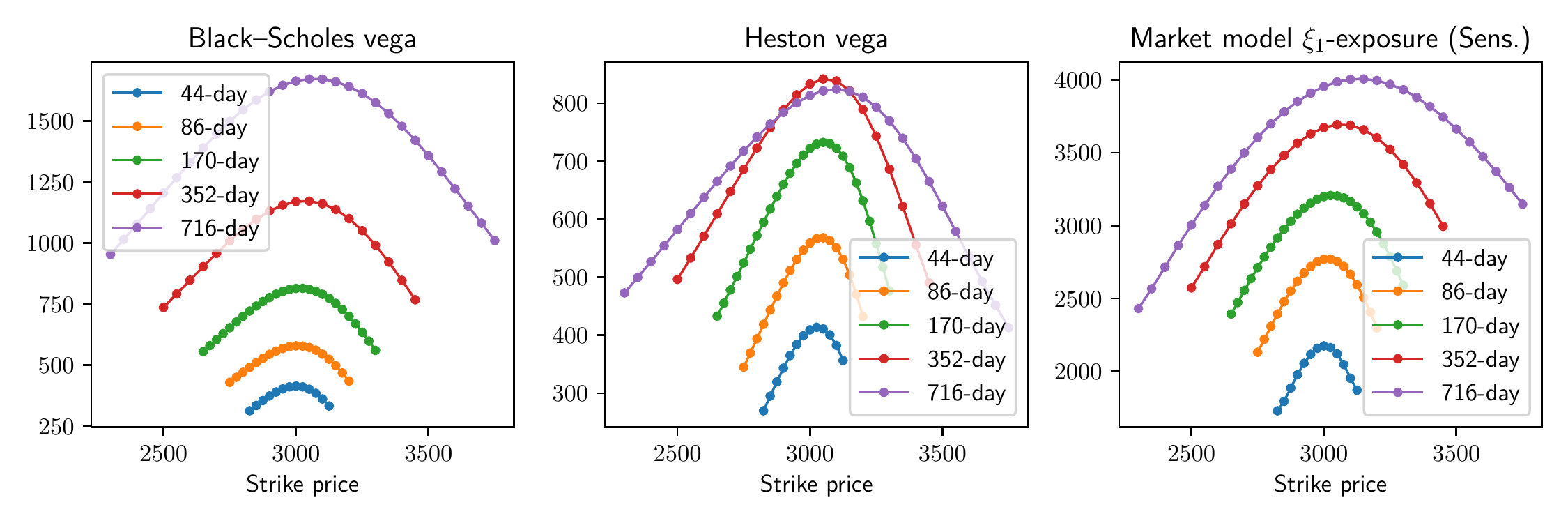}
    \caption{Black--Scholes vegas, Heston vegas and market model $\xi_1$-exposures of traded call options as of 2nd January 2019. Here we only show traded options that are within the liquid range and have selected time-to-expiries.}
    \label{fig:vega_xi1}
\end{figure}

\section{Empirical results}
\label{eq:numeric_hedging}

We assess the hedging performance using the market model specified in Section \ref{sec:specific_mdl} by examining out-of-sample hedging errors where the hedge ratios are given by \eqref{eq:hedge_ratio_pd} and \eqref{eq:hedge_ratio_mv}. In addition, we use hedge ratios derived using Black--Scholes (BS) models and Heston models as benchmarks, and examine if the market model yields better hedging performance than the benchmarks.

\subsection{Option datasets and the trained neural-SDE market model}

\paragraph{Training data.}

We use the same dataset from OptionMetrics' IvyDB Europe \cite{optionmetrics} as in \cite{cohen2022estimating} for training the neural-SDE model. To be specific, OptionMetrics offers historical daily settlement prices\footnote{These prices are calculated from implied volatilities that are interpolated using a methodology based on a kernel smoothing algorithm, according to the OptionMetrics' IvyDB Europe reference manual \cite{optionmetrics}.} for calls with expiries of 30, 60, 91, 122, 152, 182, 273, 365, 547, and 730 calendar days, at deltas of 0.2, 0.25, 0.3, 0.35, 0.4, 0.45, 0.5, 0.55, 0.6, 0.65, 0.7, 0.75, and 0.8. We collect daily call option prices from 2nd January 2002 to 31st December 2018 for the two most traded vanilla European options listed on Eurex, namely EURO STOXX 50{\small\textsuperscript\textregistered} index options and DAX{\small\textsuperscript\textregistered} options.

\begin{figure}[!ht]
    \begin{subfigure}[b]{.31\textwidth}
    \centering
        \includegraphics[scale=.66]{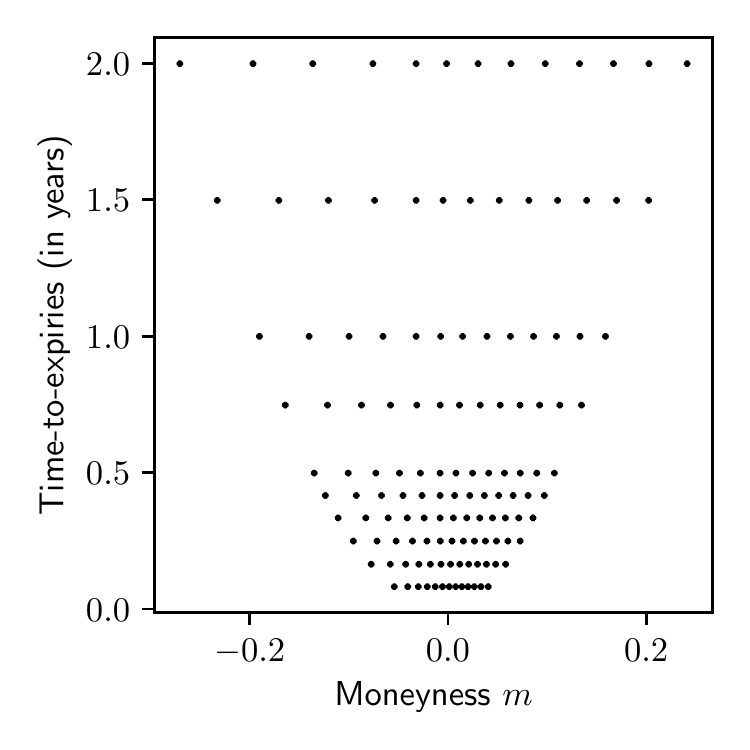}
        \caption{Liquid lattice $\mathcal{L}_\text{liq}$.}
        \label{fig:eurex_lattice}
    \end{subfigure}
    \hfill
    \begin{subfigure}[b]{.68\textwidth}
    \centering
        \includegraphics[scale=.66]{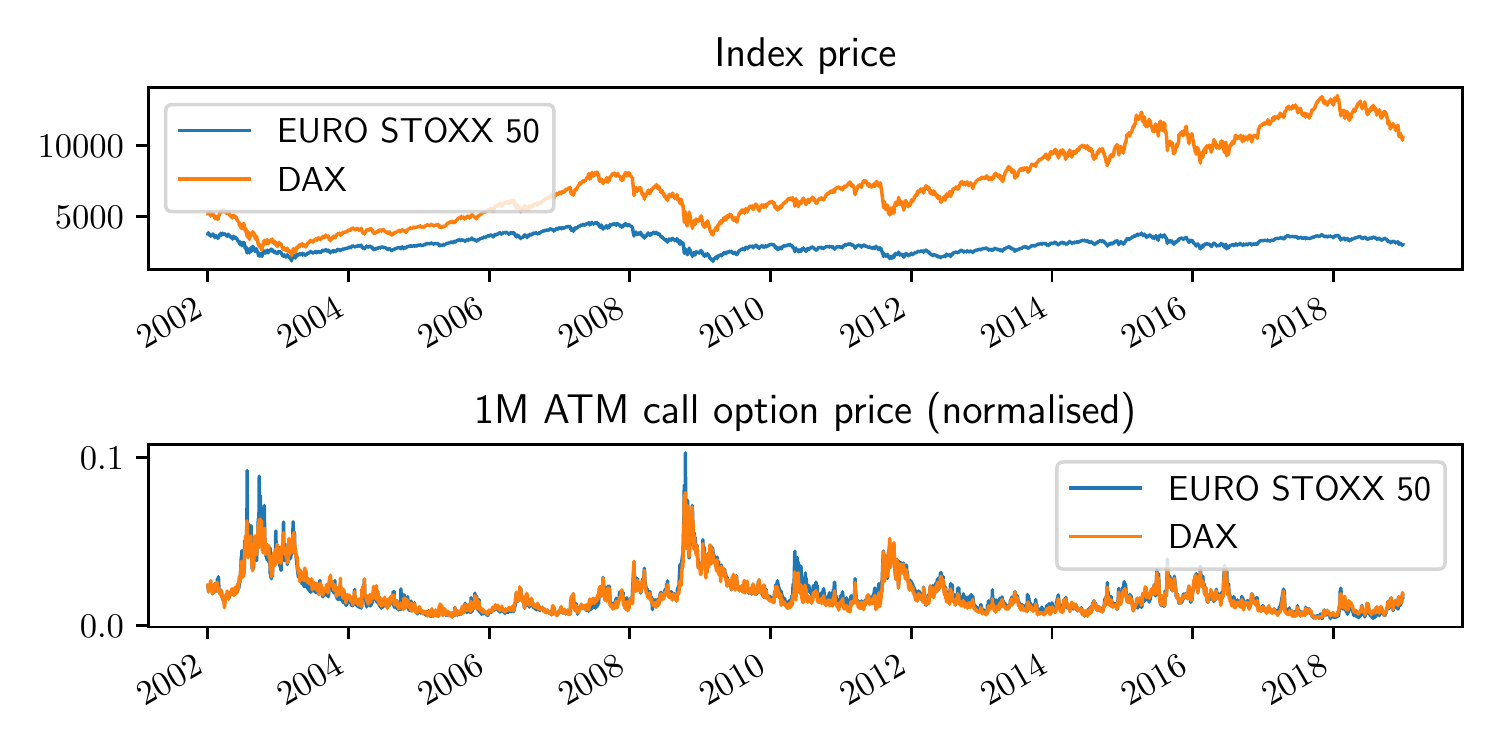}
        \caption{Historical daily prices (underlying indices and 1M ATM call options).}
        \label{fig:eurex_ts}
    \end{subfigure}
    \caption{Scattergram of liquid option lattice and historical prices (training data).}
\end{figure}

For every expiry and each delta-quoted option, we choose the median of its historical moneynesses to define our liquid option lattice $\mathcal{L}_\text{liq}$ (see details in Appendix A of \cite{cohen2022estimating});
We choose the historical \emph{median} moneynesses (for each delta-quoted option) to define our liquid option lattice $\mathcal{L}_\text{liq}$;
the liquid range $\mathcal{R}_\text{liq}$ is computed as the convex hull of the liquid lattice. We show the liquid lattice $\mathcal{L}_\text{liq}$ in Figure \ref{fig:eurex_lattice}. In addition, in Figure \ref{fig:eurex_ts}, we show historical daily prices for the EURO STOXX 50 index, the DAX index, and the 1M at-the-money (ATM) normalised call options written on the two indices. As discussed in \cite{cohen2022estimating}, we assume that the volatility processes of the two indices have similar features, and therefore concatenate the normalised call price data of the two option books when decoding factors and training the neural-SDE model for factors.

\paragraph{Testing data.}

We analyse the out-of-sample hedging performance on EURO STOXX 50 index options. 
For the testing data, we use raw end-of-day settlement prices of \textit{actually} traded contracts,
rather than using standardised option prices that are interpolated from raw prices for a standard set of expiries and deltas as in the training data. 
While the standardised prices are convenient for decoding factors and estimating statistical models, the raw prices are more accurate in reflecting options' values, which are important for calculating hedging errors. Specifically, we collect daily prices for traded call options from 2rd January 2019 to 30th September 2021 (700 trading days) for the out-of-sample hedging analysis. 
The testing period thus covers the market turmoil of early 2020 induced by the COVID-19 pandemic.

\paragraph{The trained neural-SDE market model.}

We refer the reader to Section 3 and Appendix B.2 of \cite{cohen2022estimating} for the detailed factor decoding and SDE estimation procedures. Here, we show the decoded factors, the estimated price basis functions $G$ and the trained neural-diffusion functions $\sigma$, which are ingredients for computing hedge ratios.

\begin{figure}[!ht]
    \centering
    \includegraphics[scale=.66]{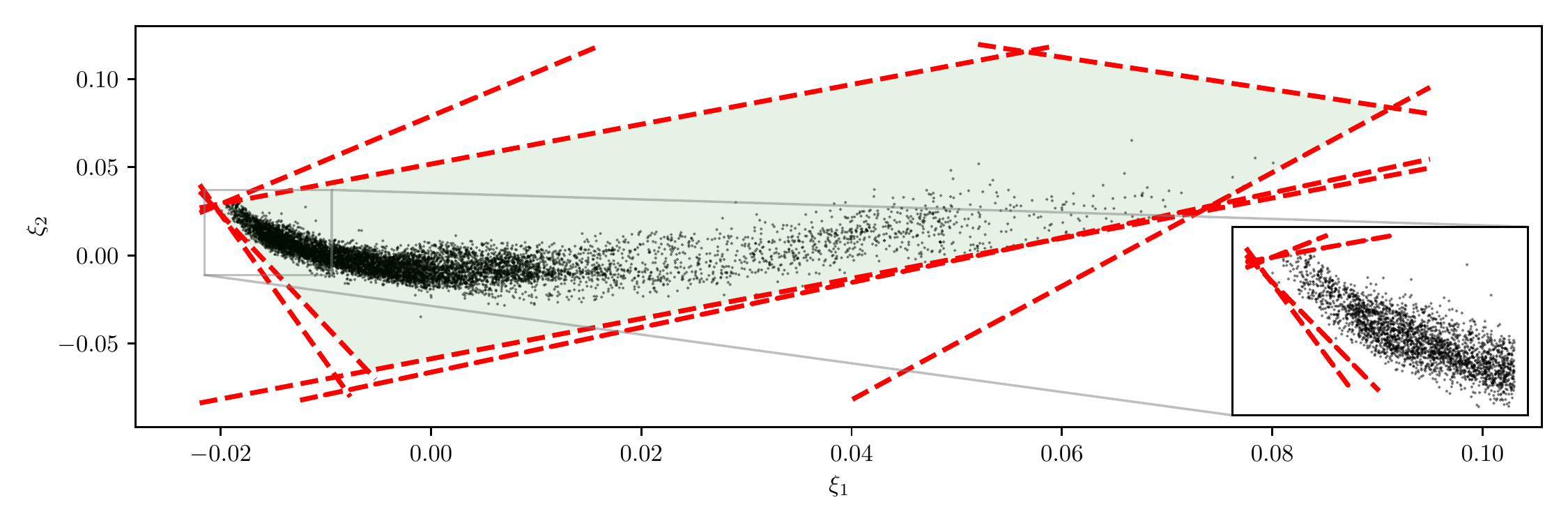}
    \caption{Trajectory (black dots) of the $\mathbb{R}^2$ factors and the corresponding static arbitrage constraints (red dashed lines) projected to the $\mathbb{R}^2$ factor space.}
    \label{fig:factors}
\end{figure}

Shown in Figure \ref{fig:factors} is the cloud of factor realisations decoded from the training data. Given these factors, we use the linear programming method (Caron, McDonald and Ponic \cite{caron1989}) to eliminate redundant constraints in the system \eqref{eq:factor_static_arbitrage}, which is the projection of the original no-arbitrage constraints, constructed in price space, to the $\mathbb{R}^2$ factor space. This results in only 8 constraints, which we indicate as red dashed lines in Figure \ref{fig:factors}. The convex polygonal domain bounded by these constraints (light green area) is the statically arbitrage-free zone for the factors, that is, provided the factor process remains in this region, we are guaranteed to have no static arbitrage in the reconstructed call prices on our liquid lattice. 

\begin{figure}[!ht]
    \centering
    \includegraphics[scale=.66]{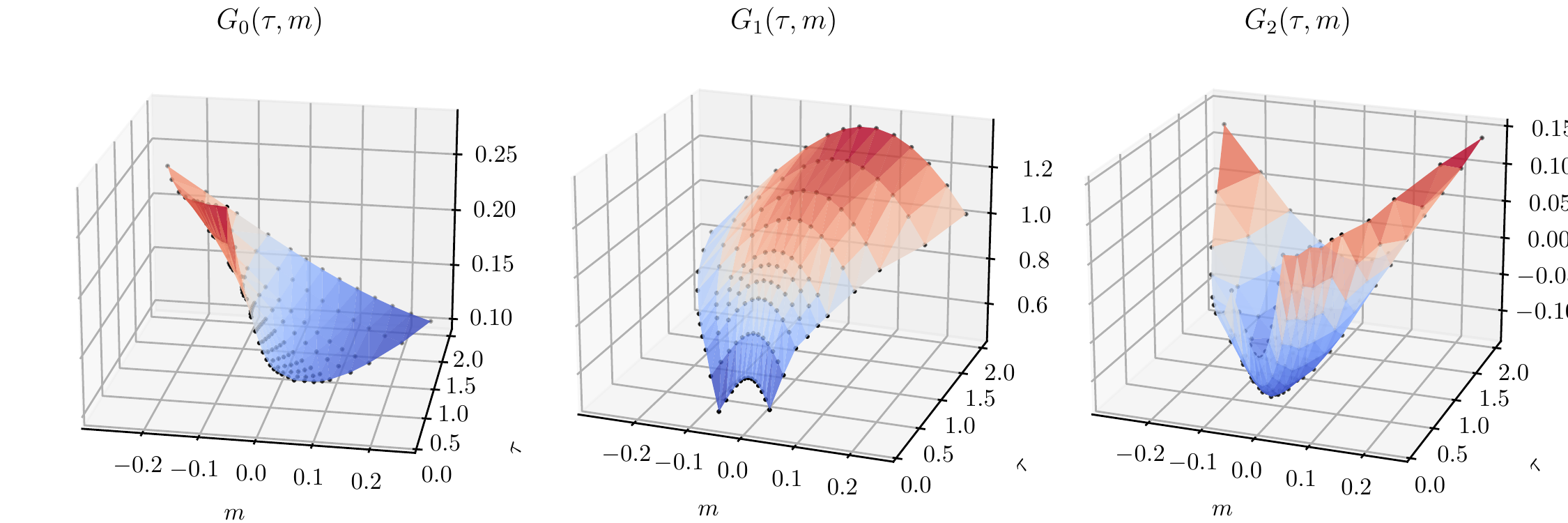}
    \caption{Price basis functions of the normalised call price surface.}
    \label{fig:Gs_primary}
\end{figure}

Next, we plot the price basis functions of these two factors, denoted as $G_1$ and $G_2$, as well as $G_0$, the constant term of $\tilde{c}$, in Figure \ref{fig:Gs_primary}. The points in the liquid lattice (in $(\tau, m)$ coordinates) are also shown on this plot. Here the real-valued functions $G_1$ and $G_2$ are obtained by interpolating the price basis vectors $\mathbf{G}_1$ and $\mathbf{G}_2$, and $G_0$ is obtained by interpolating the normalised call prices averaged over time.

\begin{figure}[!ht]
    \centering
    \includegraphics[scale=.66]{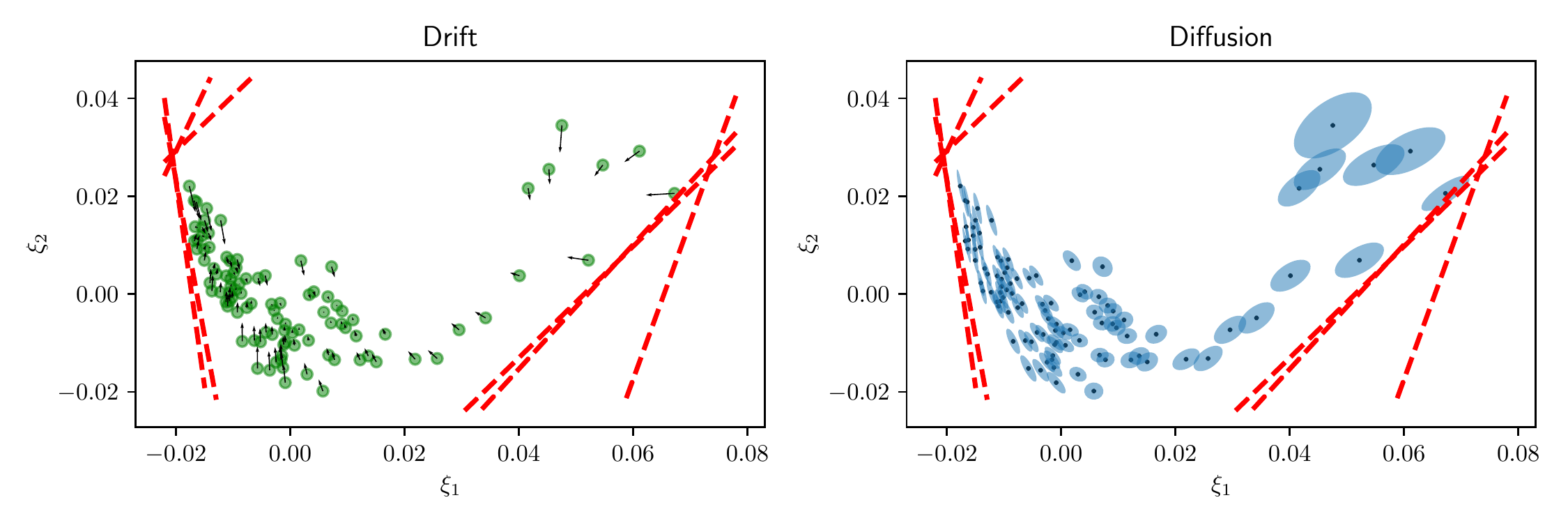}
    \caption{Drift vectors (arrows on the left plot) and diffusion matrices (ellipses representing the principal components of the diffusion on the right plot) for some randomly selected factor data points.}
    \label{fig:learnt_drift_diffusion}
\end{figure}

With the decoded factor data as input, we train a two-factor neural-SDE market model; in Appendix \ref{apd:nsde_mdl}, we demonstrate the goodness-of-fit of the trained model. Finally, we sample a few factor data points, and visualise their drift and diffusion coefficients, approximated by neural networks, in Figure \ref{fig:learnt_drift_diffusion}. As observed in the trajectory of the factor data in Figure \ref{fig:factors}, the data set is distributed around a lower-dimensional manifold. For samples on the periphery of the dataset, the learnt drifts tend to point inwards to the manifold, while the principal direction of their diffusions tend to align with the closest boundary.

\subsection{Tested option portfolios and hedging error calculation}

We assess the performance of using the trained neural-SDE market model for hedging a variety of representative option portfolios. In particular, we define the \emph{naive} portfolio on some day as the equally-weighted portfolio that consists of all traded call option contracts within the liquid range $\mathcal{R}_\textrm{liq}$ on the day. In addition, we consider a variety of portfolios from outright options to heavily hedged portfolios\footnote{In practice, some of these portfolios are constructed to trade views on volatility directions or skews, hence hedging them may contradict the intended purpose of making profits when market moves as expected. By including these portfolios in the hedging analysis, we are more motivated by examining the performance of hedging different risk profiles than giving practical implications.}. The list of tested portfolios is given in Table \ref{tab:test_portfolio}, with detailed description in Appendix \ref{apd:test_strategy}.

\begin{table}[!ht]
    \centering
    \footnotesize
\begin{tabular}{cccccccccc}
\toprule
\textbf{Portfolio} & Naive & Outright & \begin{tabular}[c]{@{}c@{}}Delta\\ spread\end{tabular} & \begin{tabular}[c]{@{}c@{}}Delta\\ butterfly\end{tabular} & Strangle & \begin{tabular}[c]{@{}c@{}}Calendar\\ spread \end{tabular} & VIX \\ \midrule
\textbf{Number} & 1 & 70 & 210 & 30 & 30 & 45 & 1 \\
\bottomrule
\end{tabular}
    \caption{Number of tested portfolios of various types.}
    \label{tab:test_portfolio}
\end{table}

Both the sensitivity-based and MV-based hedging strategies are linear with respect to the constituents of the option portfolio, because ``the hedging portfolios can be interpreted as orthogonal projections of contingent claims onto the closed linear subspace of hedgeable portfolios'', as argued by Tankov \cite{tankov2011pricing}. Consequently, to hedge a portfolio of options written on the same underlying, we compute the hedge ratios for every option in the portfolio and then add them up.

Suppose there are $L$ testing samples (between 2nd January 2019 and 30th September 2021) on trading days $\{t_l\}_{l=1,\dots,L}$, we define, for an option portfolio $\Pi$, two metrics for measuring its PnL $\Delta t$-variations:
\begin{subequations}
\begin{align}
\overline{\mathcal{E}}^2 (\Pi, \Delta t) & = \frac{1}{L-1} \sum_{l=1}^L \left(  \Pi_{t_l + \Delta t} - \Pi_{t_l}\right)^2, \label{eq:pnl_var_overall} \\
\widehat{\mathcal{E}}^2_{t_l} (\Pi, \Delta t, \lambda) & = 
\begin{cases}
\left(  \Pi_{t_1 + \Delta t} - \Pi_{t_1} \right)^2, & \textrm{ if } l = 1, \\
\lambda \widehat{\mathcal{E}}^2_{t_{l-1}} (\Pi, \Delta t, \lambda) + (1 - \lambda) \left(  \Pi_{t_l + \Delta t} - \Pi_{t_l} \right)^2, & \textrm{ for } l = 2, \dots, L.
\end{cases} \label{eq:pnl_var_now}
\end{align}
\end{subequations}
We note that $\overline{\mathcal{E}} (\Pi, \Delta t) > 0$ measures the variation of $\Pi$'s PnLs over the entire testing period, and $\widehat{\mathcal{E}}_{t_l} (\Pi, \Delta t, \lambda) > 0$ is the exponentially-weighted moving average (EWMA) PnL variation up to date $t_l$, with the weight decay factor $\lambda \in (0,1)$. On each trading day $t_l$ for the testing data samples, we suppose that an option portfolio with value $V_{t_l}$ has been written. We hedge the portfolio with liquid instruments, hold the hedged portfolio, denoted $\Pi_{t_l}$, until $t_l+\Delta t$ and then close it, and therefore measure the $\Delta t$-hedging error by the percentages:
\begin{equation}
\overline{\varepsilon} (\Delta t) = \frac{\overline{\mathcal{E}} (\Pi, \Delta t)}{\overline{\mathcal{E}} (V, \Delta t)} \times 100\%, \quad
\widehat{\varepsilon}_{t_l} (\Delta t, \lambda) = \frac{\widehat{\mathcal{E}}_{t_l} (\Pi, \Delta t, \lambda)}{\widehat{\mathcal{E}}_{t_l} (V, \Delta t, \lambda)} \times 100\%.
\label{eq:def_hedging_error}
\end{equation}
While $\overline{\varepsilon} (\Delta t)$ measures average relative hedging errors over the entire testing period, we use $\widehat{\varepsilon}_{t_l} (\Delta t, \lambda)$ for generating more time-sensitive measurements of hedging errors. In the following analysis, we will focus on studying hedging performance under two different rebalancing frequencies: daily ($\Delta t = 1$ trading day) and weekly ($\Delta t = 5$ trading days).

\subsection{Delta hedging of the naive portfolio}

We first compare performance of delta hedging using Black--Scholes models, Heston models and neural-SDE market models. In addition, because Heston models and market models take into account more risk factors than $S$, we follow the MV-based approach to compute their hedge ratios. We list the delta hedging strategies and the formulas of their hedge ratios in Table \ref{tab:hedge_ratios_delta}. 

\begin{table}[!h]
\footnotesize
\centering
\begin{tabular}{llcc}
\toprule
\multirow{2}{*}{Hedging strategy} & \multirow{2}{*}{$X^S$} & \multicolumn{2}{c}{Hedging error $\overline{\varepsilon} (\Delta t)$} \\
\cmidrule(lr){3-4}
 & & $\Delta t = 1$ trading day & $\Delta t = 5$ trading days \\
\cmidrule(lr){1-1} \cmidrule(lr){2-2} \cmidrule(lr){3-3} \cmidrule(lr){4-4}
Delta (BS) & $\Delta_\textrm{bs} (V)$ in \eqref{eq:bs_greeks} & 20.19\% & 33.48\% \\
Delta (Heston-MV) & $\Delta_\textrm{h(mv)} (V)$ in \eqref{eq:heston_delta_mv}  & 19.77\% & 33.33\% \\
Delta (nSDE-MV) & $\Delta_\textrm{mv} (V)$ in \eqref{eq:omega}  & 18.44\% & 30.35\% \\
\bottomrule
\end{tabular}
\caption{Formulas of hedge ratios under various delta hedging strategies and corresponding relative hedging errors $\overline{\varepsilon}(\Delta t)$ over the entire testing period.}
\label{tab:hedge_ratios_delta}
\end{table}

\begin{figure}[!ht]
    \centering
    \begin{subfigure}[b]{.49\textwidth}
    \centering
        \includegraphics[scale=\figscaleapp]{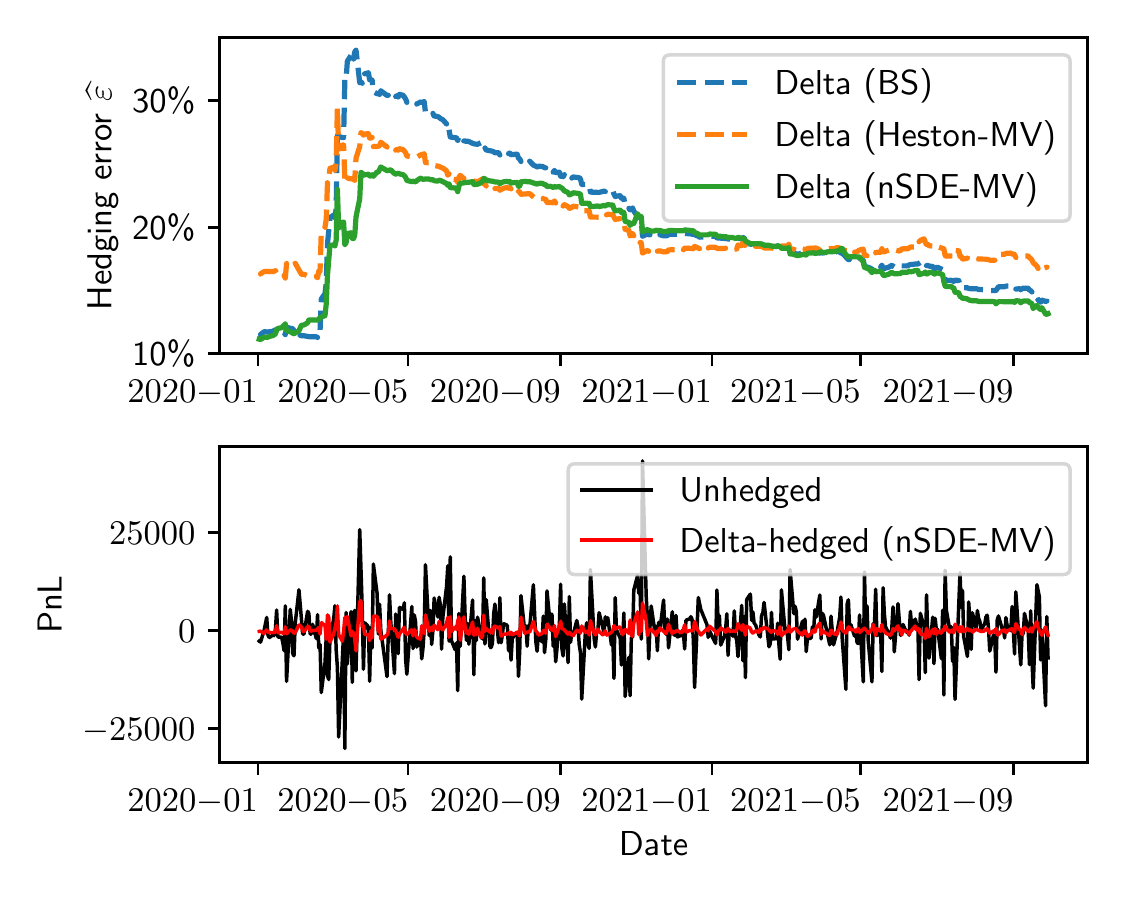}
    \caption{Daily rebalancing.}
    \end{subfigure}
    \hfill
    \begin{subfigure}[b]{.49\textwidth}
    \centering
        \includegraphics[scale=\figscaleapp]{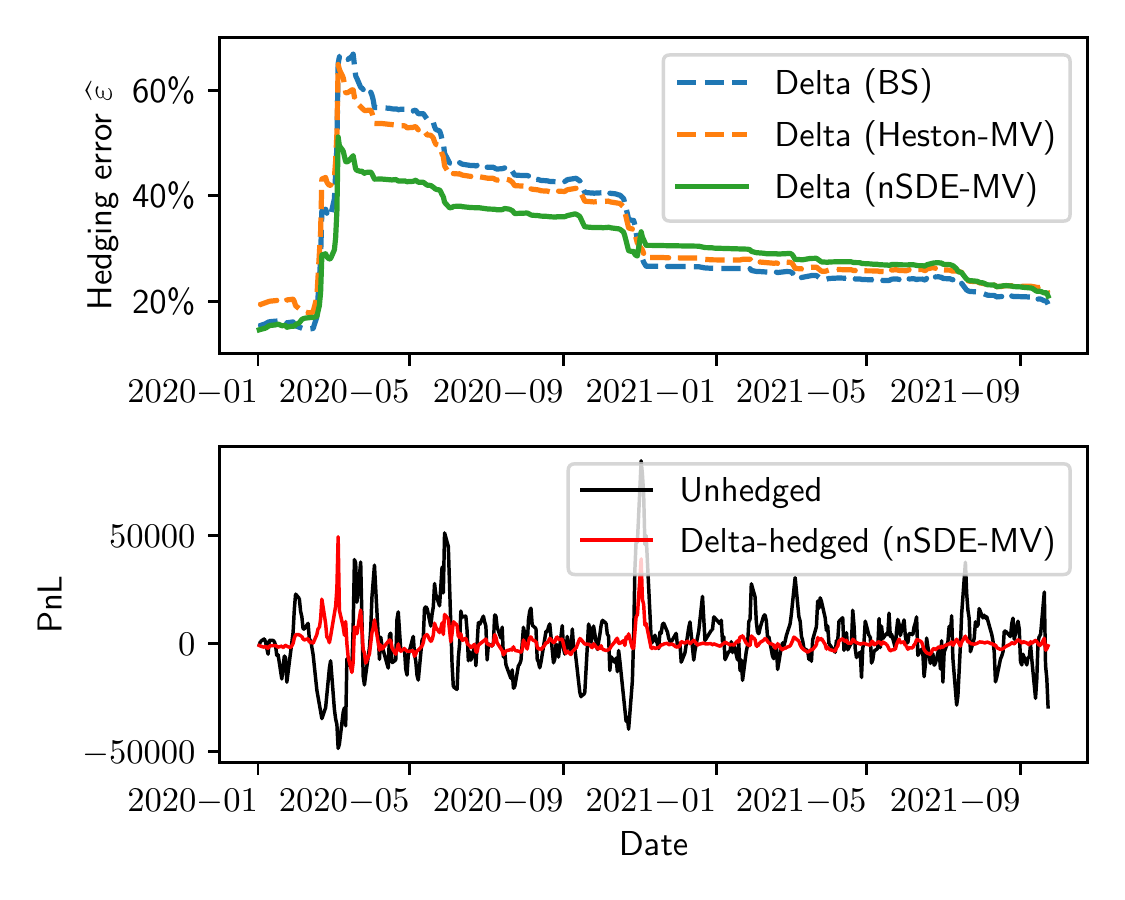}
    \caption{Weekly rebalancing.}
    \end{subfigure}
    \caption{\textit{Top} - EWMA hedging errors $\widehat{\varepsilon}_t(\Delta t, \lambda=0.99)$ for the three delta hedging strategies. \textit{Bottom} - time series of the PnLs for the unhedged naive portfolio and for the nSDE-MV delta-hedged portfolio.}
    \label{fig:h_dh_errorts}
\end{figure}

Table \ref{tab:hedge_ratios_delta} also reports $\overline{\varepsilon} (\Delta t)$, the average relative $\Delta t$-hedging errors over the testing period, for daily and weekly rebalancing cases. We observe the smallest hedging errors with nSDE-MV delta hedging, and the largest errors  with BS delta hedging. To investigate how the three delta hedging strategies perform over time, we show in Figure \ref{fig:h_dh_errorts} their relative EWMA hedging errors $\widehat{\varepsilon}_t(\Delta t, \lambda=0.99)$. We use the data of the first year, i.e. 2019, as seed and do not show them in the plots. In both daily and weekly cases, nSDE-MV delta hedging greatly reduces hedging errors, when compared with BS delta hedging, from March to September in 2020, when global equity markets were hit severely by the fear of the COVID-19 pandemic and yet to fully recover. Heston-MV delta hedging also improves BS delta hedging, but only slightly in the weekly rebalancing case. 

\begin{figure}[!ht]
\centering
\includegraphics[scale=\figscaleapp]{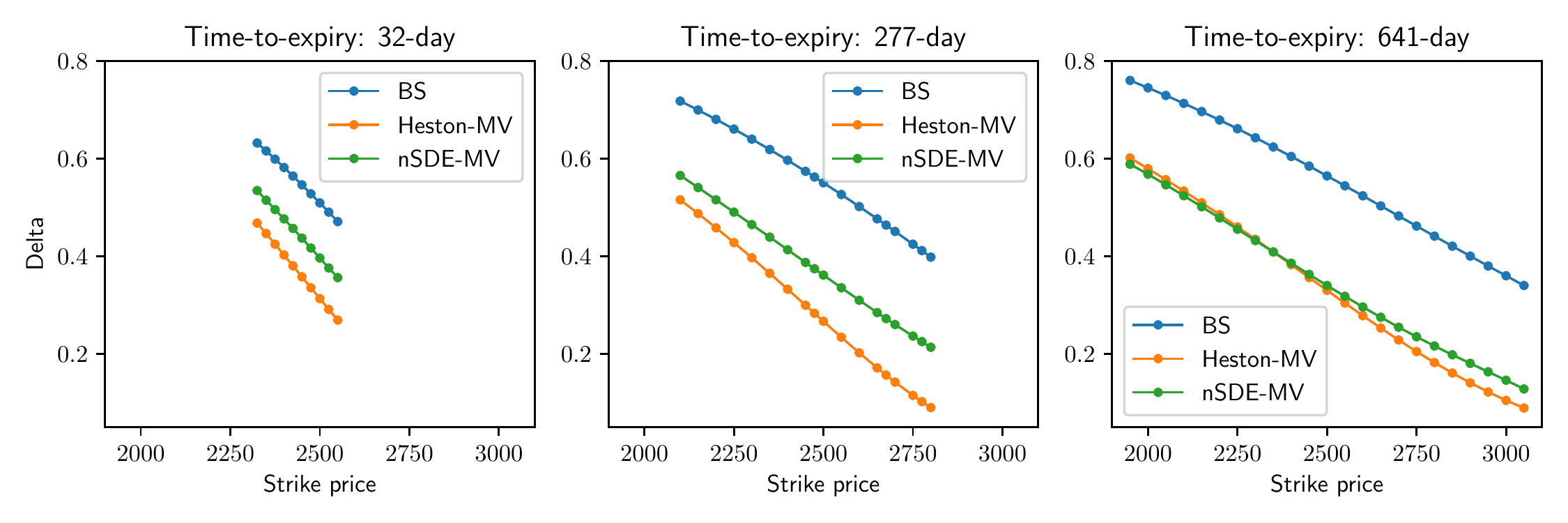}
\caption{Model-implied deltas for call options of various strike prices and expiries as of 16th March 2020.}
\label{fig:h_deltas}
\end{figure}

Neural-SDE and Heston models reduce delta hedging errors, because they are able to capture negative correlations between stock return and volatility, which lead to downward adjustments in deltas; see \eqref{eq:XS_mv1}. To verify this, in Figure \ref{fig:h_deltas}, we show the deltas calculated using the three models for call options of various strike prices and expiries, as of 16th March 2020. In particular, the neural-SDE model gives larger delta adjustments than the Heston model for short-dated options. 

\subsection{Delta-factor hedging of the naive portfolio}
\label{sec:choice_hedge_tenor}

As discussed in Section \ref{sec:specific_mdl}, we focus on examining the performance of sensitivity-based and MV-based delta-$\xi_1$ hedging using neural-SDE market models, relative to delta-vega hedging using Black--Scholes and Heston models. To construct these hedges, we assume the hedging instruments available are the underlying asset $S$ and one liquid option $C_1$.

\paragraph*{Hedging instruments of different tenors.}

Due to liquidity concerns, we restrict our choice of $C_1$, the single option used as a hedging instrument, to traded options that are closest to at-the-money. Nevertheless, we assume that we are free to choose tenors up to 1 year. We analyse the errors of hedging the naive portfolio when near ATM options of different tenors are used as the hedging instrument.

Specifically, we consider the set of 12 consecutive monthly tenors from 1M to 1Y. For each tenor, we identify the available time-to-expiry of traded options that is closest to the tenor, and then choose the closest to ATM option of the identified time-to-expiry as the hedging instrument. Thereafter, we compute the hedging errors for the naive portfolio over the out-of-sample testing period, under the four seven hedging strategies that are summarised in Table \ref{tab:hedge_ratios} (along with the formulas of corresponding hedge ratios).

\begin{table}[!h]
\footnotesize
\centering
\begin{tabular}{lllr}
\toprule
Hedging strategy & $X^S$ & $X^{C_1}$ & Relevant formulas \\
\cmidrule(lr){1-1} \cmidrule(lr){2-3} \cmidrule(lr){4-4}
Delta-vega (BS) & $\Delta_\mathrm{bs}(V) - X^{C_1} \cdot \Delta_\mathrm{bs}(C_1)$ & $\mathcal{V}_\mathrm{bs} (V) / \mathcal{V}_\mathrm{bs} (C_1)$ & \eqref{eq:bs_greeks} \\
Delta-vega (Heston) & $\Delta_\mathrm{h}(V) - X^{C_1} \cdot \Delta_\mathrm{h}(C_1)$ & $\mathcal{V}_\mathrm{h} (V) / \mathcal{V}_\mathrm{h} (C_1)$ & \eqref{eq:heston_delta_mv} \\
Delta-$\xi_1$ (Sens.) & $\Delta(V) - X^{C_1} \cdot \Delta(C_1)$ & $\Delta^1 (V) / \Delta^1  (C_1)$ & \eqref{eq:omega}, \eqref{eq:hedge_pd_mktmdl}, \eqref{eq:hedge_ratio_pd} \\
Delta-$\xi_1$ (MV) & $\Delta_\textrm{mv}(V) - X^{C_1} \cdot \Delta_\textrm{mv}(C_1)$ & $\mathcal{G} (V) / \mathcal{G}  (C_1)$ & \eqref{eq:omega}, \eqref{eq:hedge_mv_mktmdl}, \eqref{eq:hedge_ratio_mv}, \eqref{eq:mv_G} \\
\bottomrule
\end{tabular}
\caption{Formulas of hedge ratios under various hedging strategies.}
\label{tab:hedge_ratios}
\end{table}

\begin{figure}[!ht]
    \centering
    \begin{subfigure}[b]{.49\textwidth}
    \centering
        \includegraphics[scale=\figscaleapp]{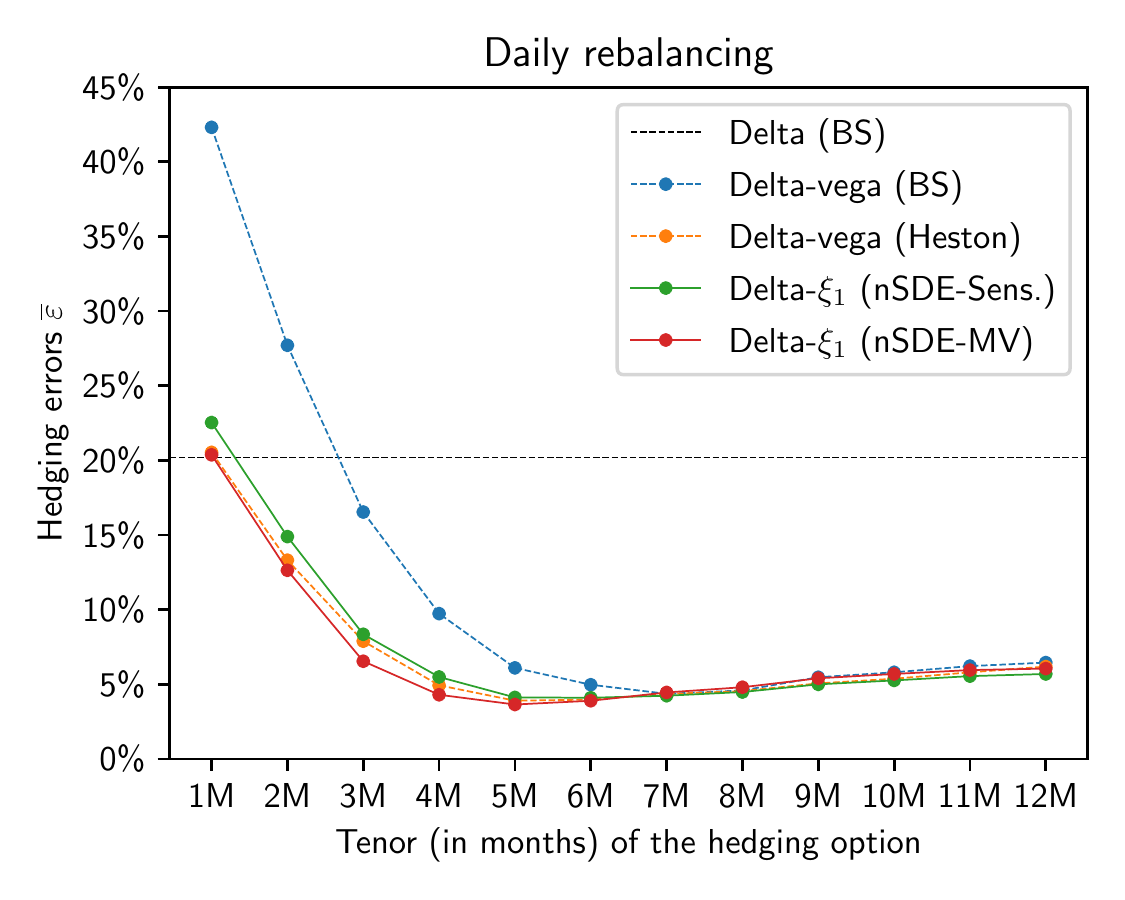}
    \end{subfigure}
    \hfill
    \begin{subfigure}[b]{.49\textwidth}
    \centering
        \includegraphics[scale=\figscaleapp]{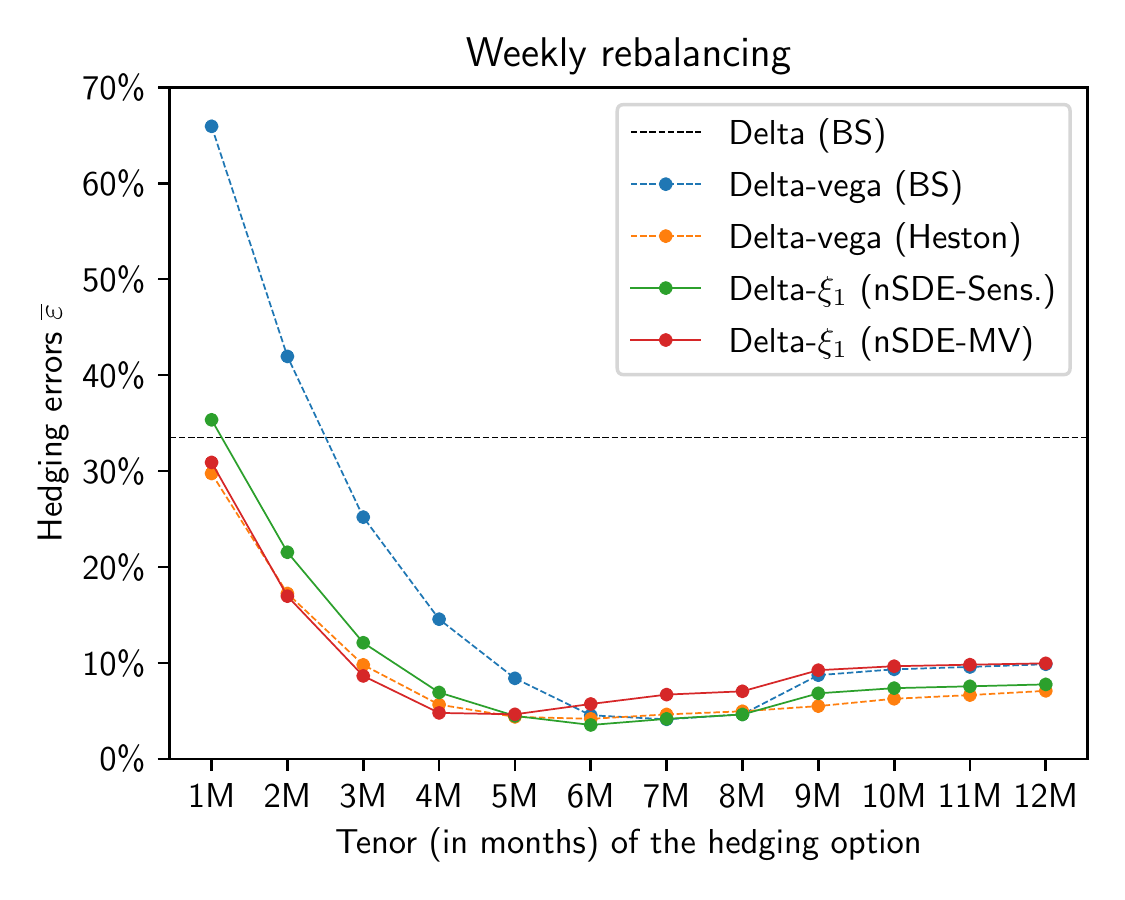}
    \end{subfigure}
    \caption{Average relative errors $\overline{\varepsilon}$ of hedging the naive option portfolio when near ATM options of different tenors are used as the hedging instrument.}
    \label{fig:h_tenor}
\end{figure}

In Figure \ref{fig:h_tenor}, we show how hedging errors change with the use of hedging options of different tenors, for both daily and weekly rebalancing cases. Overall, the performance of delta-vega (BS- and Heston-based) and delta-$\xi_1$ (sensitivity- and MV-based) hedging are much worse when short-dated options are used as the hedging instrument. In particular, using 1M ATM options as the hedging instrument gives consistently larger errors than (simpler) delta hedging. The potential reason is that, as we have seen in Figure \ref{fig:vega_xi1}, short-dated options have relatively smaller vegas and $\xi_1$-exposures, making them less effective in hedging long-dated options in the portfolio.

In the daily rebalancing case, delta-$\xi_1$ hedging seems to be at least as good as BS delta-vega hedging, regardless of which tenor of option is used as the hedging instrument. Meanwhile, Heston delta-vega hedging leads to very similar hedging errors to delta-$\xi_1$ hedging. In particular, delta-$\xi_1$ hedging and Heston delta-vega hedging outperform BS delta-vega hedging more
significantly when shorter-dated options are used as the hedging instrument. In practice, in markets where short-dated options are much more liquid than long-dated options (e.g.\ FX), we would then be more comfortable with hedging $\xi_1$-exposures or Heston vegas than BS vegas. Overall, the performance of delta-$\xi_1$ hedging and Heston delta-vega hedging is \emph{less} sensitive than that of BS delta-vega hedging to the choice of hedging option. When the hedging option's time-to-expiry is at least 7 months, the performance of the four hedging strategies is similar. Comparing sensitivity-based delta-$\xi_1$ hedging and MV-based delta-$\xi_1$ hedging, taking into account the correlation between factors does reduce the hedging errors if the hedging option's time-to-expiry is less than 7 months.

When the hedged portfolios are rebalanced weekly, we observe similar behaviours of the four hedging strategies as seen in the daily rebalancing case. MV-based delta-$\xi_1$ hedging outperforms sensitivity-based delta-$\xi_1$ hedging only when the hedging option's time-to-expiry is less than 5 months.

\paragraph*{Hedging errors over time.}

\begin{figure}[!ht]
    \centering
    \begin{subfigure}[b]{.49\textwidth}
    \centering
        \includegraphics[scale=\figscaleapp]{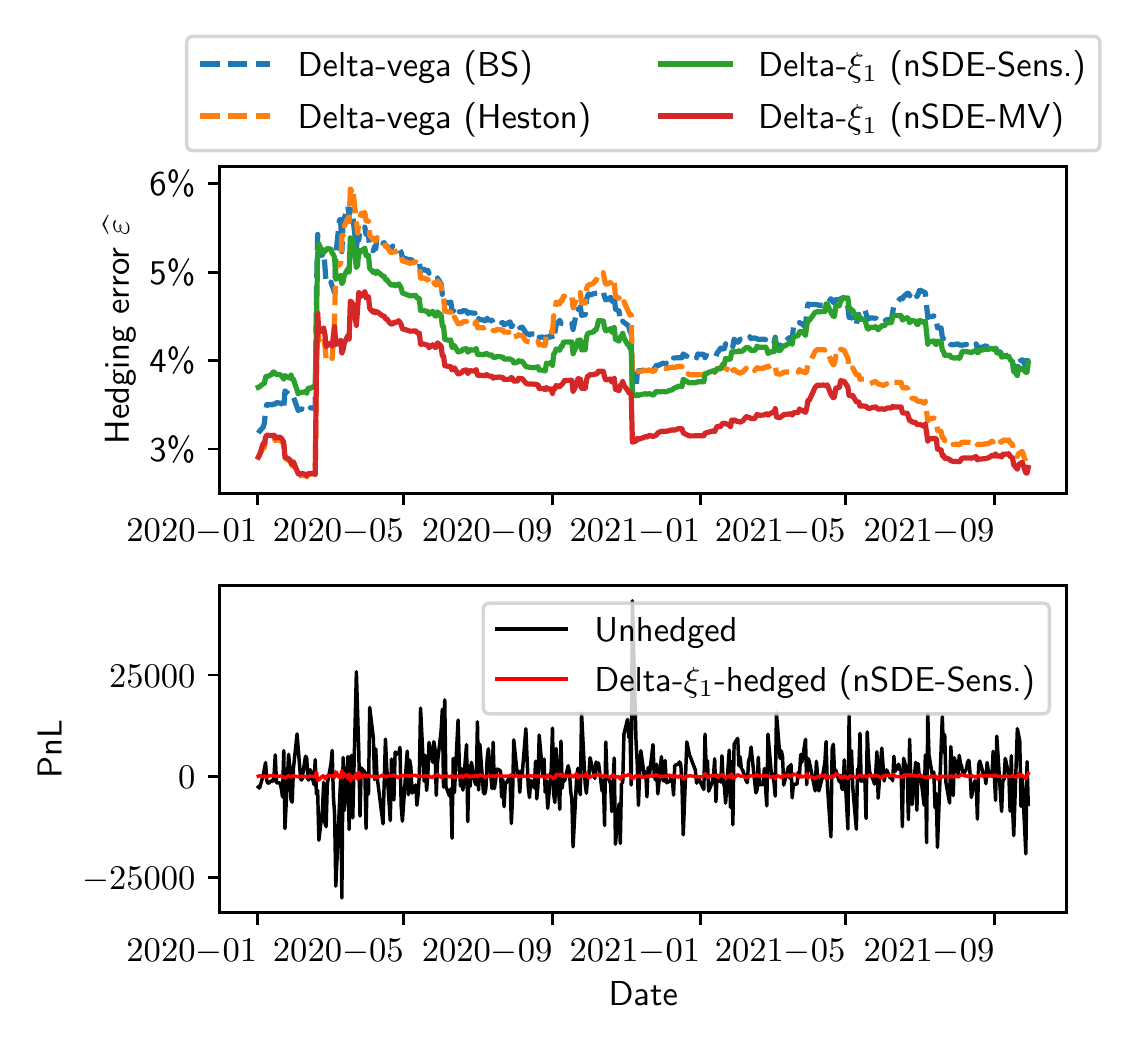}
    \caption{Daily rebalancing.}
    \label{fig:h_errorts_daily}
    \end{subfigure}
    \hfill
    \begin{subfigure}[b]{.49\textwidth}
    \centering
        \includegraphics[scale=\figscaleapp]{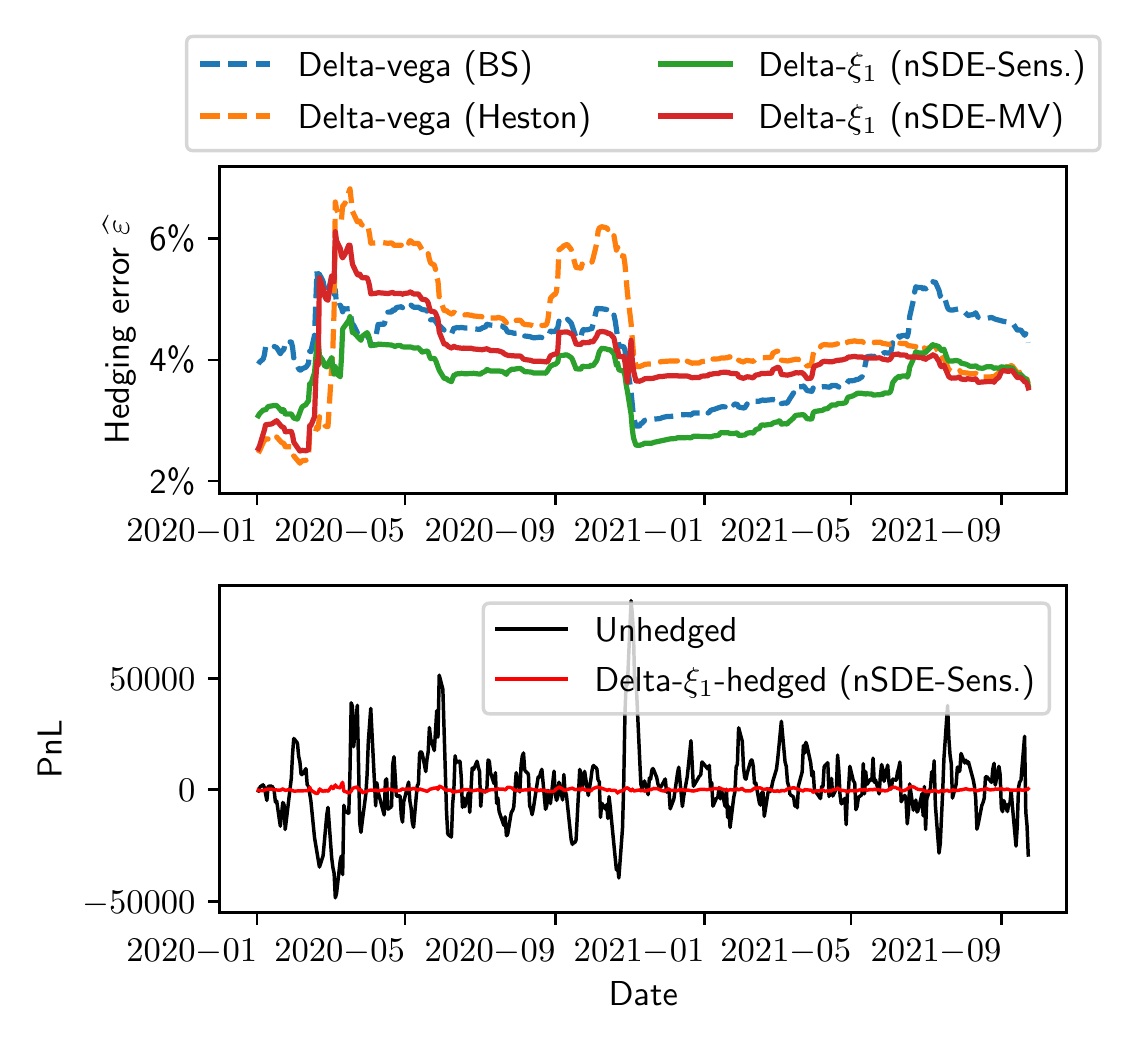}
    \caption{Weekly rebalancing.}
    \label{fig:h_errorts_weekly}
    \end{subfigure}
    \caption{\textit{Top} - EWMA hedging errors $\widehat{\varepsilon}_t(\Delta t, \lambda=0.99)$ for the four hedging strategies. \textit{Bottom} - time series of the PnLs for the unhedged naive portfolio and for the nSDE sensitivity-based delta-$\xi_1$-hedged naive portfolio.}
    \label{fig:h_errorts}
\end{figure}

Next, we investigate how the four hedging strategies perform over time. While Figure \ref{fig:h_tenor} shows the average relative hedging errors $\overline{\varepsilon} (\Delta t)$ over the entire testing period, we now present the relative EWMA hedging errors $\widehat{\varepsilon}_t(\Delta t, \lambda=0.99)$. For each hedging strategy, we choose the corresponding \emph{optimal} hedging option that gives the smallest average hedging errors in the testing period. Therefore, in the daily rebalancing case, based on the evidence in Figure \ref{fig:h_tenor}, we choose 7M ATM options as hedging instruments for BS delta-vega hedging, and 5M ATM options as hedging instruments for Heston delta-vega hedging and (sensitivity- and MV-based) delta-$\xi_1$ hedging. In the top plot of Figure \ref{fig:h_errorts_daily}, we report the EWMA hedging errors of the four hedging strategies that are rebalanced on a daily basis. For better illustrating the dynamic changes in the hedging errors, we provide the time series of the PnLs for the unhedged naive portfolio and for the sensitivity-based delta-$\xi_1$-hedged naive portfolio in the bottom plots.

First, all the investigated hedging strategies yield similar dynamics of hedging errors. Hedging errors experienced a sharp rise in March 2020, when the fear of the COVID-19 pandemic severely hit global equity markets. The hedging errors maintained a higher-than-usual level until the winter of 2020. Meanwhile, we observe concurrent larger fluctuations of the portfolio PnLs in the bottom plot. When comparing across different hedging strategies, MV-based delta-$\xi_1$ hedging is consistently the best, and maintains a 1\% advantage in hedging errors when compared with BS delta-vega hedging. Heston delta-vega hedging gives roughly the same hedging errors as MV-based delta-$\xi_1$ hedging before March 2020, but performs worse in stressed periods. Sensitivity-based delta-$\xi_1$ hedging is better than BS delta-vega hedging except before March 2020.

We report the same hedging error statistics for weekly-rebalanced hedging in Figure \ref{fig:h_errorts_weekly}. In this comparison analysis, we choose 7M ATM options as hedging options for BS delta-vega hedging, 6M ATM options for Heston delta-vega hedging and sensitivity-based delta-$\xi_1$ hedging, and 5M ATM options for MV-based delta-$\xi_1$ hedging. Sensitivity-based delta-$\xi_1$ hedging is almost always better than BS and Heston delta-vega hedging. However, unlike the daily rebalancing case, MV-based delta-$\xi_1$ hedging is worse than sensitivity-based delta-$\xi_1$ hedging except before March 2020.

\begin{remark}
Why does MV-based delta-$\xi_1$ hedging perform worse than sensitivity-based delta-$\xi_1$ hedging, in the weekly rebalancing case, after March 2020? By construction, MV-based delta-$\xi_1$ hedging is beneficial only when the neural-SDE model gives good adjustments to $\xi_1$-exposures based on accurate \emph{forecasts} of the correlation between $\xi_1$ and $\xi_2$ within the rebalancing horizon. When the rebalancing horizon gets larger (i.e.\ from one day to one week) and the instantaneous correlation between $\xi_1$ and $\xi_2$ changes drastically, the neural-SDE might give a poor model for the average correlation within the rebalancing horizon and therefore deteriorate the hedging performance. To verify these arguments, we present empirical evidence and analysis of decoded factors in Appendix \ref{apd:analysis_factors}.
\end{remark}

\paragraph*{Hedging instruments of different moneynesses.}

Although we have restricted the choice of hedging instruments to ATM options due to liquidity concerns, it is interesting to see whether ITM and OTM options could achieve lower hedging errors. Therefore, we fix the tenors and see how options of different moneynesses perform when used as hedging instruments.

\begin{figure}[!ht]
    \centering
    \begin{subfigure}[b]{.49\textwidth}
    \centering
        \includegraphics[scale=\figscaleapp]{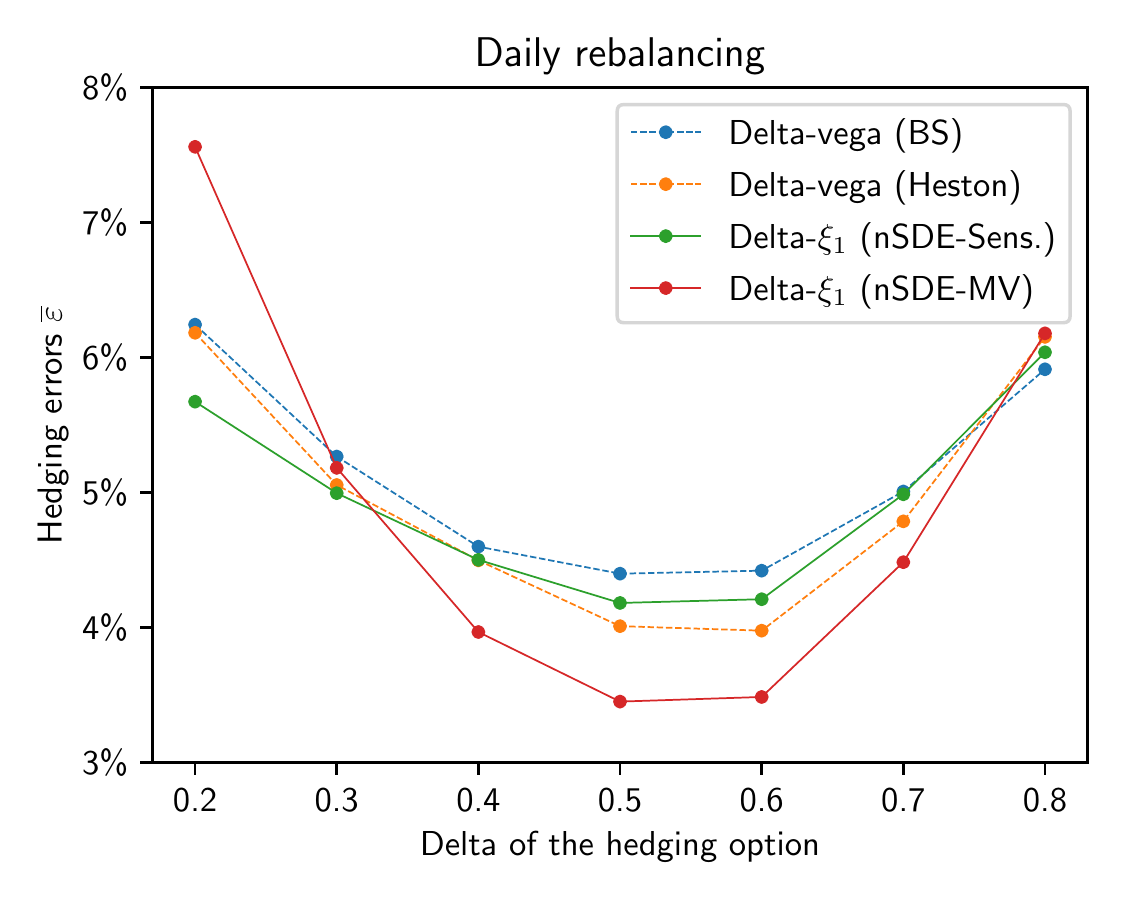}
    \end{subfigure}
    \hfill
    \begin{subfigure}[b]{.49\textwidth}
    \centering
        \includegraphics[scale=\figscaleapp]{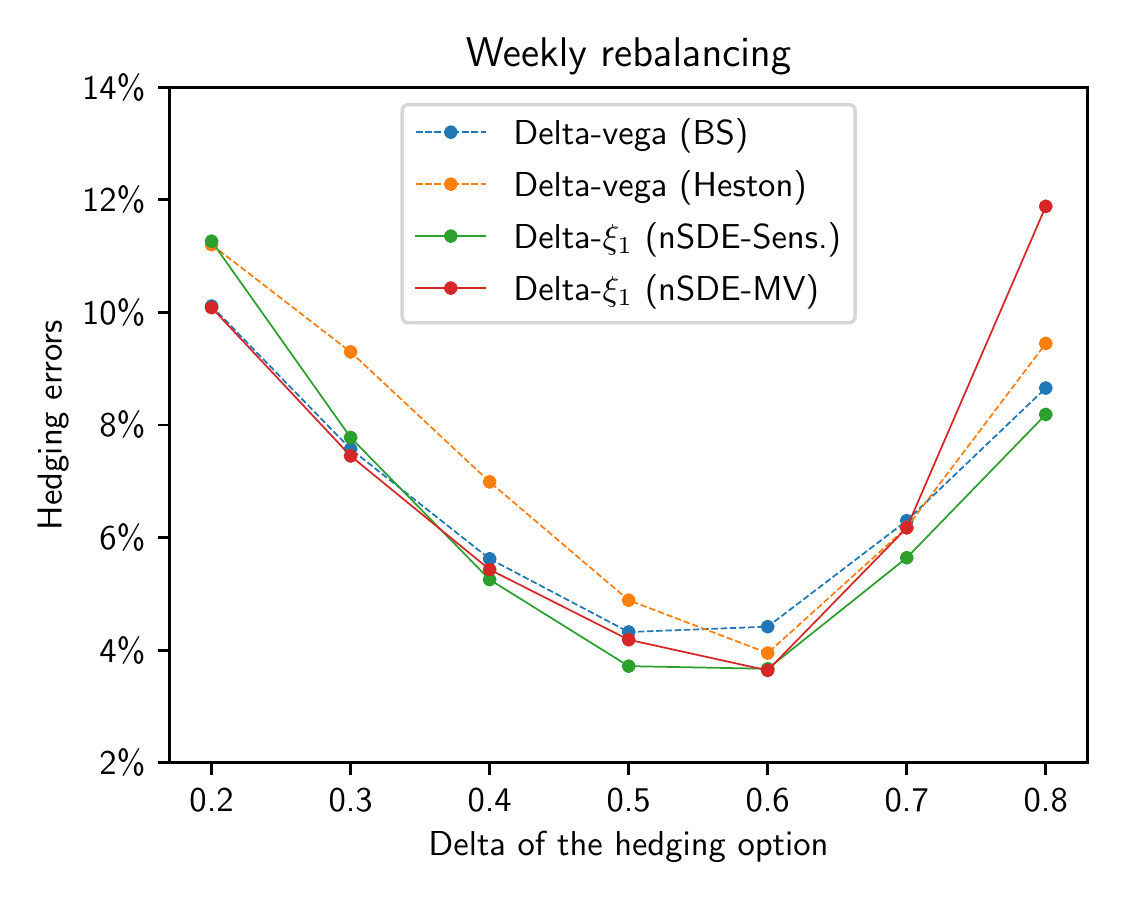}
    \end{subfigure}
    \caption{Average relative errors $\overline{\varepsilon}$ of hedging the naive option portfolio when options of different moneynesses (in deltas) are used as the hedging instrument.}
    \label{fig:h_mny}
\end{figure}

In Figure \ref{fig:h_mny}, we show how hedging errors change with the use of hedging options of different moneynesses, for both daily and weekly rebalancing cases. For the daily rebalancing case, we choose 7M options as hedging instruments for BS delta-vega hedging, and 5M options for Heston delta-vega hedging and (sensitivity- and MV-based) delta-$\xi_1$ hedging. For the weekly rebalancing case, we choose 7M options as hedging instruments for BS delta-vega hedging, 6M options for Heston delta-vega hedging and sensitivity-based delta-$\xi_1$ hedging, and 5M options for MV-based delta-$\xi_1$ hedging. For all the hedging strategies, ATM options achieve the lowest hedging errors (except the Heston delta-vega and MV-based delta-$\xi_1$ hedging in the weekly rebalancing case). This is unsurprising because both vegas and $\xi_1$-exposure reach their peak near the ATM strike level, as seen in Figure \ref{fig:vega_xi1}; hence, ATM options are more effective in reducing a portfolio's vegas and $\xi_1$-exposure.

\paragraph*{MV-based direct hedging.}

We consider an alternative MV-based hedging approach where, rather than requiring zero instantaneous co-variations between hedged portfolios and risk factors, we require a hedged portfolio to have zero instantaneous co-variations with all its hedging instruments. We call this approach MV-based \textit{direct hedging}, and compare its performance with MV-based delta-$\xi_1$ hedging.

Again, assuming the hedging instruments available are the underlying asset $S$ and one liquid call option $C_1$, we derive their hedge ratios under the two-factor neural-SDE model specified in Section \ref{sec:specific_mdl}:
\begin{equation}
    \begin{cases}
    \langle  \diff \Pi,  \diff S \rangle = 0, \\
    \langle  \diff \Pi,  \diff C_1 \rangle = 0; \\
    \end{cases}
    \Rightarrow
    \begin{bmatrix}
    1 & \Delta_\textrm{mv} (C_1) \\
    \Delta_\textrm{mv} (C_1) & \mathcal{H}(C_1, C_1) \\
    \end{bmatrix} 
    \begin{bmatrix}
    X^S \\ X^{C_1}
    \end{bmatrix} = 
    \begin{bmatrix}
    \Delta_\textrm{mv} (V) \\ \mathcal{H}(V, C_1)
    \end{bmatrix},
\end{equation}
where, for two options $U$ and $V$,
\begin{equation}
    \mathcal{H}(U,V) := \Delta(U) \Delta(V) + \frac{1}{\sigma_{1,1}^2 S^2} \begin{bmatrix}
    \Delta(U) & \Delta^1(U) & \Delta^2(U)
    \end{bmatrix} \sigma \sigma^\top \begin{bmatrix}
    \Delta(V) \\ \Delta^1(V) \\ \Delta^2(V)
    \end{bmatrix}.
\end{equation}

MV-based delta-$\xi_1$ hedging constructs hedge ratios by eliminating sensitivities to $(S, \xi_1)$ with adjustments due to correlations between $(S, \xi_1)$ and $\xi_2$. Instead, MV-based direct hedging establishes hedge ratios by eliminating a combination of exposures to all risk factors $(S, \xi_1, \xi_2)$ that affect the hedging option $C_1$.

\begin{figure}[!ht]
    \centering
    \begin{subfigure}[b]{.49\textwidth}
    \centering
        \includegraphics[scale=\figscaleapp]{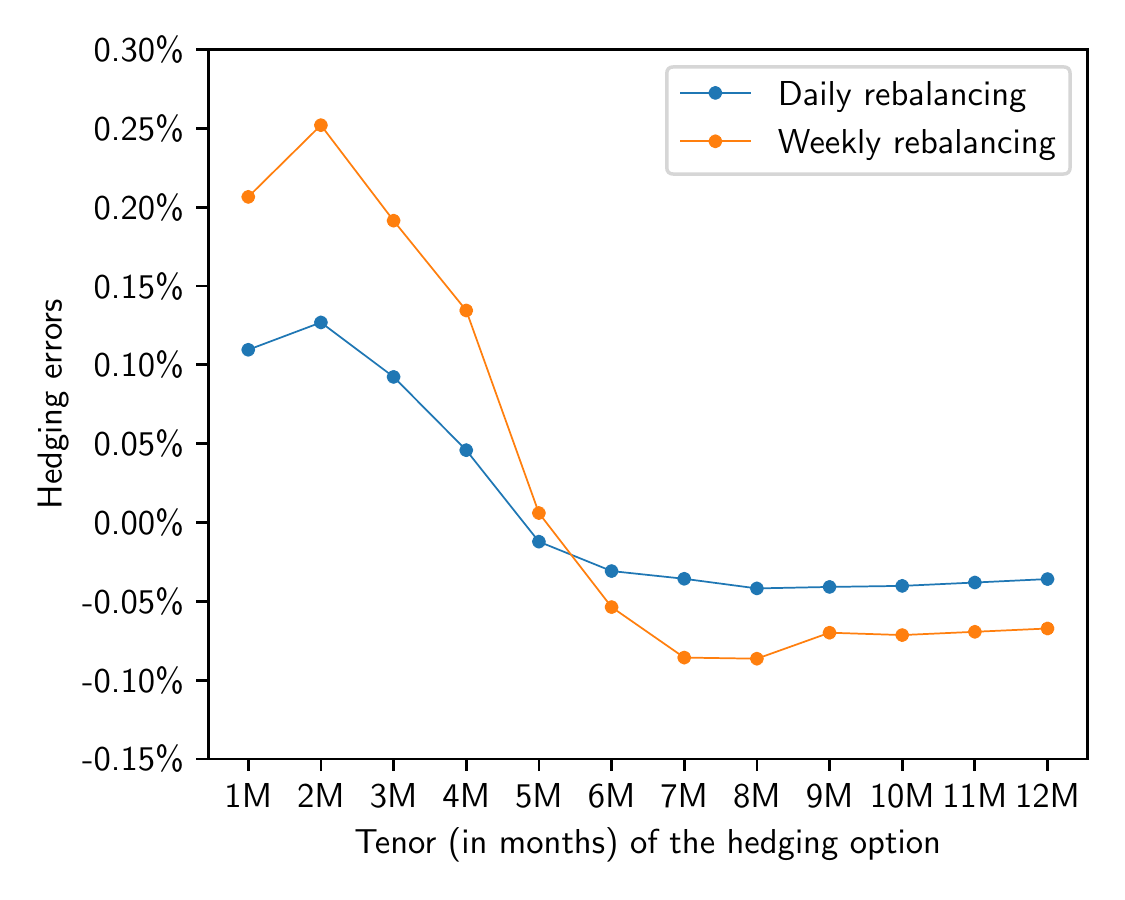}
    \end{subfigure}
    \hfill
    \begin{subfigure}[b]{.49\textwidth}
    \centering
        \includegraphics[scale=\figscaleapp]{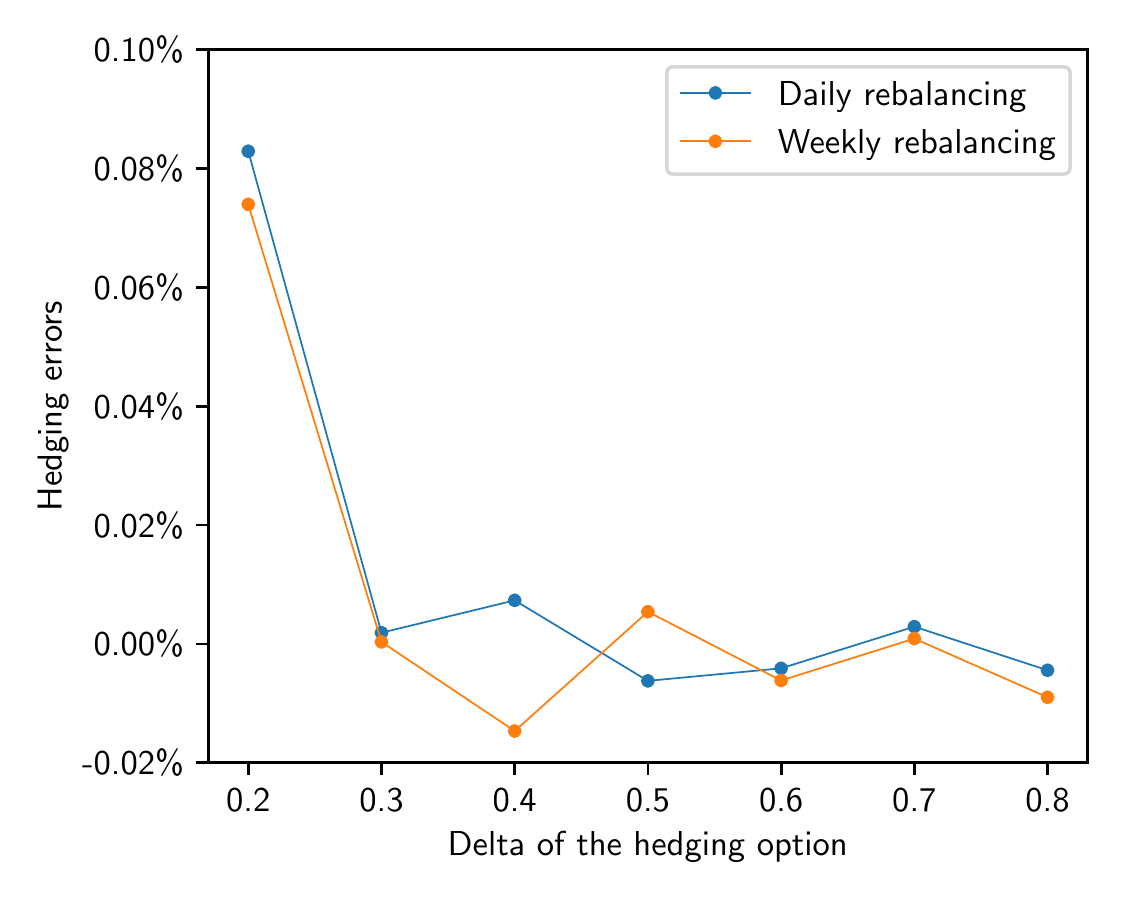}
    \end{subfigure}
    \caption{Difference in average relative hedging errors $\overline{\varepsilon}$ (errors of MV-based delta-$\xi_1$ hedging minus errors of MV-based direct hedging) for the naive option portfolio when options of different tenors (in months) and moneynesses (in deltas) are used as hedging instruments. \textit{Left} - difference in hedging errors when ATM options of different tenors are used as hedging instruments. \textit{Right} - difference in hedging errors when options of specified tenors (corresponding to Figure \ref{fig:h_mny}) and different moneynesses are used as hedging instruments.}
    \label{fig:h_direct}
\end{figure}

It turns out that \textit{MV-based direct hedging leads to quite similar hedging errors to MV-based delta-$\xi_1$ hedging, regardless of which hedging option is used}. To illustrate this, we show in Figure \ref{fig:h_direct} the difference in average relative hedging errors $\overline{\varepsilon}$ (errors of MV-based delta-$\xi_1$ hedging minus errors of MV-based direct hedging) for the naive option portfolio when options of different tenors (in months) and moneynesses (in deltas) are used as hedging instruments. The difference is close to zero in all cases, while there seems to be consistent patterns between the daily and weekly rebalancing cases.

\paragraph*{Concluding remarks.}

To conclude from the evidence presented in this section, choosing different options as a hedging instrument makes a substantial difference to the resultant hedging performance. In particular, the performance of BS delta-vega hedging is more sensitive to the choice of ATM hedging options of different tenors than that of Heston delta-vega hedging and market model delta-$\xi_1$ hedging (see Figure \ref{fig:h_tenor}). In order to allow a fair comparison of hedging performance in the following analysis (for more option portfolios), we choose 7M ATM options as the hedging instruments for all the hedging strategies because using them BS and Heston delta-vega hedging and delta-$\xi_1$ hedging yield similarly good hedging performance for the naive portfolio (see Figure \ref{fig:h_tenor}). This allows us to disentangle the effect of the different payoff characteristics in a situation when the average hedging performance is comparable.

\subsection{Hedging error analysis for more portfolios}
\label{sec:hedge_error_analysis}

We summarise the relative hedging errors $\overline{\varepsilon}$ of different categories of option portfolios in Table \ref{tab:portfolo_hedging}. When there are multiple portfolios within one category (all except the naive and VIX portfolio), the table reports the average hedging errors over all portfolios of the same category. We make the following observations:
\begin{enumerate}[leftmargin=*, label=(\roman*)]
	\item Neural-SDE MV-based delta hedging outperforms BS and Heston-MV delta hedging for the naive portfolio, outrights, strangles and the VIX portfolio, in both daily and weekly rebalancing cases.
    \item For outright and strangle portfolios, (sensitivity-based and MV-based) delta-$\xi_1$ hedging consistently outperforms BS delta-vega hedging by at least $3\%$. Meanwhile, MV-based delta-$\xi_1$ hedging slightly improves sensitivity-based hedging, though it is worse than Heston delta-vega hedging. Compared with other portfolio categories, these two types only involve options of one direction (i.e.\ either all-long or all-short).
    \item For delta spread portfolios, there is no significant difference in hedging their vegas or $\xi_1$-exposures. However, Heston delta-vega hedging reduces at least $7\%$ errors when compared with the other strategies of hedging butterfly portfolios in the weekly rebalancing case.
    \item Sensitivity-based delta-$\xi_1$ hedging achieves the lowest hedging errors for the VIX portfolio, while MV-based delta-$\xi_1$ hedging improves over BS and Heston delta-vega hedging by at least 5\%. This seems to imply that $\xi_1$-exposure is a better indicator of volatility risk than BS or Heston vega.
    \item All the given hedging strategies perform poorly for calendar spread portfolios. The best strategy is the simpler delta hedging, which only manages a reduction of less than 4\% of the PnL fluctuations. 
\end{enumerate}

\begin{table}[!ht]
\centering
\carr
\begin{tabular}{lccccccc}
\toprule
    & \multicolumn{7}{c}{Daily rebalancing ($\Delta t = 1$ trading day)} \\
    \cmidrule(lr){2-8}
Portfolio category & \begin{tabular}[c]{@{}c@{}} Delta \\ (BS) \end{tabular} 
	& \begin{tabular}[c]{@{}c@{}} Delta \\ (Heston-MV) \end{tabular}
	& \begin{tabular}[c]{@{}c@{}} Delta \\ (nSDE-MV) \end{tabular} 
    & \begin{tabular}[c]{@{}c@{}} Delta-vega \\ (BS) \end{tabular} 
    & \begin{tabular}[c]{@{}c@{}} Delta-vega \\ (Heston) \end{tabular} 
    & \begin{tabular}[c]{@{}c@{}} Delta-$\xi_1$ \\ (nSDE-Sens.) \end{tabular}
    & \begin{tabular}[c]{@{}c@{}} Delta-$\xi_1$ \\ (nSDE-MV) \end{tabular} \\ 
 \cmidrule(lr){1-1} \cmidrule(lr){2-4} \cmidrule(lr){5-8}
Naive           & 20.19\% & 19.77\% & \textbf{18.44}\% & 4.36\%  & 4.35\%  & \textbf{4.22}\%  & 4.33\% \\
Outright        & 29.00\% & 28.66\% & \textbf{27.32}\% & 14.89\% & \textbf{10.68}\% & 11.34\% & 11.06\% \\
Delta spread    & \textbf{12.96}\% & 19.05\% & 15.21\% & 11.45\% & 11.47\% & 11.69\% & \textbf{11.38}\% \\
Delta butterfly & \textbf{62.54}\% & 85.41\% & 73.67\% & 48.77\% & \textbf{47.07}\% & 51.84\% & 49.67\% \\
Strangle        & 24.47\% & 23.46\% & \textbf{22.66}\% & 12.05\% & \textbf{8.49}\%  & 9.02\%  & 8.78\%  \\
Calendar spread & \textbf{96.21}\% & 100.3\% & 104.4\% & 164.4\% & \textbf{110.9}\% & 114.2\% & 111.1\% \\
VIX             & 81.54\% & 82.49\% & \textbf{77.24}\% & 59.46\% & 51.59\% & \textbf{51.12}\% & 53.39\% \\
\bottomrule
\end{tabular}
\begin{tabular}{lccccccc}
\toprule
    & \multicolumn{7}{c}{Weekly rebalancing ($\Delta t = 5$ trading days)} \\
    \cmidrule(lr){2-8}
Portfolio category & \begin{tabular}[c]{@{}c@{}} Delta \\ (BS) \end{tabular} 
	& \begin{tabular}[c]{@{}c@{}} Delta \\ (Heston-MV) \end{tabular}
	& \begin{tabular}[c]{@{}c@{}} Delta \\ (nSDE-MV) \end{tabular} 
    & \begin{tabular}[c]{@{}c@{}} Delta-vega \\ (BS) \end{tabular} 
    & \begin{tabular}[c]{@{}c@{}} Delta-vega \\ (Heston) \end{tabular} 
    & \begin{tabular}[c]{@{}c@{}} Delta-$\xi_1$ \\ (nSDE-Sens.) \end{tabular}
    & \begin{tabular}[c]{@{}c@{}} Delta-$\xi_1$ \\ (nSDE-MV) \end{tabular} \\ 
 \cmidrule(lr){1-1} \cmidrule(lr){2-4} \cmidrule(lr){5-8}
Naive           & 33.48\% & 33.33\% & \textbf{30.35}\% & 4.13\%  & 4.64\%  & \textbf{4.09}\%  & 6.13\% \\
Outright        & 43.27\% & 44.13\% & \textbf{39.30}\% & 22.68\% & \textbf{13.50}\% & 16.68\% & 15.41\% \\
Delta spread    & 21.47\% & 28.30\% & \textbf{21.24}\% & 18.48\% & 18.73\% & 19.47\% & \textbf{17.30}\% \\
Delta butterfly & 72.52\% & 97.63\% & \textbf{68.83}\% & 49.21\% & \textbf{38.34}\% & 45.58\% & 50.37\% \\
Strangle        & 38.35\% & 38.02\% & \textbf{34.05}\% & 18.30\% & \textbf{9.48}\%  & 12.41\% & 12.18\% \\
Calendar spread & 96.77\% & \textbf{96.08}\% & 106.0\% & 226.5\% & \textbf{109.7}\% & 140.1\% & 132.2\% \\
VIX             & 87.35\% & 88.74\% & \textbf{82.74}\% & 56.59\% & 51.05\% & \textbf{45.20}\% & 49.71\% \\
\bottomrule
\end{tabular}
\caption{Average hedging errors $\overline{\varepsilon} (\Delta t)$ of different categories of option portfolios under different hedging strategies. The lowest hedging error of each category is highlighted in \textbf{bold}, separately for delta hedging strategies and delta-vega/$\xi_1$-hedging strategies.}
\label{tab:portfolo_hedging}
\end{table}

\begin{figure}[!ht]
    \centering
    \includegraphics[width=\textwidth]{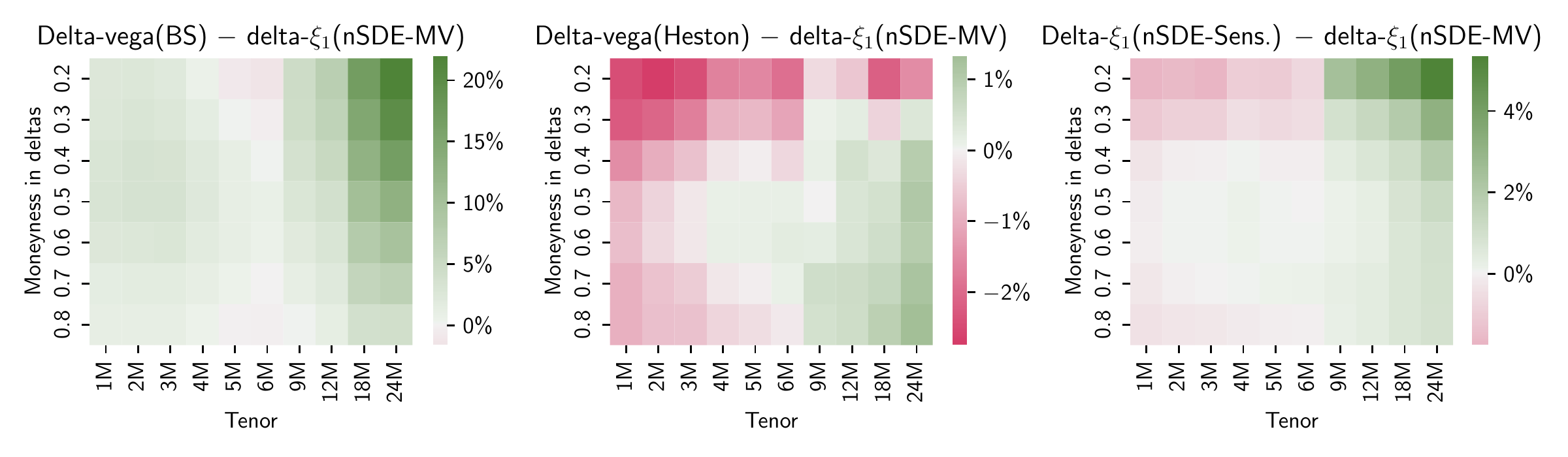}
    \caption{Comparison of hedging errors $\overline{\varepsilon}(\Delta t = 1$ trading day$)$ for positions of outright options with various tenors (i.e.\ time-to-maturities) and moneynesses (measured in deltas), hedged using 7M ATM options.}
    \label{fig:h_heoutrights_daily}
\end{figure}

Next, we compare the hedging performance of the four delta-factor hedging strategies when applied to outright options of different expiries and moneynesses. In Figure \ref{fig:h_heoutrights_daily}, we show the heatmaps of three pairwise hedging error differences, when the hedges are rebalanced on a daily basis. In each plot, given values indicate the second listed strategy performs better, i.e.\ less hedging errors. The leftmost plot shows that MV-based delta-$\xi_1$ hedging outperforms BS delta-vega hedging for almost all combinations of tenors and moneynesses. In particular, the advantage of MV-based delta-$\xi_1$ hedging gets more pronounced for long-dated (18M and 24M) OTM options. It is also interesting to see that BS delta-vega hedging slightly outperforms delta-$\xi_1$ hedging for some 5M and 6M options. The reason may be that the hedging instrument we choose is a 7M near ATM option, while BS delta-vega hedging gives more efficient hedge ratios for options with specifications nearer the hedging instrument. This is consistent with our observation in Section \ref{sec:choice_hedge_tenor} that the performance of BS delta-vega hedging is more sensitive to the choice of the hedging option than that of market model delta-$\xi_1$ hedging. 

The right plot shows that, when hedging $\xi_1$-exposures, it is beneficial to follow the MV-based approach and take into account the correlation between factors only when the options to hedge have long expiries (typically more than 9 months). When compared with Heston delta-vega hedging, as shown in the middle plot, MV-based delta-$\xi_1$ hedging is better for long-dated ATM and ITM options, and is worse for all short-dated options.

\begin{figure}[!ht]
    \centering
    \includegraphics[width=\textwidth]{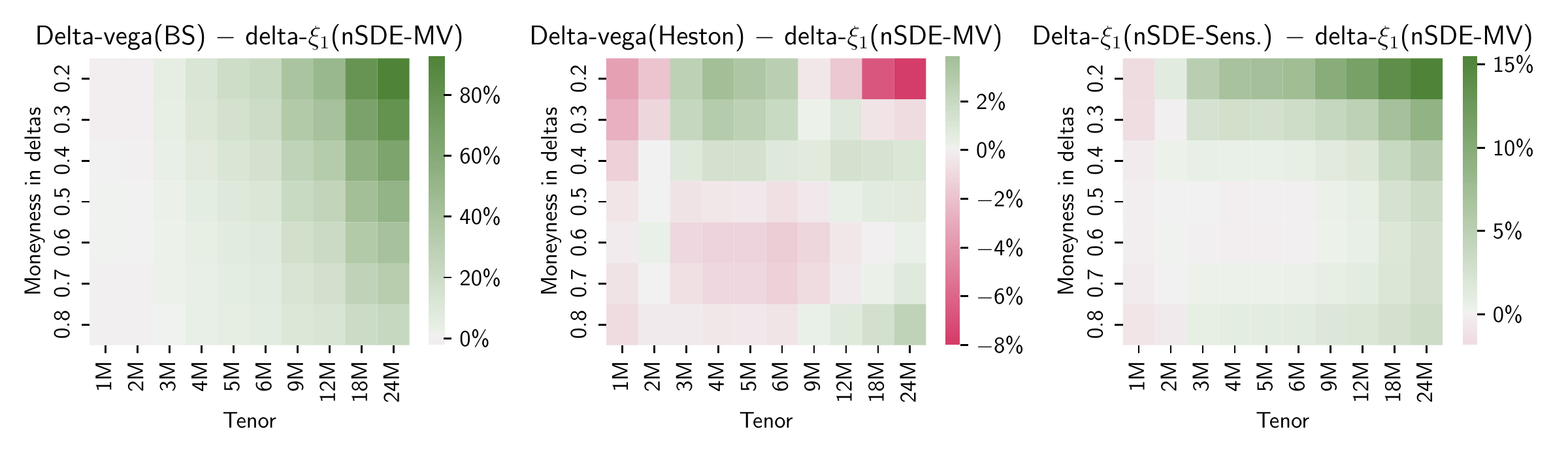}
    \caption{Comparison of hedging errors $\overline{\varepsilon}(\Delta t = 1$ trading day$)$ for positions of outright options with various tenors (i.e.\ time-to-maturities) and moneynesses (measured in deltas), hedging using 2M ATM options.}
    \label{fig:h_heoutrights_daily_2M}
\end{figure}

The disadvantage of MV-based delta-$\xi_1$ hedging over Heston delta-vega hedging for short-dated option is partly due to the tenor choice of the hedging options. Compared with Figure \ref{fig:h_heoutrights_daily}, where we use 7M ATM options to hedge, we show in Figure \ref{fig:h_heoutrights_daily_2M} a similar analysis but with 2M ATM options as hedging instruments. We observe that the relative performance of hedging short-dated options by MV-based delta-$\xi_1$ gets improved. Nevertheless, hedging OTM long-dated options gets worse as shorter-tenor options are used as hedging instruments.

\begin{figure}[!ht]
    \centering
    \includegraphics[width=\textwidth]{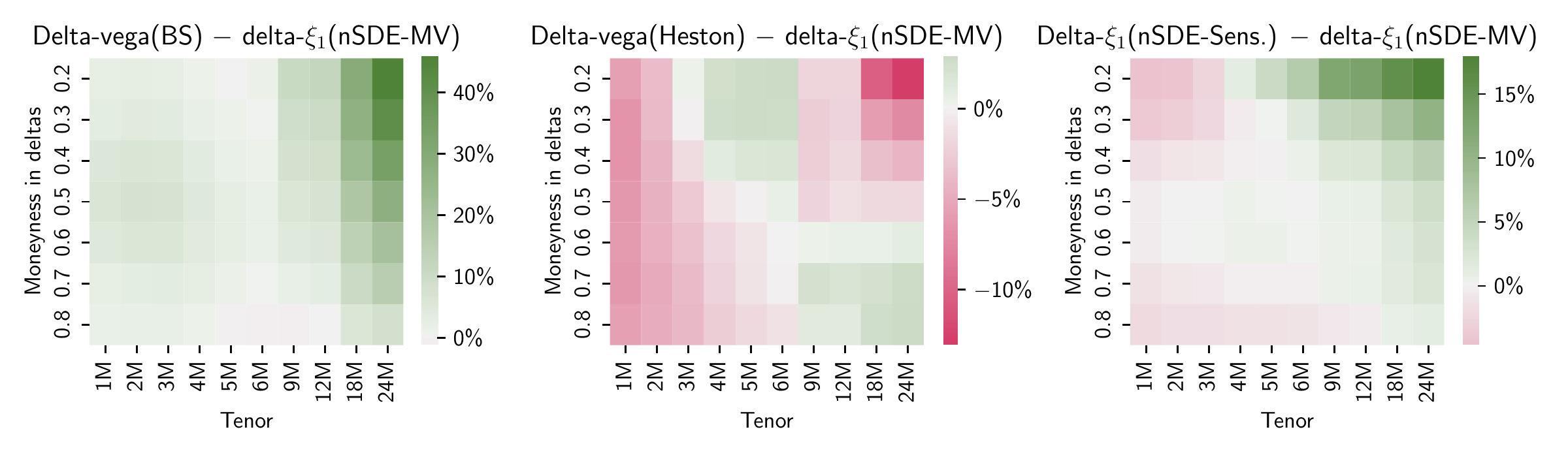}
    \caption{Comparison of hedging errors $\overline{\varepsilon}(\Delta t = 5$ trading days$)$ for positions of outright options with various tenors (i.e.\ time-to-maturities) and moneynesses (measured in deltas), hedging using 7M ATM options.}
    \label{fig:h_heoutrights_weekly}
\end{figure}

For completeness, we show in Figure \ref{fig:h_heoutrights_weekly} the heatmaps of the three pairwise hedging error differences when the hedges are rebalanced on a weekly basis. We have consistent observations with the daily rebalancing case, while the size of the differences gets larger.

\subsubsection*{Revisiting the hedging of the VIX portfolio}

When hedging the VIX portfolio, the best hedging strategy, i.e.\ sensitivity-based delta-$\xi_1$ hedging, only manages to reduce its PnL variations by half (see Table \ref{tab:portfolo_hedging}). This is likely due to the fact that the VIX portfolio consists of approximately 1M options \cite{vix} while the hedging options used in all the hedging strategies have 7M tenor. Hence, to find the optimal hedging instrument to use for the VIX portfolio, we perform a similar hedging instrument analysis as for the naive portfolio in Figure \ref{fig:h_tenor}.

\begin{figure}[!ht]
    \centering
    \begin{subfigure}[b]{.49\textwidth}
    \centering
        \includegraphics[scale=\figscaleapp]{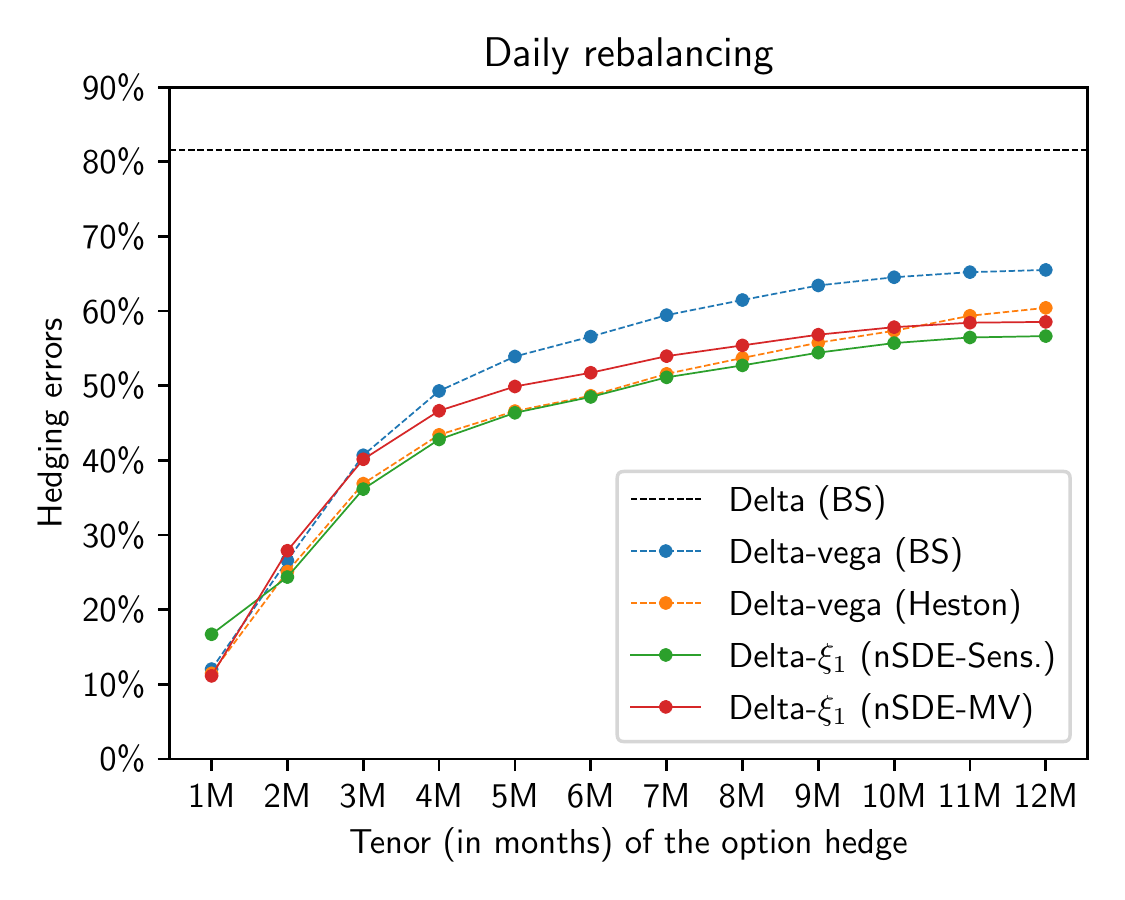}
    \end{subfigure}
    \hfill
    \begin{subfigure}[b]{.49\textwidth}
    \centering
        \includegraphics[scale=\figscaleapp]{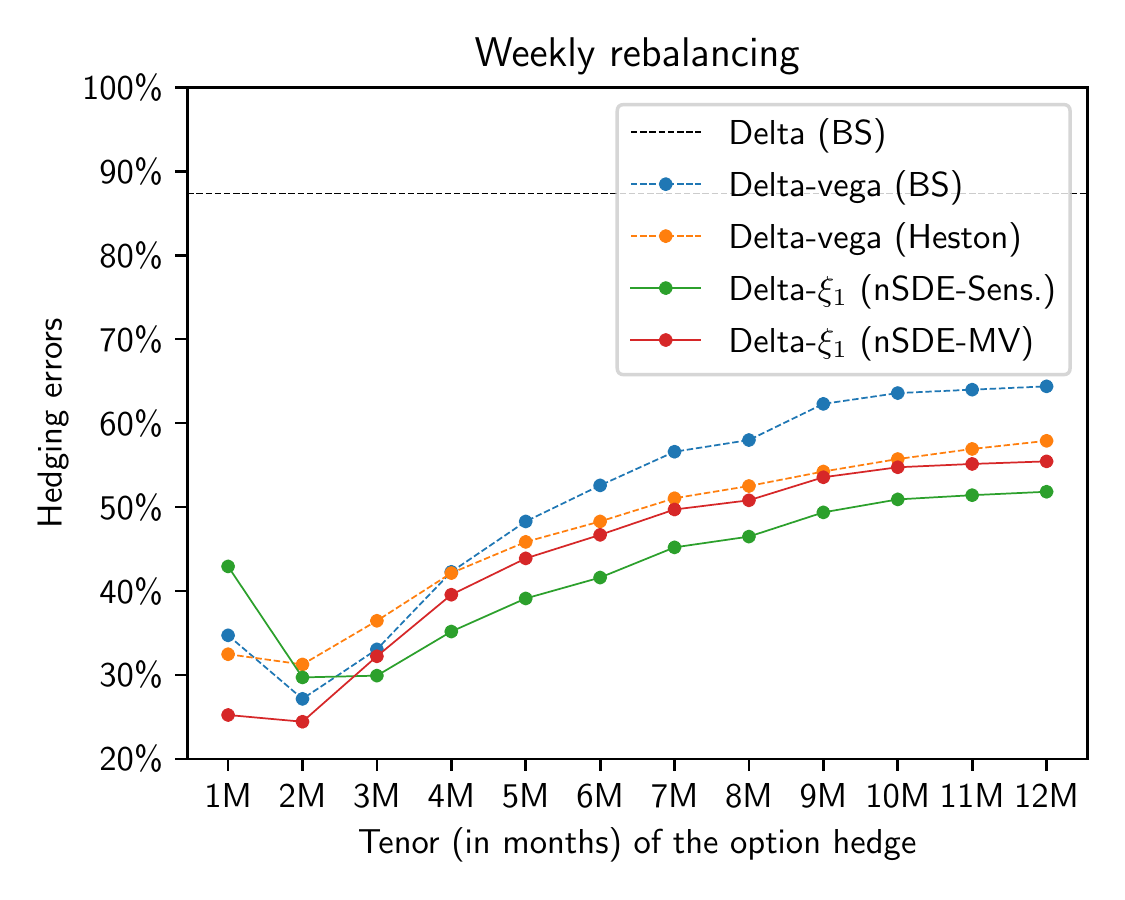}
    \end{subfigure}
    \caption{Average relative errors $\overline{\varepsilon}$ of hedging the VIX option portfolio when near ATM options of different tenors are used as the hedge.}
    \label{fig:h_vix_tenor}
\end{figure}

In Figure \ref{fig:h_vix_tenor}, we show how hedging errors change with the use of hedging options of different tenors, for both daily and weekly rebalancing. In both cases, MV-based delta-$\xi_1$ hedging is at least as good as BS and Heston delta-vega hedging, regardless of which tenor of option is used as the hedge. In practice, it might be more sensible to hedge the VIX with 1M options, which are usually more liquid; and when 1M options are used as hedges, MV-based delta-$\xi_1$ hedging performs the best and significantly improves sensitivity-based delta-$\xi_1$ hedging. Similar to hedging the naive portfolio, the performance of delta-$\xi_1$ hedging is \emph{less} sensitive than that of BS delta-vega hedging to the choice of the hedging option for the VIX portfolio.

\begin{figure}[!ht]
    \centering
    \begin{subfigure}[b]{.49\textwidth}
    \centering
        \includegraphics[scale=\figscaleapp]{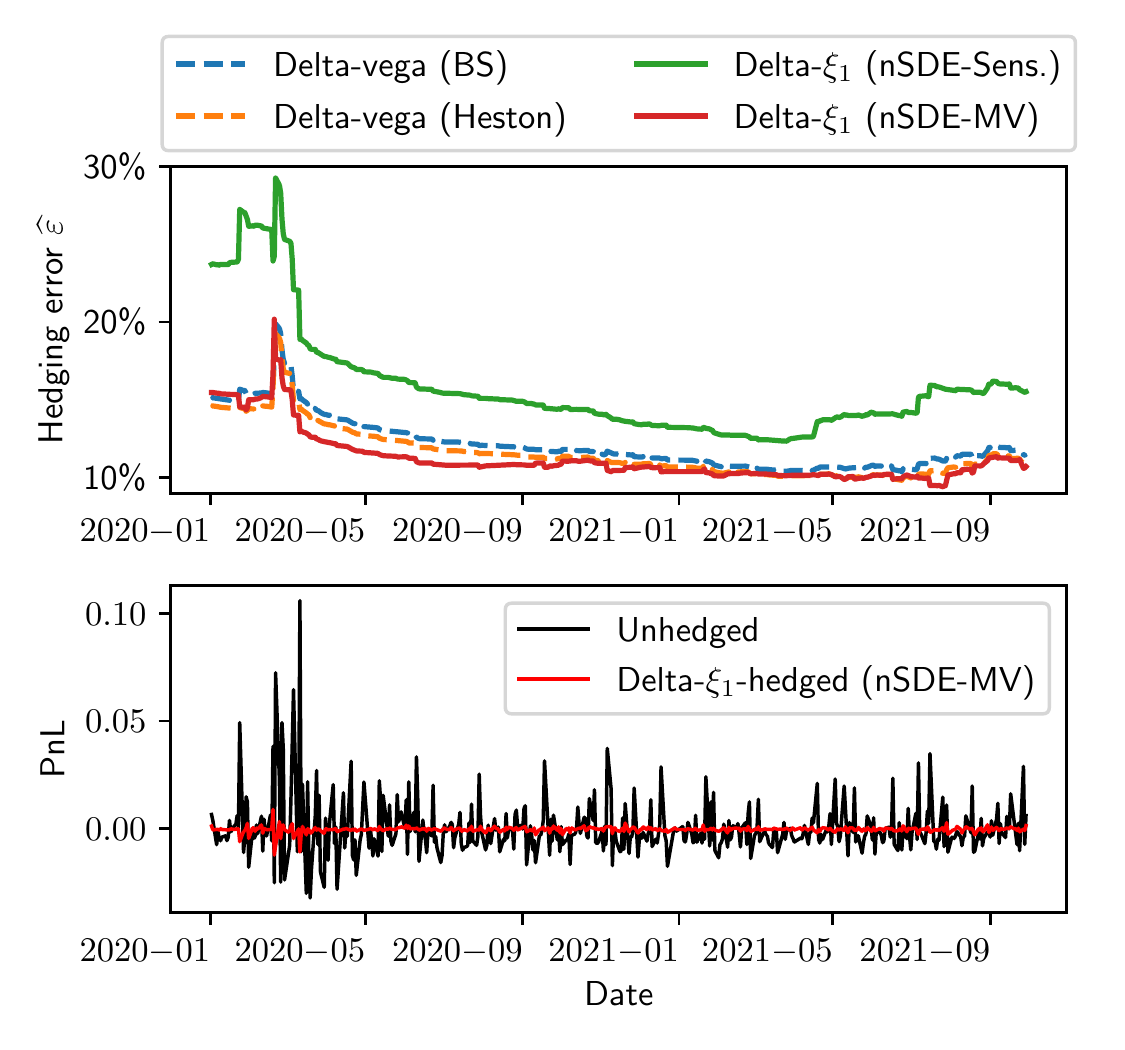}
    \caption{Daily rebalancing.}
    \label{fig:h_vix_errorts_daily}
    \end{subfigure}
    \hfill
    \begin{subfigure}[b]{.49\textwidth}
    \centering
        \includegraphics[scale=\figscaleapp]{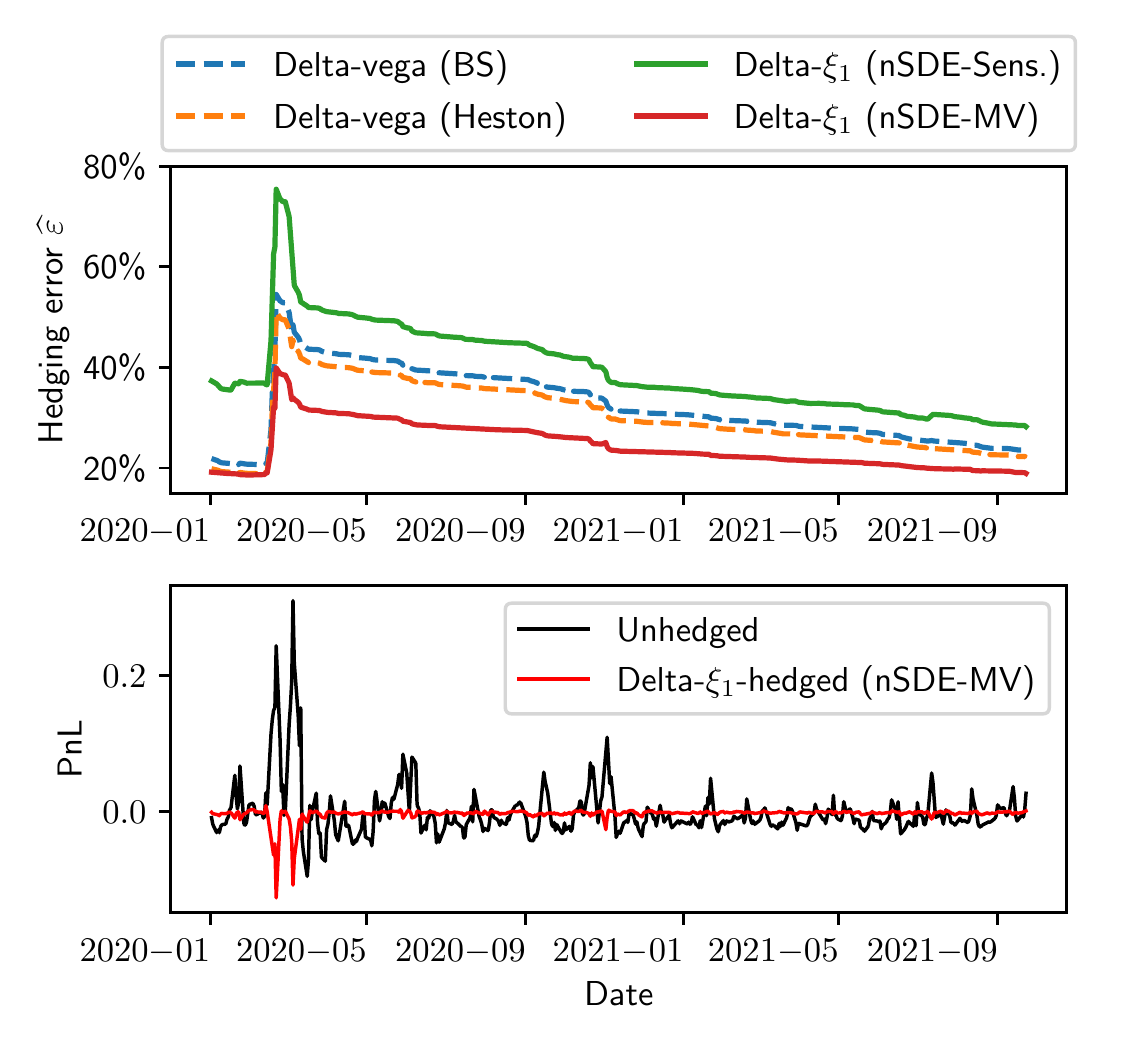}
    \caption{Weekly rebalancing.}
    \label{fig:h_vix_errorts_weekly}
    \end{subfigure}
    \caption{\textit{Top} - EWMA hedging errors $\widehat{\varepsilon}_t(\Delta t, \lambda=0.99)$ for the four delta hedging strategies. \textit{Bottom} - time series of the PnLs for the unhedged VIX portfolio and for the nSDE MV-based delta-$\xi_1$-hedged VIX portfolio.}
    \label{fig:h_vix_errorts}
\end{figure}

We report the relative EWMA hedging errors $\widehat{\varepsilon}_t(\Delta t, \lambda=0.99)$ of the four hedging strategies using 1M ATM options as the hedge (that are rebalanced on a daily basis and on a weekly basis respectively) in Figure \ref{fig:h_vix_errorts_daily} and \ref{fig:h_vix_errorts_weekly}. For better illustrating the dynamic changes in the hedging errors, we append the time series of the PnLs for the unhedged VIX portfolio and for the MV-based delta-$\xi_1$-hedged VIX portfolio at the bottom. MV-based delta-$\xi_1$ hedging consistently improves sensitivity-based delta-$\xi_1$ hedging over the course of the testing period, where the largest improvement amounts to about 20\%. MV-based delta-$\xi_1$ hedging is also consistently better than BS and Heston delta-vega hedging most of time, while the advantage is more apparent in the weekly rebalancing case.

\section{Conclusions}
\label{sec:conclusion}

We derive sensitivity-based and MV-based hedging strategies in neural-SDE market models
that satisfy generic systems of linear equations.
In particular, we focus on a two-factor market model and examine the performance of its derived (1) MV-based delta hedging and (2) single-option hedging strategies, called delta-$\xi_1$ hedging, when applied to various portfolios of EURO STOXX 50 index options over typical and stressed market periods. We compare market model delta(-$\xi_1$) hedging with Black--Scholes and Heston delta(-vega) hedging, along with considering the use of different hedging instruments and rebalancing frequencies. The empirical results address the following three questions:
\begin{enumerate}[leftmargin=*, label=(\roman*)]
	\item \textit{Does MV-based delta hedging improve sensitivity-based delta hedging? In addition, since MV-based hedge ratios rely on a model-specific correlation between $S$ and other modelled risk factors, do market models predict more accurate correlations than Heston models?}
	The statistics displayed in Table \ref{tab:hedge_ratios_delta} and Figure \ref{fig:h_dh_errorts} give positive answers to both questions. In fact, the observation that MV hedge ratios provide better hedges on average is consistent with empirical findings in the literature \cite{alexander2007model, alexander2012does}. We find that neural-SDE market models yield particularly lower hedging errors during stressed market periods, indicating their capability of better predicting correlation between $S$ and other modelled risk factors.
    \item \textit{Does market model delta-$\xi_1$ hedging outperform Black--Scholes and Heston delta-vega hedging?} We have shown that the hedging performance of a strategy highly depends on the choice of hedging instruments, and therefore we answer this question as follows:
    \begin{enumerate}
        \item BS delta-vega hedging seems to be more sensitive to the choice of the hedging instrument, in particular its tenor, than Heston delta-vega hedging and delta-$\xi_1$ hedging, both when hedging the naive portfolio and the VIX portfolio. Traders might favour less sensitive hedging strategies as they will have more flexibility in choosing the hedging instrument and can better accommodate liquidity and other practical restrictions.
        \item When optimal hedges are chosen, delta-$\xi_1$ hedging is shown to outperform BS delta-vega hedging consistently over time and at different rebalancing frequencies, for the naive and the VIX portfolios. In particular, delta-$\xi_1$ hedging leads to the largest improvements for hedging long-dated and OTM options. However, there is no significant difference in their hedging performance when hedging delta spread and delta butterfly portfolios.
        \item Heston delta-vega hedging shares many similarities with MV-based delta-$\xi_1$ hedging, in terms of the sensitivity of hedging errors with respect to hedging options (Figure \ref{fig:h_tenor}) and hedging error dynamics (Figure \ref{fig:h_errorts} and \ref{fig:h_vix_errorts}). Overall, MV-based delta-$\xi_1$ hedging outperforms Heston delta-vega hedging for the naive and VIX portfolio. However, the analysis on outright options suggests that Heston delta-vega hedging is better suited for short-dated options.
    \end{enumerate}
    
    \item \textit{In the market model, does the dependence between factors that is estimated by the neural-SDE improve delta-$\xi_1$ hedging?
    } There is not a definite answer to this question, as it depends on which hedging instrument is used. From the evidence in the study of the naive and VIX portfolios, MV-based delta-$\xi_1$ hedging can improve on the sensitivity-based counterpart when short-dated options are used as the hedge. Specifically, it is worth highlighting that when hedging the VIX using 1M options, MV-based delta-$\xi_1$ hedging can improve sensitivity-based delta-$\xi_1$ hedging by as much as 20\%, and BS and Heston delta-vega hedging by 10\%.
\end{enumerate}

Overall, the hedging performance of ``PCA-factor'' hedging is impressive: though $\xi_1$ is a factor decoded statistically from historical option prices using a simple, model-free PCA-based method, hedging its exposure is at least as good as BS vega hedging and comparable with Heston vega hedging. Meanwhile, since MV-based delta-$\xi_1$ hedging is not necessarily better than the model-free sensitivity-based counterpart, in some cases it may not be necessary to establish and estimate a model for factor dynamics at all.

\appendix

\section{The trained neural-SDE market model}
\label{apd:nsde_mdl}

We establish a neural-SDE model for the joint dynamics of $(\ln S, \xi_1, \xi_2)$ and derive hedge ratios based on this model. In this section, we demonstrate the goodness-of-fit of this model and its capability as a realistic market simulator.

\begin{figure}[!ht]
    \centering
    \includegraphics[scale=\figscaleapp]{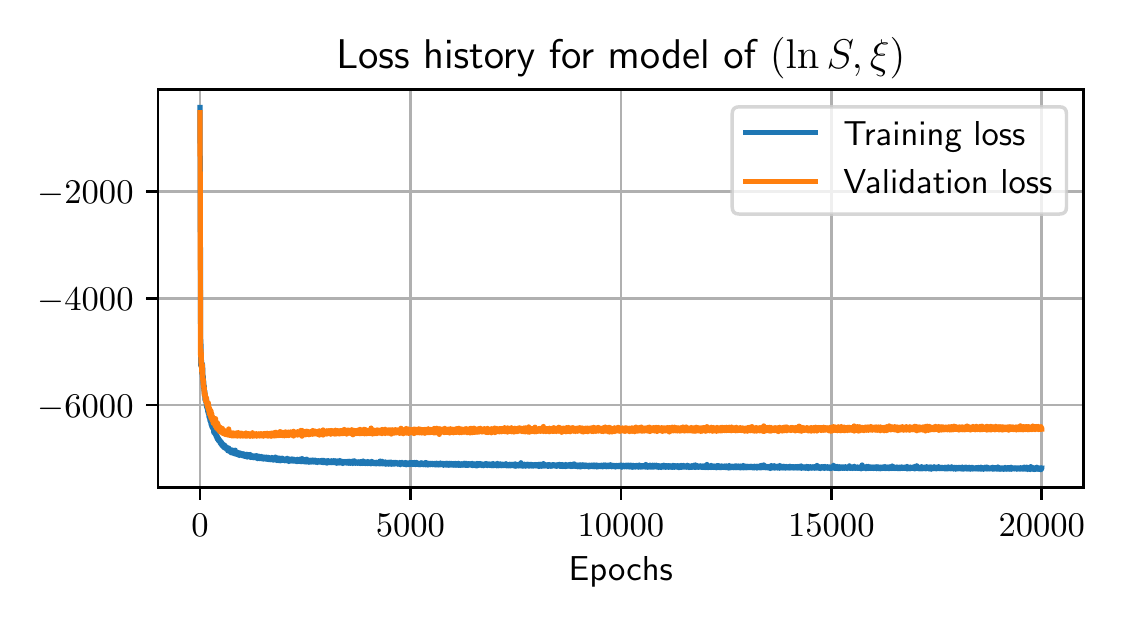}
    \caption{Evolution of training losses and validation losses for the joint model of $\ln S$ and $\xi$.}
    \label{fig:loss_hist_xiS}
\end{figure}

In Figure \ref{fig:loss_hist_xiS}, we show the evolution of training losses and validation losses over epochs during the training of the model. Both the training and validation losses decline rapidly over the first 1000 epochs and then converge gradually.

\begin{figure}[!ht]
    \centering
    \includegraphics[scale=\figscaleapp]{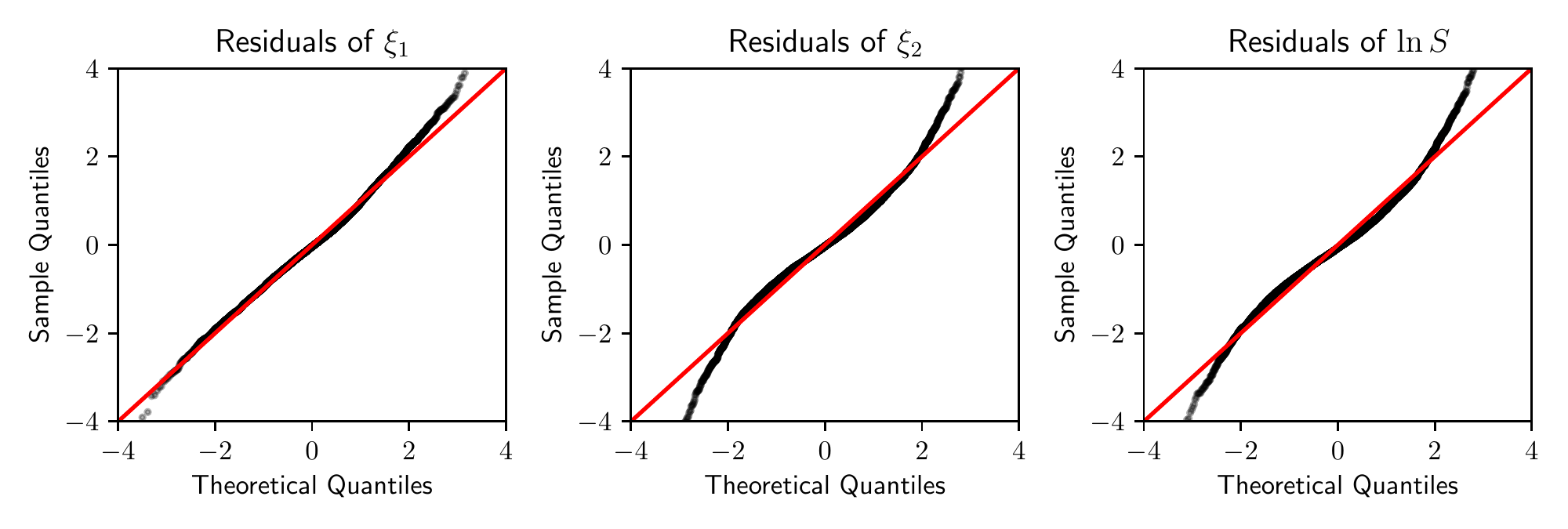}
    \caption{QQ-plots of historical in-sample model residuals.}
    \label{fig:h_hist_res_marginaldensity_xiS}
\end{figure}

The loss function is constructed based on the likelihood of observing historical factor paths, subject to arbitrage constraints. Our approximate likelihood corresponds to an assumption that the model residuals are independent standard normal white noise. Therefore, it is worth checking if the model residuals appear to be independent standard normals, within our training data. In Figure \ref{fig:h_hist_res_marginaldensity_xiS}, we show the normal QQ-plots of the empirical marginal distributions for the residuals. For the underlying stock indices and $\xi_2$, their residuals show mild fat tails. For $\xi_1$, its residuals appear slightly right-skewed, implying higher-than-expected probabilities for large volatilities.

\begin{figure}[!ht]
    \centering
    \includegraphics[scale=\figscaleapp]{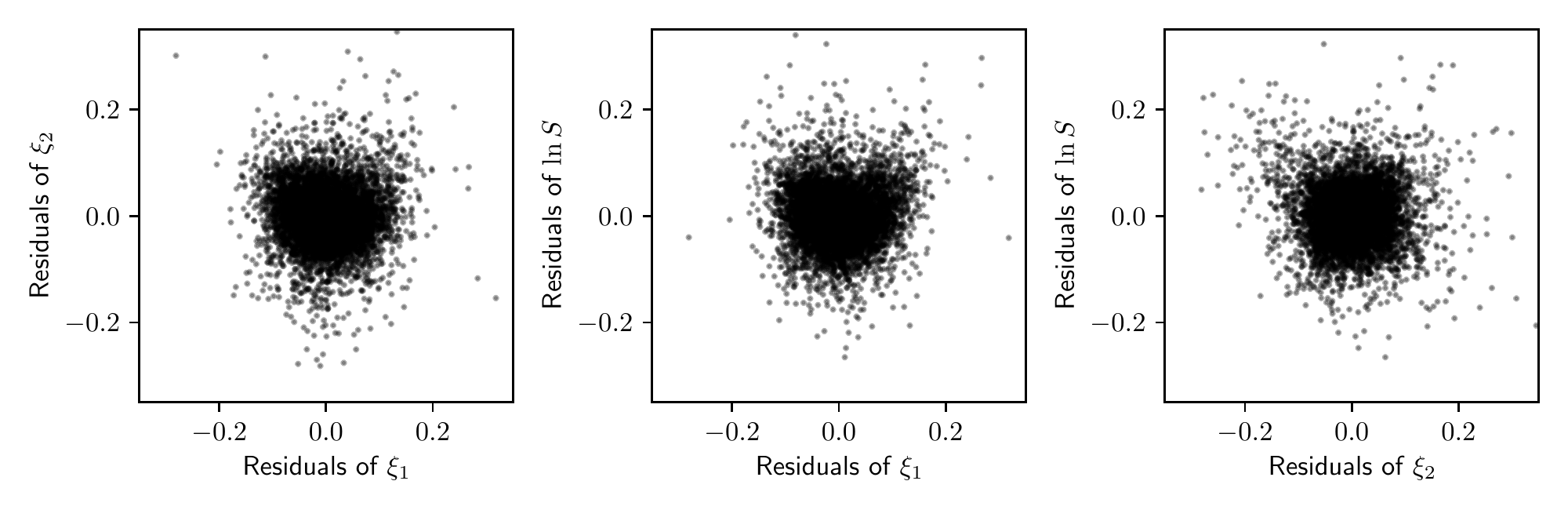}
    \caption{Scattter plots of historical in-sample model residuals.}
    \label{fig:hist_res_jointdensity_xiS}
\end{figure}

To investigate dependence between the model residuals, we show in Figure \ref{fig:hist_res_jointdensity_xiS} the scatter plots of each pair of residuals. The residuals for each pair look fairly uncorrelated, indicating that the trained neural nets have well captured factor covariances over time, which is important for MV-based hedging under the neural-SDE market model.

\begin{figure}[!ht]
    \centering
    \includegraphics[scale=\figscaleapp]{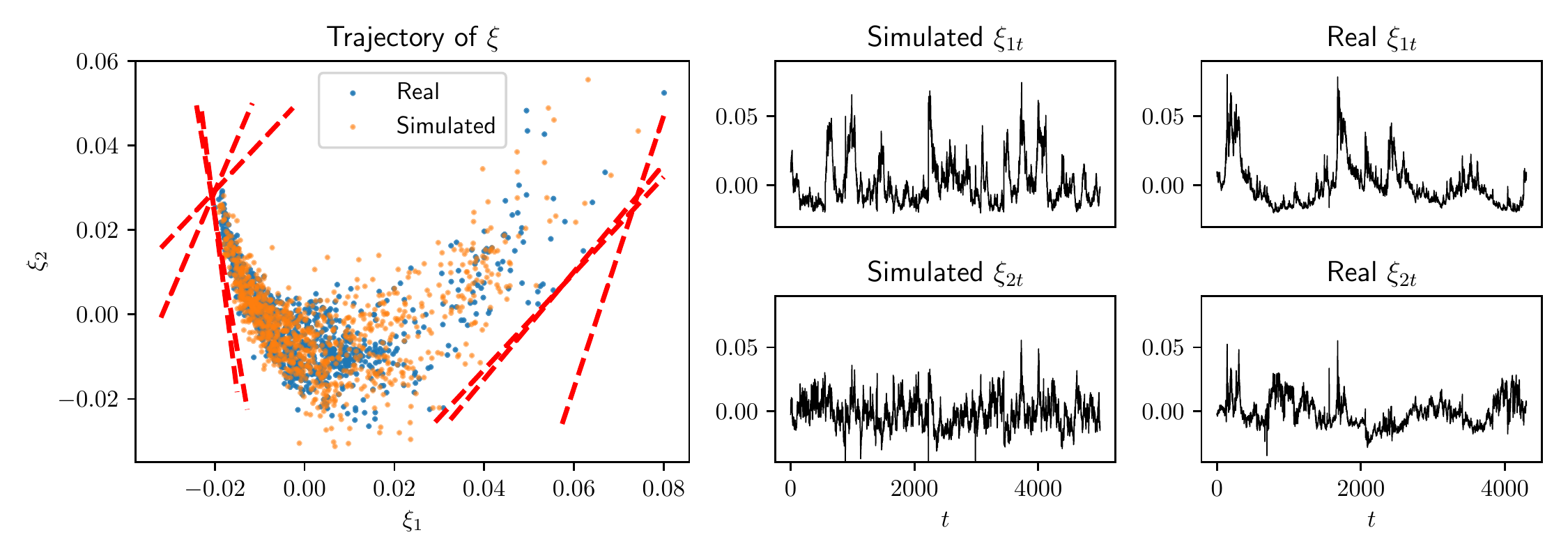}
    \caption{Simulation of $\xi = (\xi_1, \xi_2)$ from the learnt neural-SDE model, compared with the real data.}
    \label{fig:sim_factors_xiS}
\end{figure}

Finally, we assess the learnt model’s ability to simulate time series data that are like the input real data. We take the trained model and simulate sample paths using a tamed Euler scheme. Rather than generating the innovations from normal distributions, we randomly sample them from historical residuals of the trained models. There are two major advantages of drawing innovations from historical residuals rather than normal distributions. First, historical residuals have fatter tails than normal (see Figure \ref{fig:h_hist_res_marginaldensity_xiS}). Second, the joint historical residuals implicitly preserve the (potentially higher-order) dependence of factors that may fail to be captured by the covariance terms (see Figure \ref{fig:hist_res_jointdensity_xiS}). We show the time series of $\xi$ in Figure \ref{fig:sim_factors_xiS} (right). In the scatter plot on the left of Figure \ref{fig:sim_factors_xiS}, we see that the dependence structure between $\xi_1$ and $\xi_2$ is well captured. In addition, the simulated factors remain within the no-arbitrage region, due to the hard constraints imposed on the drift and diffusion functions. 

\section{Tested option portfolios for hedging analysis}
\label{apd:test_strategy}

To assess performance of a hedging strategy, we evaluate its hedging errors for a variety of representative trading strategies. In particular, defining a set of deltas $\mathcal{D} = \{0.2, 0.3, 0.4, 0.5, 0.6, 0.7, 0.8\}$ and a set of time-to-expiries $\mathcal{T} = \{30, 60, 91, 122, 152, 182, 273$, $365, 547, 730\} $ (in calendar days)\footnote{These portfolios are constructed from options with specified time-to-expiries and moneynesses (deltas), which are in general impossible to match perfectly with traded options. Therefore, we find the options that have closest time-to-expiries and moneynesses to the desired specifications when constructing portfolios.}, we consider:

\begin{enumerate}[label=(\arabic*)]
    \item \emph{Outright call} option $C(\tau, \Delta)$ with all deltas $\Delta \in \mathcal{D}$ and all time-to-expiries $\tau \in \mathcal{T}$. There are in total $7 \times 10 = 70$ outright test strategies.
    \item \emph{Delta spread} $C(\tau, \Delta_1) - C(\tau, \Delta_2)$ with all distinguished delta pairs $\Delta_1 > \Delta_2 \in \mathcal{D}$ and all time-to-expiries $\tau \in \mathcal{T}$. There are in total ${7 \choose 2} \times 10 = 210$ delta spread test strategies.
    \item \emph{Delta butterfly} $C(\tau, \Delta) + C(\tau, 1-\Delta) - 2 \times C(\tau, 0.5)$ with all deltas $\Delta \in \mathcal{D} \setminus \{0.5\}$ and all time-to-expiries $\tau \in \mathcal{T}$. There are in total $3 \times 10 = 30$ delta butterfly spread test strategies.
    \item \emph{Strangle} $C(\tau, \Delta) + C(\tau, 1 - \Delta)$ with all deltas $\Delta \in \mathcal{D} \setminus \{0.5\}$ and all time-to-expiries $\tau \in \mathcal{T}$. There are in total $3 \times 10 = 30$ strangle test strategies.
    \\Note that a typical straggle involves simultaneously buying a call and a put, i.e. $C(\tau, \Delta) + P(\tau, -\Delta)$. Here we use put-call parity to express such straddles in terms of only calls, ignoring the underlying and cash components.
    \item \emph{Calendar spread} $C(\tau_1, \Delta) - C(\tau_2, \Delta)$  with $\Delta = 0.5$ and all distinguished time-to-expiry pairs $\tau_1 > \tau_2 \in \mathcal{T}$. There are in total $1 \times {10 \choose 2} = 45$ calendar spread test strategies.
    \item \emph{VIX}, the square of which is a linear combination of OTM call and put option prices, and can be further written as a linear combination of call prices only, provided that put-call parity holds under no-arbitrage; see details in \cite{vix}.
\end{enumerate}

\section{Heston model calibration results}
\label{apd:heston_calib}

\begin{figure}[!ht]
    \centering
    \includegraphics[scale=.66]{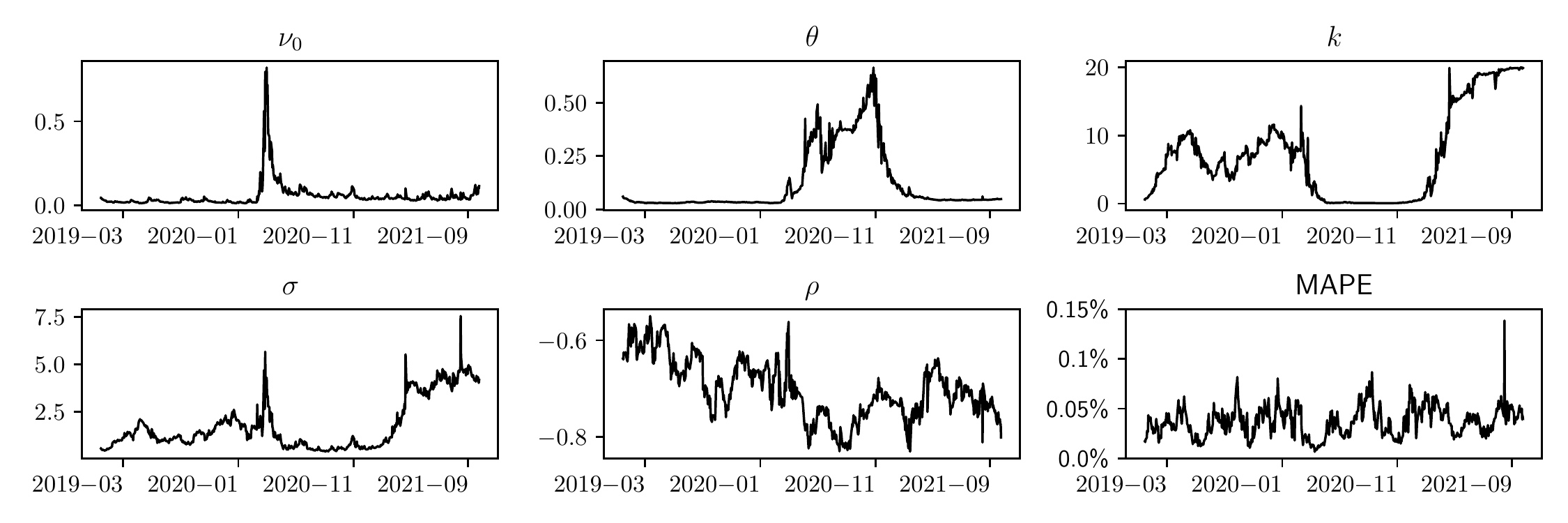}
    \caption{Heston model calibrated parameters and calibration errors (measured as mean absolute percentage error) for historical price data of EURO STOXX 50 index options.}
    \label{fig:heston_calib}
\end{figure}

Given the Heston model specified by the SDE \eqref{eq:heston_sde}, we characterise option price surfaces by the parameters $\Theta = (S_0, \nu_0, \theta, k, \sigma, \rho)$. On each testing date $t$, we calibrate parameters by solving
\begin{equation*}
    \widehat{\Theta}_t = \underset{\Theta}{\arg\min} \sum_{j=1}^N \frac{1}{\mathcal{V}_{\textrm{bs},t}(\tau_j, m_j)} \left( c^\Theta (\tau_j, m_j) - \tilde{c}_t (\tau_j, m_j) \right)^2,
\end{equation*}
where $\mathcal{V}_\textrm{bs}$ is the Black--Scholes option vega calculated through \eqref{eq:bs_greeks}. In Figure \ref{fig:heston_calib}, we show the time series of the calibrated parameters and calibration errors, measured by MAPE, for historical price data of EURO STOXX 50 index options during the testing period. Since $\nu_0$, $\theta$, $k$ and $\sigma$ are positive by construction, we include the positivity constraint for these parameters during the calibration. In addition, we constrain the correlation parameter $\rho$ between $-1$ and $1$. Over time, we observe less than $0.15\%$ calibration errors measured by MAPE (and typically below 0.05\%).

\section{Analysis of decoded factors, COVID-19 and the 2008 GFC}
\label{apd:analysis_factors}

\begin{figure}[!ht]
    \centering
    \begin{subfigure}[b]{.49\textwidth}
    \centering
        \includegraphics[scale=\figscaleapp]{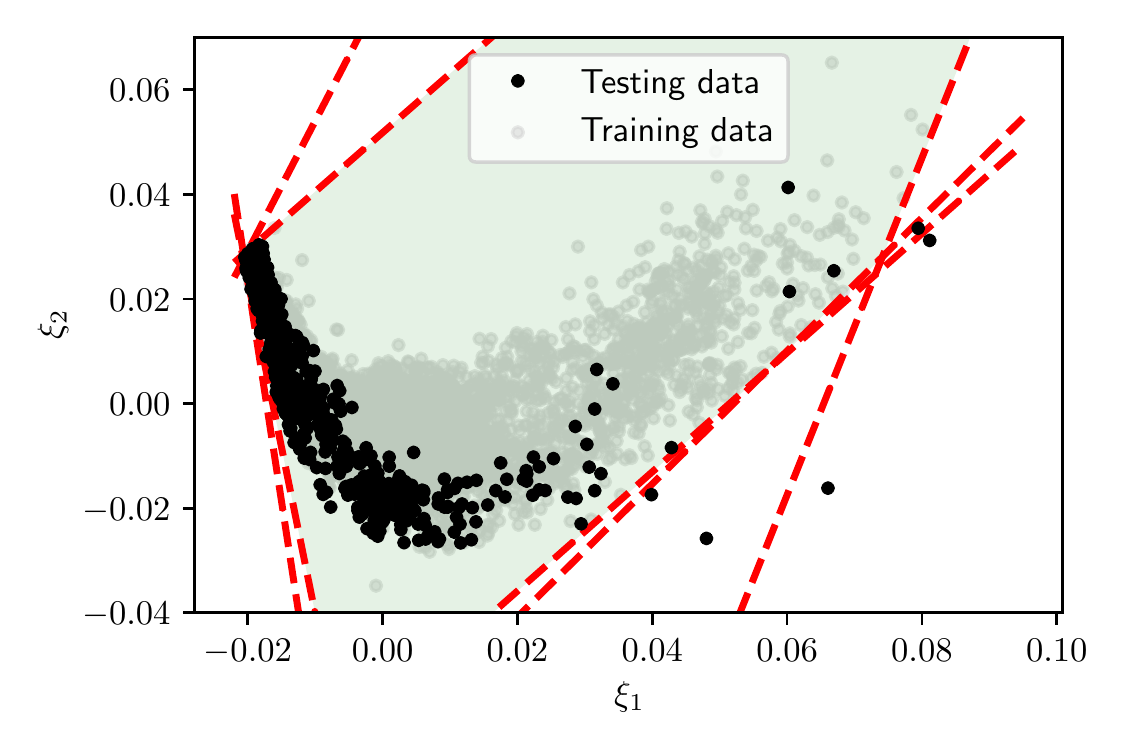}
    \end{subfigure}
    \hfill
    \begin{subfigure}[b]{.49\textwidth}
    \centering
        \includegraphics[scale=\figscaleapp]{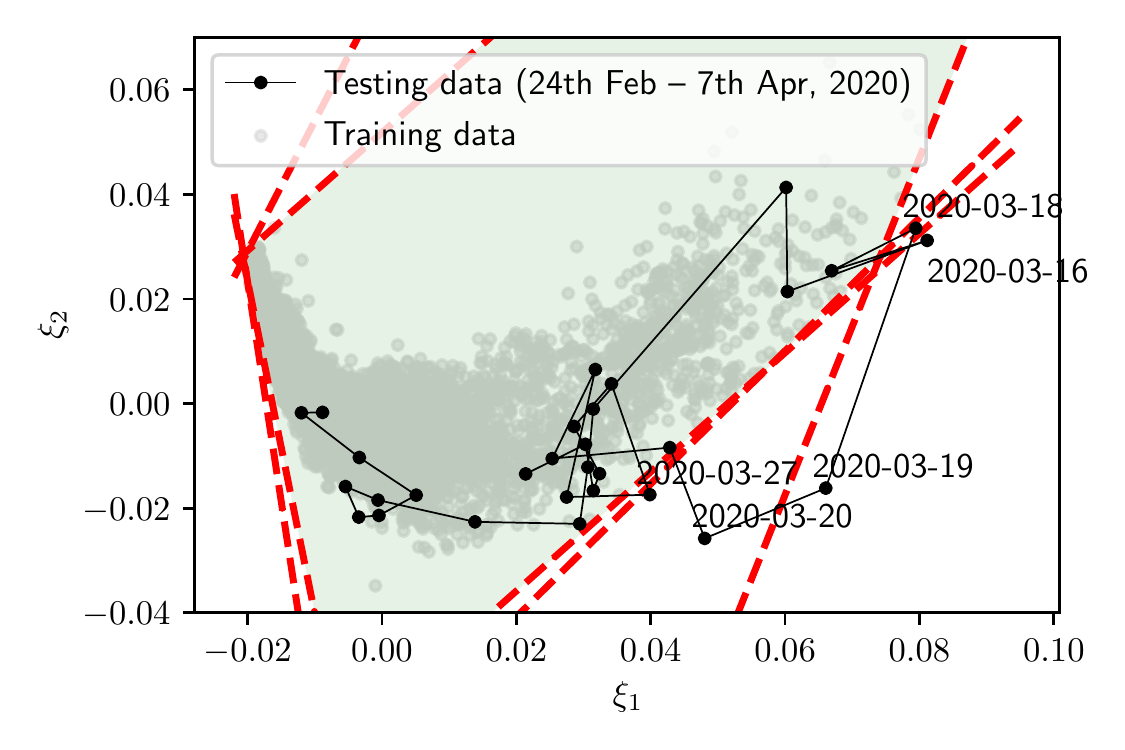}
    \end{subfigure}
    \begin{subfigure}[b]{.49\textwidth}
    \centering
        \includegraphics[scale=\figscaleapp]{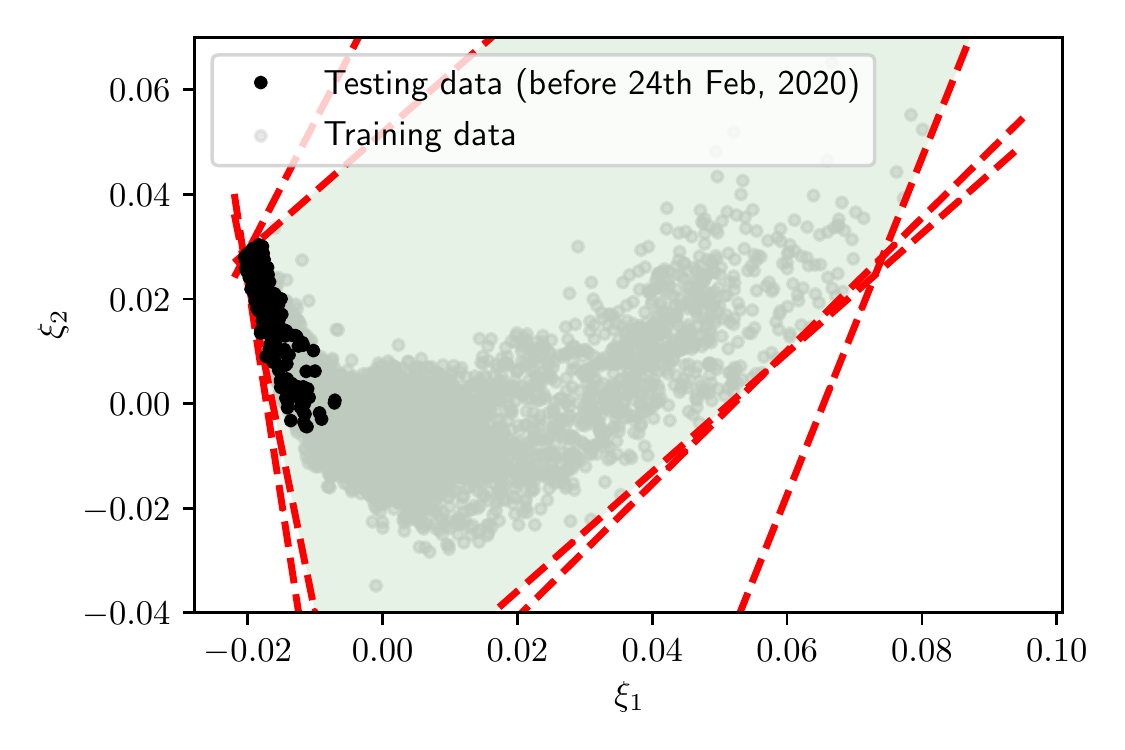}
    \end{subfigure}
    \hfill
    \begin{subfigure}[b]{.49\textwidth}
    \centering
        \includegraphics[scale=\figscaleapp]{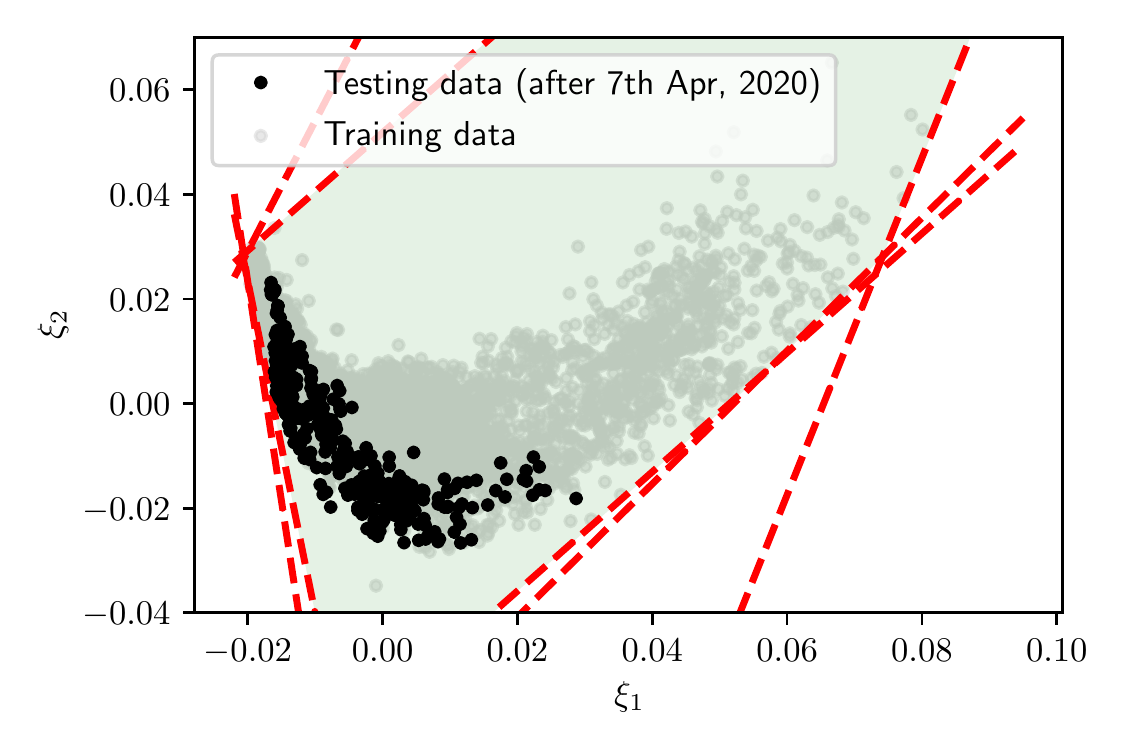}
    \end{subfigure}
    \caption{Scatter plots of the decoded factors $(\xi_1, \xi_2)$ for the testing data (from 2nd January 2019 to 30th September 2021). The decoded factors for the training data are shown as light grey dots, and the corresponding static arbitrage constraints (projected to the $\mathbb{R}^2$ factor space) are given as red dashed lines. While the top left plot shows factors of all testing data, the other plots show factors of testing data within the specified periods.}
    \label{fig:factors_real_test}
\end{figure}

Since the diffusion matrix of a neural-SDE is a (smooth) function of $(\xi_1, \xi_2)$, we now examine if $(\xi_1, \xi_2)$ behaves differently before and after March 2020. We show the trajectory of the decoded factors $\xi_1$ and $\xi_2$ for the testing data in Figure \ref{fig:factors_real_test} by black dots. The top left plot shows the factors for all the testing data (from 2nd January 2019 to 30th September 2021).

\begin{figure}[!ht]
    \begin{subfigure}[b]{.49\textwidth}
    \centering
        \includegraphics[scale=\figscaleapp]{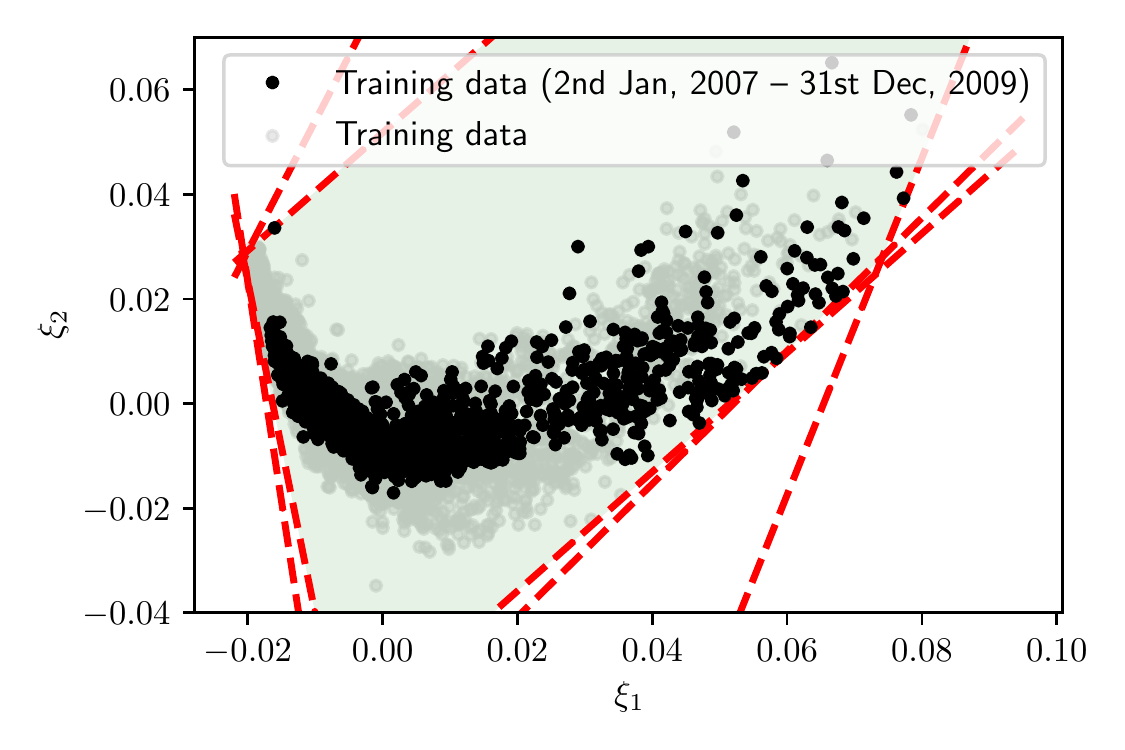}
    \end{subfigure}
    \hfill
    \begin{subfigure}[b]{.49\textwidth}
    \centering
        \includegraphics[scale=\figscaleapp]{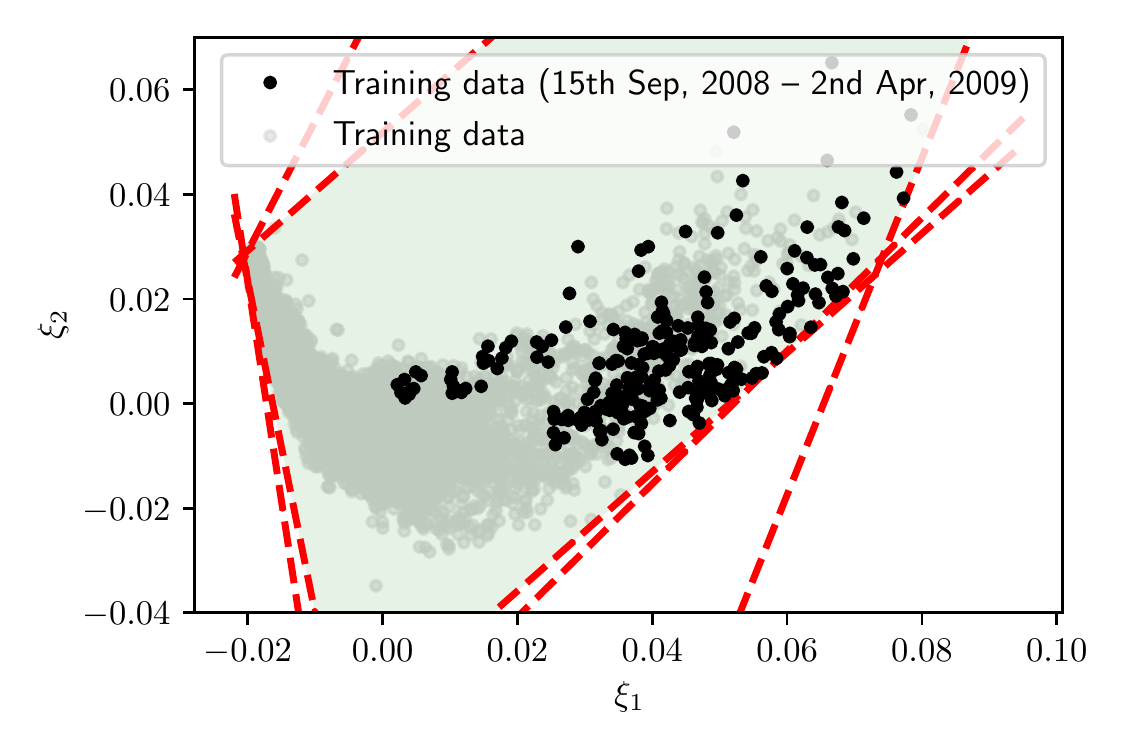}
    \end{subfigure}
    \caption{Trajectories of the decoded factors $(\xi_1, \xi_2)$ for the option price data from 2007 to 2009.}
    \label{fig:h_2008}
\end{figure}

From 24th February to 7th April in 2020, equity markets across the world suffered from a series of crashes due to panic selling following the onset of the COVID-19 pandemic. In particular, on the Black Thursday (12th March), EURO STOXX 50 and DAX closed down 12\% and 11\% respectively. In the US market, on 16th March, the sudden and dramatic drop in stock prices triggered multiple trading halts (also known as ``circuit breakers'') in a single day. We divide the testing period into three sub-periods that are before, between and after the market crashes. Accordingly, we show the trajectories of $(\xi_1, \xi_2)$ during the three sub-periods in the other subplots. During the market crashes (24th Feb -- 7th April), as seen in the top right plot, the trajectory of $(\xi_1, \xi_2)$ spreads over a wide range of the conventional area (light grey dots) where the training data usually reside. In particular, we mark on the plot the corresponding dates of the five data points that are out side of the arbitrage-free region\footnote{This does not indicate that arbitrage was present in the market on these dates, but rather that our two-factor model fails to represent prices accurately, and so the best fitting factor values do not represent the prices in an arbitrage-free way. This was not observed in our training data (which includes the 2007-8 financial crisis, see Figure \ref{fig:h_2008}).} (confined by the red dashed line and highlighted in green). Note that the 16th of March is among these dates.

The most visited areas of the factors seem to be distinct \emph{before} and \emph{after} the start of the COVID-19 pandemic. Specifically, before the pandemic, as seen in the bottom left plot, factors tended to cluster and have small variations in $\xi_1$. In contrast, after the pandemic started, factors tended to spread out with much larger variations in $\xi_1$, as seen in the top and bottom right plots. Furthermore, comparing with the diffusion matrices visualised by ellipses in Figure \ref{fig:learnt_drift_diffusion}, factors before the pandemic have more ill-conditioned diffusion matrices (i.e. with flatter ellipses), implying that the instantaneous correlation between $\xi_1$ and $\xi_2$ is stable and close to $-1$. Hence, since factors change more drastically post-pandemic, the neural-SDE model fails to give good forecasts of the average factor correlation over weekly rebalancing horizons. We conjecture this to be a contributing factor why MV-based hedging becomes worse than sensitivity-based hedging after the outbreak of the pandemic.

\begin{figure}[!ht]
    \centering
    \begin{subfigure}[b]{.49\textwidth}
    \centering
        \includegraphics[scale=\figscaleapp]{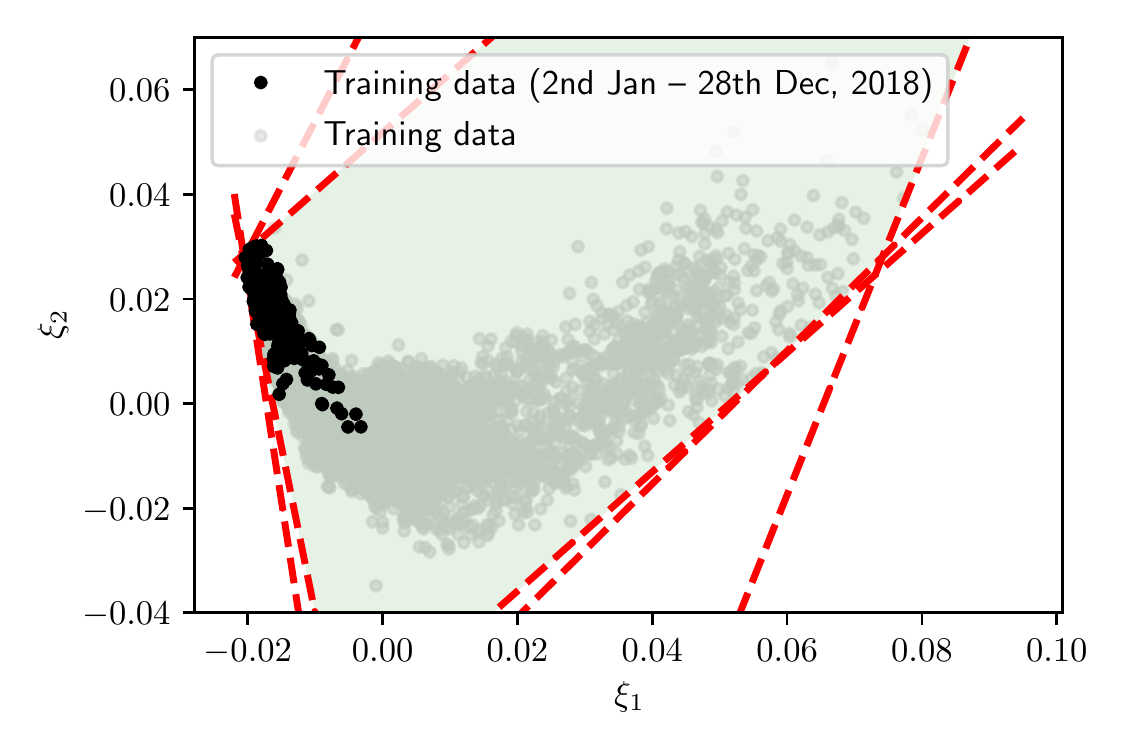}
    \end{subfigure}
    \hfill
    \begin{subfigure}[b]{.49\textwidth}
    \centering
        \includegraphics[scale=\figscaleapp]{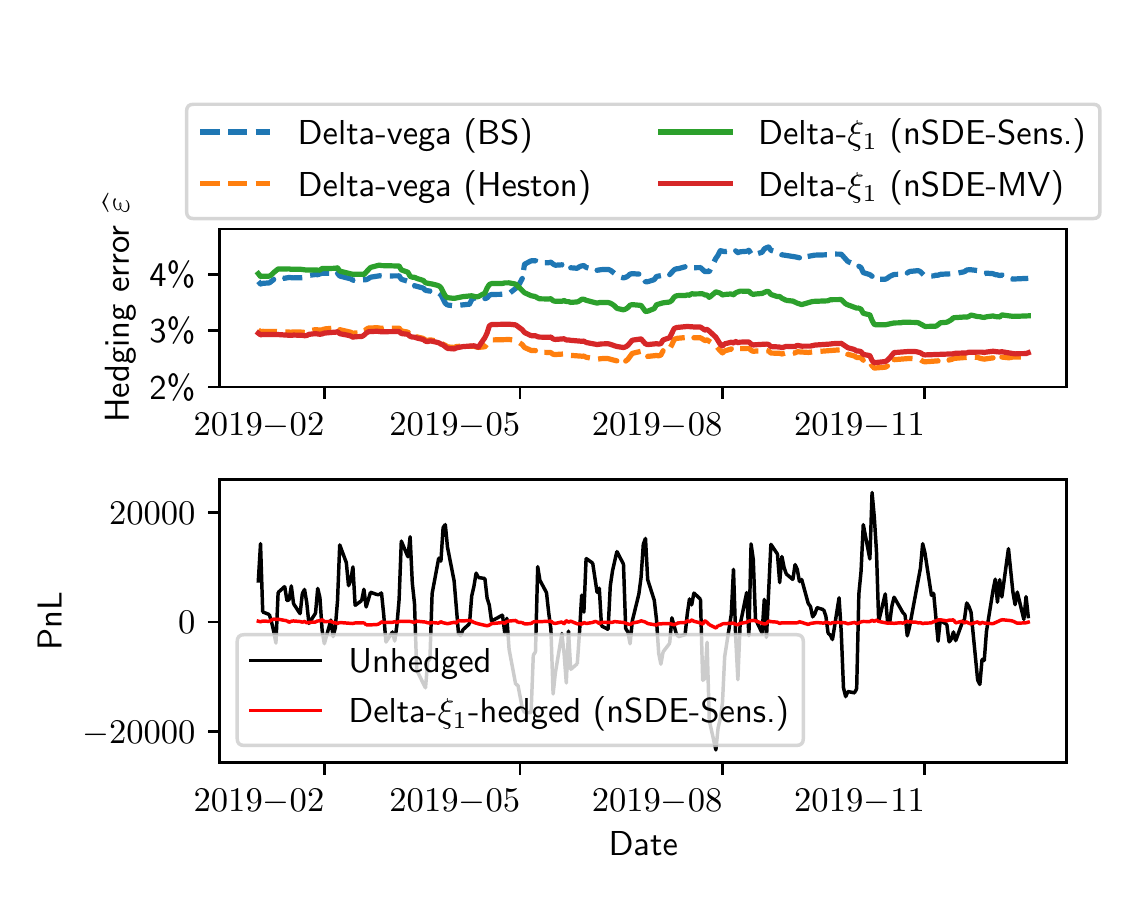}
    \end{subfigure}
    \caption{\textit{Left} - trajectories of the decoded factors $(\xi_1, \xi_2)$ for the price data in 2018. \textit{Right} - relative EWMA hedging errors $\widehat{\varepsilon}_t (\Delta t = 5$ trading days, $\lambda=0.99)$ of the four hedging strategies and time series of the PnLs for the unhedged naive portfolio and for the sensitivity-based delta-$\xi_1$-hedged naive portfolio.}
    \label{fig:h_2018}
\end{figure}

To support the above arguments with more data, we evaluate the relative EWMA hedging errors $\widehat{\varepsilon}_t(\Delta t, \lambda=0.99)$ back to the year ending January 2019. Note that all delta-$\xi_1$ hedge ratios computed before 2019 are based on in-sample estimation of price basis functions $G$ and the factor diffusion matrix $\sigma$. In Figure \ref{fig:h_2018}, we see that factors in 2018 cluster around similar regions to factors in 2019 and early 2020 before the pandemic. In addition, since the instantaneous correlations estimated within the regions are stable, the neural-SDE model tends to produce reasonable forecasts of the average factor correlation over weekly rebalancing horizons. The corresponding MV-based delta-$\xi_1$ hedge yields lower hedging errors than sensitivity-based delta-$\xi_1$ hedging, as seen on the right plot of Figure \ref{fig:h_2018}.

\vspace*{-.2cm}

\paragraph*{The 2008 Global Financial Crisis.}

Finally, as a comparative analysis, we investigate neural-SDE market model hedging performance during the 2008 Global Financial Crisis, a similar stressed market period caused by the US subprime crisis. 

First, in Figure \ref{fig:h_2008} on the left, we show the trajectory of the decoded factors $\xi_1$ and $\xi_2$ for option price data in 2007, 2008 and 2009. In particular, on the right we highlight the trajectory between 15th September 2008 (when Lehman Brothers filed bankruptcy) and 2nd April 2009 (when G-20 finance ministers announced a coordinated response to the crisis), when global equity markets were hit most severely. We observe that the trajectory spreads over a wide range during 2007--2009, and both $\xi_1$ and $\xi_2$ tend to be unusually large during the most stressed period.

\begin{figure}[!ht]
    \centering
    \begin{subfigure}[b]{.49\textwidth}
    \centering
        \includegraphics[scale=\figscaleapp]{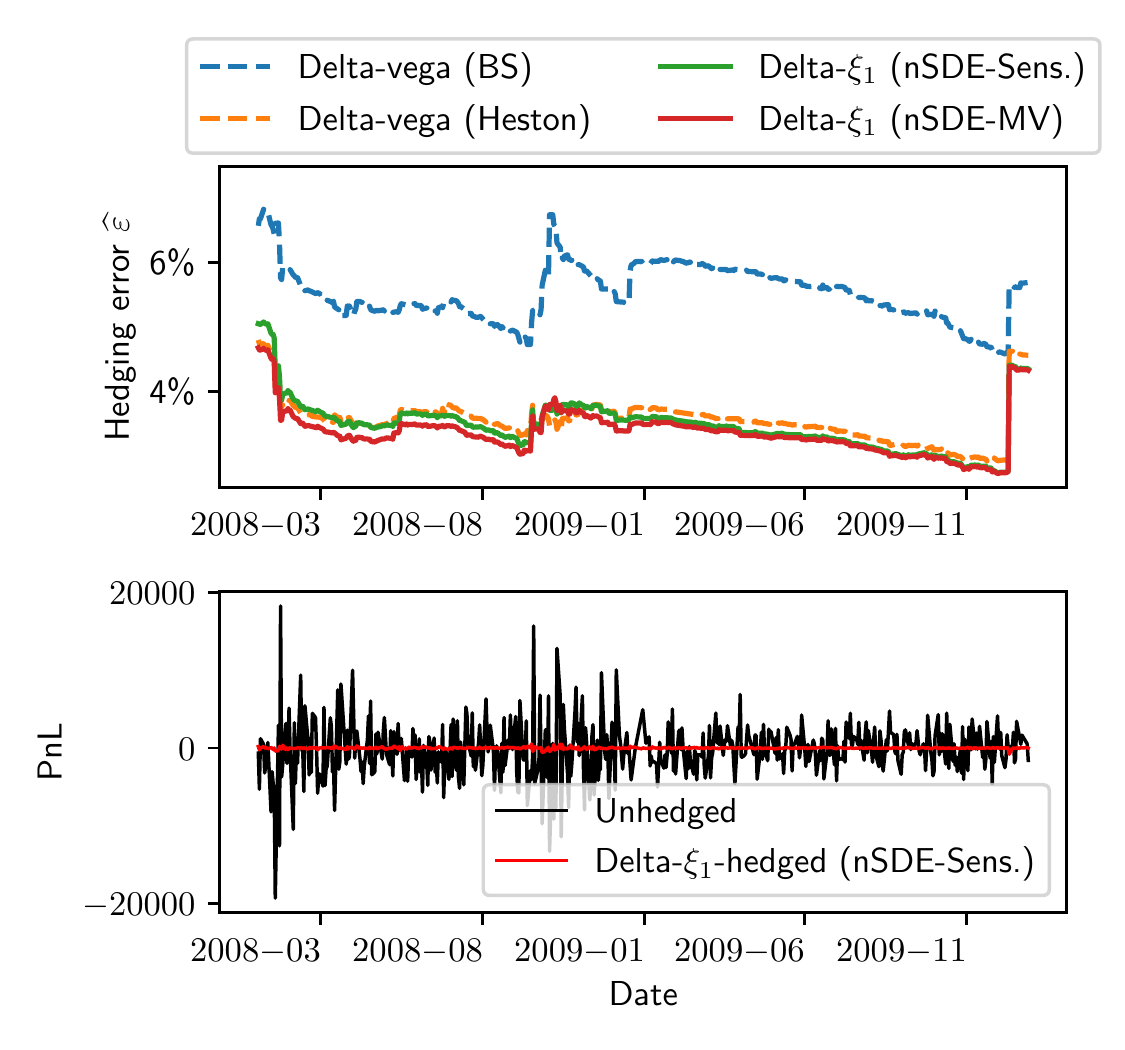}
    \caption{Daily rebalancing.}
    \label{fig:h_errorts_daily_2008}
    \end{subfigure}
    \hfill
    \begin{subfigure}[b]{.49\textwidth}
    \centering
        \includegraphics[scale=\figscaleapp]{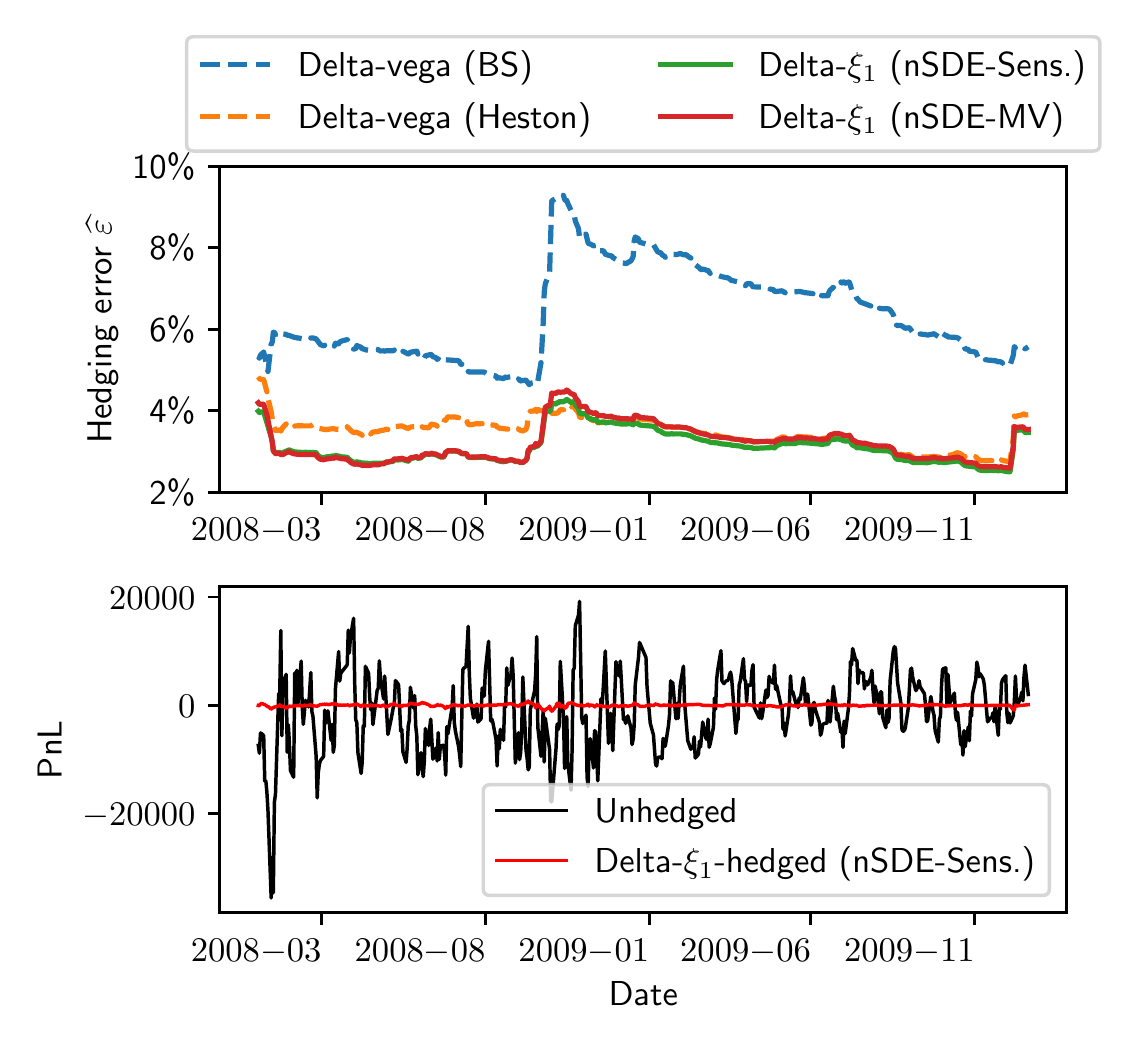}
    \caption{Weekly rebalancing.}
    \label{fig:h_errorts_weekly_2008}
    \end{subfigure}
    \caption{\textit{Top} - EWMA hedging errors $\widehat{\varepsilon}_t(\Delta t, \lambda=0.99)$ for the four hedging strategies. \textit{Bottom} - time series of the PnLs for the unhedged naive portfolio and for the nSDE sensitivity-based delta-$\xi_1$-hedged naive portfolio.}
    \label{fig:h_errorts_2008}
\end{figure}

Next, we examine the performance of hedging the naive portfolio during the crisis using the four hedging strategies. Note that all delta-$\xi_1$ hedge ratios computed here are based on in-sample estimation of the price basis functions $G$ and the factor diffusion matrix $\sigma$. We report the relative EWMA hedging errors $\widehat{\varepsilon}_t(\Delta t, \lambda=0.99)$ of the four hedging strategies using 7M ATM options as the hedge (that are rebalanced on a daily basis and on a weekly basis respectively) in Figure \ref{fig:h_errorts_daily_2008} and \ref{fig:h_errorts_weekly_2008}.

Heston and neural-SDE market models produce obviously better performing hedging strategies than Black--Scholes model, in both daily and weekly rebalancing cases. Over time, Heston delta-vega hedging yields very similar hedging errors to delta-$\xi_1$ hedging. We observe that MV-based delta-$\xi_1$ hedging becomes worse than sensitivity-based hedging after the outbreak of the COVID-19 pandemic in the 2019--2021 testing data. A similar phenomenon is observed after the 2008 financial crisis was triggered. In the daily rebalancing case, MV-based delta-$\xi_1$ hedging is consistently the best over time. Nevertheless, in the  weekly rebalancing case, sensitivity-based delta-$\xi_1$ overtakes its MV-based counterpart after September 2008, though the hedging error reduction is marginal.

\section*{Acknowledgements}
This publication is based on work supported by the EPSRC Centre for Doctoral Training in Industrially Focused Mathematical Modelling (EP/L015803/1) in collaboration with CME Group.

Samuel Cohen and Christoph Reisinger acknowledge the support of the Oxford-Man Institute for Quantitative Finance, and Samuel Cohen also acknowledges the support of the UKRI Prosperity Partnership Scheme (FAIR) under the EPSRC Grant EP/V056883/1, and the Alan Turing Institute.

\setlength{\bibsep}{0.0pt}

\bibliographystyle{abbrv}
\bibliography{reference}

\end{document}